\documentclass[prb,twocolumn,showpacs,preprintnumbers,amsmath,amssymb]{revtex4}
\usepackage{graphicx,hyperref}
\usepackage{bm}
\usepackage{subfigure}
\usepackage{color}

\def\nbC{{\mathchoice {\setbox0=\hbox{$\displaystyle\rm C$}%
\hbox{\hbox to0pt{\kern0.4\wd0\vrule height0.9\ht0\hss}\box0}}
{\setbox0=\hbox{$\textstyle\rm C$}\hbox{\hbox
to0pt{\kern0.4\wd0\vrule height0.9\ht0\hss}\box0}}
{\setbox0=\hbox{$\scriptstyle\rm C$}\hbox{\hbox
to0pt{\kern0.4\wd0\vrule height0.9\ht0\hss}\box0}}
{\setbox0=\hbox{$\scriptscriptstyle\rm C$}\hbox{\hbox
to0pt{\kern0.4\wd0\vrule height0.9\ht0\hss}\box0}}}}
\def\nbQ{{\mathchoice {\setbox0=\hbox{$\displaystyle\rm
Q$}\hbox{\raise
        0.15\ht0\hbox to0pt{\kern0.4\wd0\vrule height0.8\ht0\hss}\box0}}
{\setbox0=\hbox{$\textstyle\rm Q$}\hbox{\raise
0.15\ht0\hbox to0pt{\kern0.4\wd0\vrule height0.8\ht0\hss}\box0}}
{\setbox0=\hbox{$\scriptstyle\rm Q$}\hbox{\raise
0.15\ht0\hbox to0pt{\kern0.4\wd0\vrule height0.7\ht0\hss}\box0}}
{\setbox0=\hbox{$\scriptscriptstyle\rm Q$}\hbox{\raise
0.15\ht0\hbox to0pt{\kern0.4\wd0\vrule height0.7\ht0\hss}\box0}}}}
\def\nbT{{\mathchoice {\setbox0=\hbox{$\displaystyle\rm
T$}\hbox{\hbox to0pt{\kern0.3\wd0\vrule height0.9\ht0\hss}\box0}}
{\setbox0=\hbox{$\textstyle\rm T$}\hbox{\hbox
to0pt{\kern0.3\wd0\vrule height0.9\ht0\hss}\box0}}
{\setbox0=\hbox{$\scriptstyle\rm T$}\hbox{\hbox
to0pt{\kern0.3\wd0\vrule height0.9\ht0\hss}\box0}}
{\setbox0=\hbox{$\scriptscriptstyle\rm T$}\hbox{\hbox
to0pt{\kern0.3\wd0\vrule height0.9\ht0\hss}\box0}}}}
\def\nbS{{\mathchoice
{\setbox0=\hbox{$\displaystyle     \rm S$}\hbox{\raise0.5\ht0%
\hbox to0pt{\kern0.35\wd0\vrule height0.45\ht0\hss}\hbox
to0pt{\kern0.55\wd0\vrule height0.5\ht0\hss}\box0}}
{\setbox0=\hbox{$\textstyle        \rm S$}\hbox{\raise0.5\ht0%
\hbox to0pt{\kern0.35\wd0\vrule height0.45\ht0\hss}\hbox
to0pt{\kern0.55\wd0\vrule height0.5\ht0\hss}\box0}}
{\setbox0=\hbox{$\scriptstyle      \rm S$}\hbox{\raise0.5\ht0%
\hboxto0pt{\kern0.35\wd0\vrule height0.45\ht0\hss}\raise0.05\ht0%
\hbox to0pt{\kern0.5\wd0\vrule height0.45\ht0\hss}\box0}}
{\setbox0=\hbox{$\scriptscriptstyle\rm S$}\hbox{\raise0.5\ht0%
\hboxto0pt{\kern0.4\wd0\vrule height0.45\ht0\hss}\raise0.05\ht0%
\hbox to0pt{\kern0.55\wd0\vrule height0.45\ht0\hss}\box0}}}}
\def\nbZ{{\mathchoice {\hbox{$\sf\textstyle Z\kern-0.4em Z$}}
{\hbox{$\sf\textstyle Z\kern-0.4em Z$}}
{\hbox{$\sf\scriptstyle Z\kern-0.3em Z$}}
{\hbox{$\sf\scriptscriptstyle Z\kern-0.2em Z$}}}}

\begin{document}

\title{Nonperturbative Functional Renormalization Group for Random Field Models. III: Superfield formalism and ground-state dominance.}

\author{Matthieu Tissier} \email{tissier@lptl.jussieu.fr}
\affiliation{LPTMC, CNRS-UMR 7600, Universit\'e Pierre et Marie Curie,
bo\^ite 121, 4 Place Jussieu, 75252 Paris c\'edex 05, France}

\author{Gilles Tarjus} \email{tarjus@lptl.jussieu.fr}
\affiliation{LPTMC, CNRS-UMR 7600, Universit\'e Pierre et Marie Curie,
bo\^ite 121, 4 Place Jussieu, 75252 Paris c\'edex 05, France}

\date{\today}

\begin{abstract}
We reformulate the nonperturbative functional renormalization group for the random field Ising model in a superfield formalism, extending the supersymmetric description of the critical behavior of the system first proposed by Parisi and Sourlas [Phys. Rev. Lett. \textbf{43}, 744 (1979)]. We show that the two crucial ingredients for this extension are the introduction of a weighting factor, which accounts for ground-state dominance when multiple metastable states are present, and of multiple copies of the original system, which allows one to access the full functional dependence of the cumulants of the renormalized disorder and to describe rare events. We then derive exact renormalization group equations for the flow of the renormalized cumulants associated with the effective average action.
\end{abstract}

\pacs{11.10.Hi, 75.40.Cx}

\maketitle

\section{Introduction}
\label{sec:introduction}

Theoretical investigations of the critical behavior of the random field Ising model (RFIM) have a decades-long history.\cite{imry-ma75,nattermann98} One of their central themes concerns the validity of the ``dimensional reduction'' property, according to which the long-distance behavior of the RFIM in $d$ dimensions  is identical to that of the system without disorder in $d-2$. First pointed out by Grinstein,\cite{grinstein76} the property was shown to emerge from perturbation expansion to all orders by Aharony \textit{et al.}\cite{aharony76} and Young.\cite{young77} Finally, Parisi and Sourlas, in a beautiful $2$-page letter,\cite{parisi79} related the critical behavior of the RFIM to a supersymmetric scalar field theory and showed that the supersymmetry leads to dimensional reduction. Phenomenological,\cite{imry-ma75,bray85} numerical,\cite{nattermann98} and rigorous studies\cite{imbrie84,bricmont87} have however established that the dimensional-reduction prediction is wrong, at least in low dimensions ($d\leq 3$). It has also been understood that the superfield construction of Ref.~[\onlinecite{parisi79}] loses its validity when multiple metastable states are present, which is the case in the region of interest.\cite{parisi84b} In addition, it has been shown that dimensional reduction follows from supersymmetry not only in perturbative expansions as in Ref.~[\onlinecite{parisi79}] but in a nonperturbative manner.\cite{cardy83,klein83,klein84}

Despite the elegance of the formalism, no steps beyond the original superfield formulation have so far proven useful,\cite{footnote0} and the goal of the present work is to provide a solution by combining superfield formalism and functional renormalization group (FRG). This is part of our ongoing program\cite{tarjus04,tissier06,tarjus08,tissier08} to build a consistent and comprehensive theory of the long-distance physics of random field models, and more generally of disordered systems, based on the nonperturbative functional renormalization group (NP-FRG). In papers I\cite{tarjus08} and II\cite{tissier08} of this series, we showed that the breakdown of dimensional reduction is related to the appearance of a nonanalytic dependence of the effective action (or Gibbs free-energy functional in the terminology of magnetic systems) in the dimensionless fields. However, our approach, which was based on a ``replica method'' for handling the average over the random field, could not address the pending question of supersymmetry and its breaking. (In consequence, there was no means to avoid explicit breaking of the supersymmetry in the regulators and in the approximation scheme.) As the effective Hamiltonian (or bare action) of the model in the presence of a random field has always multiple minima in the region of interest (near the critical point), it would seem that a superfield approach and the  dimensional-reduction predictions can never be valid. Our objective is nonetheless to give a meaning to the superfield formalism even in cases where multiple minima are present and/or supersymmetry is broken, and to investigate the validity of the dimensional-reduction results as one varies spatial dimension. As will be shown, a resolution of the problem requires an FRG formulation.

Our starting point is that the fundamental flaw of the Parisi-Sourlas supersymmetric construction\cite{parisi79} has in fact a two-fold origin:

(1) the presence of metastable states, which can be described at zero temperature as the multiplicity of solutions of the stochastic field equation associated with the extremization of the bare action of the RFIM, is not counterbalanced by a means to select the ground state and 

(2) the use of a single copy of the original disordered system does not give access to the full functional dependence of the cumulants of the renormalized disorder and is thereby unable to account for the rare events, such as avalanches and droplets, that characterize random-field systems. 

The second aspect has already been addressed in our previous investigation through the NP-FRG and it requires introducing copies or replicas of the original system with the same disorder but different applied sources (resulting in an \textit{explicit} breaking of the permutational symmetry between replicas). We will show in the companion paper that the spontaneous breaking of supersymmetry (more precisely of  ``superrotational invariance'')  that comes with the breakdown of dimensional reduction is precisely linked to the emergence of a strong enough nonanalytic dependence of the cumulants of the renormalized disorder. Curing the first problem on the other hand implies a way to properly select the ground state among all solutions of the stochastic field equation. We propose a resolution of this problem through the introduction of a weighting factor and the construction of a superfield theory in a curved superspace. In the present paper, we relate  ground-state dominance to a formal property that we call ``Grassmannian ultralocality''. This finding may have value for other problems in which a generating functional is built from the solutions of a stochastic field equation, as in other disordered or glassy systems,\cite{glasses} in turbulence\cite{turbulence} or in nonabelian gauge field theories.\cite{gribov78,esposito04,zinnjustin89}

The outline of the article is as follows. In Sec.~\ref{sec:model} we introduce the superfield formalism for the $T=0$ equilibrium long-distance properties of the RFIM, as proposed by Parisi and Sourlas,\cite{parisi79} and we discuss the symmetries, supersymmetries and the associated Ward-Takahashi identities. 

In Sec.~\ref{sec:ultralocality} we discuss the property of ``Grassmannian ultralocality'' and its connection to the fact that a unique solution of the stochastic field equation is taken into account in the computation of the generating functionals; we next formulate a procedure to properly select the ground state, which is relevant for the $T=0$ equilibrium long-distance properties of the RFIM, through the insertion of a weighting factor and the consideration of a curved superspace. 

We show in Sec.~\ref{sec:cumulants} that a complete description of the renormalized random functional that describes the physics of the RFIM, including rare events such as avalanches and droplets, requires introducing copies of the original system with independently controlled sources so that the hierarchy of cumulants with their full functional dependence can be generated. We then consider the properties of the superfield theory with multiple copies, its (super)symmetries and the associated Ward-Takahashi identites. 

In the following section, Sec.~\ref{sec:ultralocal}, we explore the formal consequences of the property of ``Grassmannian ultralocality''. 

We next describe in Sec.~\ref{sec:NPFRG} the generalization  to superfields in a curved superspace of our previously developed NP-FRG approach in the effective action formalism. We carefully discuss the issue of the infrared regulator and we derive the exact FRG equations for the cumulants of the renormalized disorder; we also consider the implication for these equations of the hypothesis of ``Grassmannian ultralocality''. 

In Sec.~\ref{sec:ground-state_dominance}, we detail how our formalism allows one to describe ground-state dominance at long distance in the NP-FRG. We show that the limit of infinite curvature, which corresponds to a vanishing auxiliary temperature, gives back the  FRG equations for the cumulants of the renormalized disorder under the property of ``Grassmannian ultralocality''. 

We finally conclude with a summary and a discussion. Additional information is provided in several appendices.

In the companion paper, we apply the NP-FRG in the superfield formalism to describe supersymmetry and its spontaneous breaking in the critical behavior of the RFIM. We introduce a nonperturbative approximation scheme to the exact FRG equations for the cumulants and solve the flow equations numerically to determine the critical exponents and the fixed-point properties as a function of space dimension $d$. This provides a resolution of the long-standing puzzles associated with the long-distance physics of the RFIM.

A short account of this work has appeared in Ref.[\onlinecite{tissier11}].

\section{The Parisi-Sourlas superfield formalism for the RFIM}
\label{sec:model}
\subsection{Model and generating functionals}

The starting point is the field-theoretical description of the RFIM in terms of a scalar field
$\varphi(x)$ in a $d$-dimensional space and a bare action $S[\varphi;h]$ given by
\begin{equation}
\label{eq_ham_dis}
S= \int_{x} \bigg\{ \frac
{1}{2}  \left(\partial_{\mu} \varphi(x) \right) ^2 + U_B(\varphi(x)) -
h(x) \varphi(x) \bigg\} ,
\end{equation}
where $ \int_{x} \equiv \int d^d x$,  $U_B(\varphi)= (\tau/2) \varphi^2  +  (u/4!) \varphi^4$, and $h(x)$ is a
random source (a random magnetic field in the language of magnetic systems) that is taken with a Gaussian distribution characterized by a zero mean and a variance $\overline{h(x)h(y)}=\Delta_B \ \delta^{(d)}(x-y)$. A (ultra-violet) momentum cutoff $\Lambda$,
associated with an inverse microscopic lengthscale such as a lattice
spacing, is also implicitly considered.

The Parisi-Sourlas construction goes as follows.\cite{parisi79} Taking advantage of the fact that at long-distance the thermal fluctuations are negligible compared to those induced by disorder (formally, the critical behavior is controlled by a zero-temperature fixed point\cite{villain84,fisher86,nattermann98}), one can focus on the ground-state configuration which is solution of the stochastic field equation
\begin{equation}
\label{eq_stochastic}
\dfrac{\delta S[\varphi;h]}{\delta \varphi(x)} = J(x),
\end{equation}
where we have added an external source (a magnetic field) $J$ conjugate to the $\varphi$ field.

Provided the solution of Eq.~(\ref{eq_stochastic}) is unique, the equilibrium (Green's) correlation functions of the $\varphi$ field are obtained from appropriate derivatives of the generating functional
\begin{equation}
  \label{eq_generating_func1}
\begin{aligned}
\mathcal Z_h[\hat{J},J]=\int& \mathcal D\varphi \; \delta\left[\dfrac{\delta S_B[\varphi]}{\delta \varphi}-h-J\right] \; \left|\det \dfrac{\delta^2 S_B[\varphi]}{\delta \varphi \delta \varphi}\right| \\& \times \exp  \int_{x}  \hat{J}(x)  \varphi(x) ,
\end{aligned}
\end{equation}
where $S_B[\varphi] = \int_{x}\{(1/2) (\partial_{\mu} \varphi(x))^2 + U_B(\varphi(x))\}$; the delta functional $\delta[\;]$ enforces that $\varphi$ is solution of Eq.~(\ref{eq_stochastic}) and the absolute value of the functional determinant of $\delta^2 S_B[\varphi]/(\delta \varphi(x) \delta \varphi(y))$ is the associated jacobian. Due to the postulated uniqueness of the solution, the absolute value can be dropped and the functional can be built through standard field-theoretical techniques.\cite{zinnjustin89} One first introduces  auxiliary fields: a bosonic ``response'' field $\hat{\varphi}(x)$ associated with the integral representation of the delta functional and two fermionic ``ghost'' fields  $\psi(x)$ and  $\bar{\psi}(x)$ [satisfying $\psi^2=\bar{\psi}^2=\psi \bar\psi +\bar\psi \psi=0$] to exponentiate the determinant. We also introduce  two fermionic sources $\bar{K}(x),K(x)$ linearly coupled to the ghost fields. This leads to
\begin{equation}
  \label{eq_generating_func2}
\begin{aligned}
&\mathcal Z_h[\hat{J},J, \bar{K},K] =  \exp \left( \mathcal{W}_{h}[\hat{J},J, \bar{K},K]\right) \\&= \mathcal N \int \mathcal D\varphi \mathcal D\hat{\varphi} \mathcal D\psi D\bar{\psi} \exp \big\{ \int_{x} - \hat{\varphi}(x) \dfrac{\delta S_B[\varphi]}{\delta \varphi(x)}\\& + \hat{\varphi}(x) \left[ h(x) + J(x)\right]  + \hat{J}(x)  \varphi(x) + \psi(x) \bar{K}(x) \\&+ K(x) \bar{\psi}(x) + \int_{x}\int_{y}\bar{\psi}(x) \dfrac{\delta^2 S_B[\varphi]}{\delta \varphi(x) \delta \varphi(y)} \psi(y)\big\},
\end{aligned}
\end{equation}
where $\mathcal N$ is an irrelevant constant factor that will be dropped in the following and $\mathcal{W}_h$ is a random functional that depends on the bare quenched disorder (\textit{i.e.} the bare random field $h$). One then performs the average of the partition function (and not of its logarithm as in the standard approach to disordered systems) over the Gaussian disorder, which provides
\begin{equation}
\begin{aligned}
  \label{eq_generating_func3}
\mathcal Z&[\hat{J},J, \bar{K},K]= \overline{\mathcal Z_h[\hat{J},J, \bar{K},K]}\\&= \int \mathcal D\varphi \mathcal D\hat{\varphi} \mathcal D\psi D\bar{\psi} \exp \big\{-S_{ss}[\varphi,\hat{\varphi},\psi,\bar{\psi}] + \\&  \int_{x} \left( \hat{J}(x)  \varphi(x) + \psi(x)  \bar{K}(x)+K(x) \bar{\psi}(x) + J(x)  \hat{\varphi}(x)\right) \big\},
\end{aligned}
\end{equation}
where
\begin{equation}
\label{eq_susyaction1}
\begin{aligned}
S_{ss}= & \int_{x}\hat{\varphi}(x) \dfrac{\delta S_B[\varphi]}{\delta \varphi(x)}   -  \int_{x}\int_{y}\bar{\psi}(x) \dfrac{\delta^2 S_B[\varphi]}{\delta \varphi(x) \delta \varphi(y)} \psi(y) \\&- \frac{\Delta_B}{2}\int_{x}\hat{\varphi}(x)^2\,.
\end{aligned}
\end{equation}
We define $W[\hat{J},J, \bar{K},K]= \log\mathcal Z[\hat{J},J, \bar{K},K]$, which is the generating functional of the connected Green's functions. An important feature is that due to the identity $Z[\hat{J}=0,J, \bar{K}=0,K=0]=1$, the $\varphi$-field connected correlation functions of the original problem are obtained by functional derivatives of $W$ with respect to $\hat{J}$ that are further evaluated for $K=\hat{K}=\hat{J}=0$. For instance,
\begin{equation}
\label{eq_physical_disconn}
\overline{\left\langle \varphi(x)\right\rangle \left\langle \varphi(y)\right\rangle} - \overline{\left\langle \varphi(x)\right\rangle}\;\overline{\left\langle \varphi(y)\right\rangle} = \frac{\delta^2 W}{\delta \hat{J}(x) \delta \hat{J}(y)}\rvert_{K=\hat{K}=\hat{J}=0},
\end{equation}
where, since we consider the zero-temperature limit, $\left\langle \varphi(x)\right\rangle$ is equal to the solution of the stochatic field equation, Eq.~(\ref{eq_stochastic}).

The next step of the construction is to introduce a superspace by adding to the $d$-dimensional Euclidean space with coordinates $x=\left\lbrace x^\mu\right\rbrace $ two anti-commuting Grassmann coordinates $\theta,\bar{\theta}$ (satisfying $\theta^2=\bar{\theta}^2=\theta \bar{\theta}+\bar{\theta}\theta=0$).\cite{zinnjustin89} By letting $\underline{x}=(x,\theta,\bar{\theta})$ denote the coordinates, the associated (super)metric is given by
\begin{equation}
  \label{eq_metric}
d\underline{x}^2 = dx^2 + \frac{4}{\Delta_B}d\bar{\theta} d\theta = g_{mn}dx^m dx^n,
\end{equation}
where $\{m\}\equiv \{\mu,\theta,\bar{\theta}\}$, $dx^2=dx^{\mu} dx^{\mu}$, and a summation over repeated indices is implied; the metric tensor $g_{mn}$ satisfies: $g_{\mu \nu}=\delta_{\mu \nu},\; g_{\theta \bar{\theta}}= - g_{\bar{\theta} \theta}=-2/\Delta_B $, with all other components equal to zero. With the notations $\partial_m= \partial/\partial x^m, \partial_{\mu}= \partial/\partial x^{\mu}, \partial_{\theta}=\partial / \partial \theta$, etc..., one can express the corresponding ``super-Laplacian'' as
\begin{equation}
  \label{eq_superlaplacian}
\Delta_{ss}= g^{mn}\partial_m \partial_n = \partial_\mu \partial_\mu+\Delta_B \partial_\theta \partial_{\bar{\theta}},
\end{equation}
where $g^{mp}g_{pn}=\delta^m_n$. After introducing the superfield
\begin{equation}
  \label{eq_superfield}
\Phi(\underline{x})=\varphi(x) + \bar{\theta} \psi(x)+ \bar{\psi}(x) \theta + \bar{\theta}\theta \hat{\varphi}(x)
\end{equation}
and the supersource
\begin{equation}
  \label{eq_supersource}
\mathcal J(\underline x)=J(x) + \bar{\theta} K(x)+ \bar{K}(x) \theta + \bar{\theta}\theta \hat{J}(x),
\end{equation}
the generating functional can be cast in the form\cite{parisi79}
\begin{equation}
\begin{aligned}
  \label{eq_part_func_PS}
\mathcal Z[\mathcal J]&=\int\mathcal D\Phi  \exp \left(-S_{ss}[\Phi] +   \int_{\underline x}  \mathcal J(\underline x)  \Phi(\underline x)\right) \\&=\exp (W[\mathcal J])
\end{aligned}
\end{equation}
with
\begin{equation}
\label{eq_susyaction2}
S_{ss}[\Phi]=\int_{\underline{x}}[-\frac 12 \Phi(\underline{x})\Delta_{ss}\Phi(\underline{x})+U_{B}(\Phi(\underline{x}))],
\end{equation}
and with $ \int_{\underline{x}} \equiv \int_x  \iint d\theta d\bar{\theta}$. By a Legendre transform, one  introduces the effective action $\Gamma[\Phi]$ which is the generating functional of the $1$-particle irreducible (1PI) correlation functions or proper vertices:\cite{zinnjustin89}
\begin{equation}
\label{eq_legendre_1copy}
\Gamma[\Phi] = -W[\mathcal J] + \int_{\underline x}  \mathcal J(\underline x) \Phi(\underline x),
\end{equation}
where the (classical) superfield $\Phi$ and the supersource $\mathcal J$ are related through
\begin{equation}
\label{eq_legendre_phi}
\Phi(\underline x)= \frac{\delta W[\mathcal J ] }{\delta  \mathcal J(\underline x)}
\end{equation}
and
\begin{equation}
\label{eq_legendre_J_1copy}
\mathcal{J}(\underline x)=\frac{\delta \Gamma[\Phi] }{\delta \Phi(\underline x)},
\end{equation}
which can also easily be expressed in terms of the components of the classical superfield, $\phi, \hat{\phi}, \psi, \bar{\psi}$, and of the supersource, $\hat{J}, J, \bar{K}, K$. (Note that for simplicity we keep the same notation for the superfield $\Phi$ and its classical value, and similarly for the ghost fields $\psi, \bar \psi$; we only make a distinction for the bosonic fields, \textit{i.e.} $\phi=\langle\varphi\rangle$ and $\hat \phi=\langle\hat \varphi\rangle$.)

\subsection{Symmetries, supersymmetries, and Ward-Takahashi identities}

The action $S_{ss}[\Phi]$ is invariant under a large group of transformations:

(i) sign changes associated with a $Z_2$ discrete symmetry $\Phi \rightarrow - \Phi$ (the critical behavior we aim at describing is associated with a spontaneous breaking of this $Z_2$ symmetry, \textit{i.e.} a paramagnetic-to-ferromagnetic transition in the language of magnetism);

(ii) rotations and translations in the $d$-dimensional Euclidean space with infinitesimal generators $L_{\mu \nu}=x^\mu \partial_{\nu}-x^\nu \partial_{\mu}$ and $\partial_{\mu}$, where $\mu, \nu=1,...,d$, respectively;

(iii) transformations of the symplectic group with generators acting on the Grassmann subspace: $\bar t= \bar{\theta}\partial_{\theta}$, $t=\theta \partial_{\bar{\theta}}$ (associated with ``rotations'' in the $2$-dimensional Grassmann subspace) and $N=\bar{\theta}\partial_{\bar{\theta}}-\theta \partial_{\theta}$ (corresponding to the ``ghost-number conservation''); the generators satisfy the commutation relations $[\bar t,t]=N$, $[N,\bar t]=2 \bar t$, and $[N, t]= -2 t$;

(iv) translations in the $2$-dimensional Grassmann subspace with generators $\partial_{\theta},\partial_{\bar{\theta}}$: these are linear, BRS-type, symmetries;\cite{zinnjustin89}

(v) ``superrotations'' that preserve the supermetric and can be represented by the generators $\bar {\mathcal Q}_\mu=-x^\mu \partial_\theta + \frac{2}{\Delta_B} \bar{\theta} \partial_{\mu}$ and $\mathcal Q_\mu=x^\mu \partial_{\bar{\theta}}+ \frac{2}{\Delta_B} \theta \partial_{\mu}$.

The last two sets of transformations mix bosonic and fermionic fields and for this reason represent ``fermionic symmetries''.\cite{zinnjustin89} They are associated with supersymmetries: the translations in the $2$-dimensional Grassmann subspace (iv) form a supergroup and the Euclidean rotations (ii), the symplectic group (iii) and the superrotations (v) form another supergroup known as the orthosymplectic supergroup $OSp(2,d)$.\cite{osp2d} 

The linearly realized continuous symmetries and supersymmetries (ii)-(v) lead to a set of Ward-Takahashi (WT) identities that can be expressed either at the level of the generating functional of the Green's functions or that of the effective action.\cite{zinnjustin89} For instance, the invariance of the action under the superrotations gives rise to the following WT identities:
\begin{equation}
\label{eq_ward_0}
\int_{\underline{x}}\Phi(\underline{x})\mathcal Q_{\underline{x}}\Gamma_{\underline{x}}^{(1)}[\Phi]=0
\end{equation}
and for $p > 1$
\begin{equation}
\begin{aligned}
\label{eq_ward_p}
\bigg(\mathcal Q_{\underline{x}_1}+...&+ \mathcal Q_{\underline{x}_p} \bigg)\Gamma_{\underline{x}_1...\underline{x}_{p}}^{(p)}[\Phi] \\&+ \int_{\underline{x}_{p+1}}\Phi(\underline{x}_{p+1})\mathcal Q_{\underline{x}_{p+1}}\Gamma_{\underline{x}_1...\underline{x}_{p+1}}^{(p+1)}[\Phi]=0,
\end{aligned}
\end{equation}
where we have used the notation $\mathcal Q_{\underline{x}}$ to indicate a generic component $\mathcal Q_{\mu}$ of the generator of the superrotation acting on the superspace coordinate $\underline{x}$; the proper (1PI) vertices are defined as
\begin{equation}
\label{eq_derivative_supervertex}
\Gamma_{\underline{x}_1, ... ,\underline{x}_p}^{(p)}[\Phi]=\frac{\delta^p \Gamma [\Phi]}{\delta \Phi(\underline{x}_1)... \delta \Phi(\underline{x}_p)}.
\end{equation}
Similar identites are obtained with the generator $\bar {\mathcal Q}_{\underline{x}}$ (and additional ones with the generators $\partial_{\mu}$, $L_{\mu \nu}$, $\partial_{\theta},\partial_{\bar{\theta}}$, $\bar t, t$, and $N$).

At this point, it is important to stress that it is the existence of the superrotation invariance (v) which leads to the property of dimensional reduction in the critical behavior of the RFIM, with the two Grassmann dimensions acting as negative dimensions.\cite{parisi79,cardy83}  One should however keep in mind that the whole formal construction collapses when the stochastic field equation, Eq.~(\ref{eq_stochastic}), has more than one solution, which is usually the case in the region of interest.\cite{parisi84b} In the following, we show how to generalize the Parisi-Sourlas construction in order to preserve the relevance of the superfield formalism.

\section{``Grassmannian ultralocality'' and ground-state dominance}
\label{sec:ultralocality}

\subsection{On the ``ultralocality'' of the random generating functional in the Grassmann subspace}

Before presenting the method to select the ground-state configuration, which should dominate at $T=0$, we discuss an important property of the random generating functional in the superfield framework. To alleviate the notations we consider the case of a $d=0$ Euclidean space, but the same considerations equally apply to any number of Euclidean dimensions. When the stochastic field equation [Eq.~(2)], which can be rewritten as
\begin{equation}
\label{eq_stochastic_app}
\frac{\partial S_B(\varphi)}{\partial \varphi} = h+J \, ,
\end{equation}
has a unique solution, say $\varphi_*(h+J)$, the generating functional of the (Green's) correlation functions of the $\varphi$ field is simply expressed as
\begin{equation}
\label{eq_generating1_app}
\mathcal Z_h(\hat J,J)= e^{\hat J \varphi_*(h+J)} \, , \end{equation}
or, after adding the fermionic sources $K, \bar K$ and using the auxiliary fields [see Eq.~(4)], as
\begin{equation}
\label{eq_generating2_app}
\mathcal Z_h(\hat J,J, \bar K, K) \equiv \mathcal Z_h[\mathcal J]= e^{\hat J \varphi_*(h+J) - \bar K K S_B^{(2)} (\varphi_*(h+J))^{-1}}
\end{equation}
where $S_B^{(2)} (\varphi)$ is the second derivative of the bare action and $\mathcal J$ is the supersource introduced in Eq.~(\ref{eq_supersource}). As a consequence, the (random) generating functional $\mathcal W_h=\log \mathcal Z_h$ is equal to
\begin{equation}
\label{eq_generating3_app}
\mathcal W_h[\mathcal J]= \hat J \, \varphi_*(h+J)-\bar K K \, S_B^{(2)} (\varphi_*(h+J))^{-1} \, .
\end{equation}
It is straighforward to show that this precisely corresponds to an ``ultralocal'' form in the Grassmann subspace,\cite{footnote001}
\begin{equation}
\label{eq_ultralocal_1}
\mathcal W_h[\mathcal J]=\int_{\underline{\theta}}W_h(\mathcal J(\underline{\theta}))
\end{equation}
where $\underline \theta$ collects the two Grassmann coordinates $\bar \theta$ and $\theta$ and $\int_{\underline \theta}\equiv\int \int d\theta d\bar \theta$; indeed,
\begin{equation}
\label{eq_ultralocal_app}
\int_{\underline{\theta}}W_h(\mathcal J(\underline{\theta}))= \hat J \, W_h^{(1)}(J)-\bar K K \, W_h^{(2)} (J)
\end{equation}
with $W_h^{(1)}(J)=\varphi_*(h+J)$ and $W_h^{(2)}(J)=\partial \varphi_*(h+J)/ \partial J=S_B^{(2)} (\varphi_*(h+J))^{-1}$ as it should be. The whole formal construction is fully justified in this case.

What happens when the stochastic field equation has several solutions ? Let us sort the solutions and label them with an index $\alpha$;  $\alpha=0$ denotes the ground state, which is unique for a continuous distribution of the random field, aside from exceptional values of $h+J$ for which a coexistence of several ground states may take place.

The sought-for generating functional that describes the equilibrium situation at $T=0$ only takes into account the contribution of the ground state. As a result, one has exactly the same relations as above, with $\varphi_*(h+J)$ replaced by the ground-state solution $\varphi_0(h+J)$. Actually, the notation $\varphi_0(h+J)$ is a little misleading as the ground state is generally a piecewise function of $h+J$ with discontinuities occuring at special values of the latter. These discontinuities are known as ``avalanches" (or ``shocks") and have been observed for instance in numerical computations of the ground state of the RFIM on $2$- and $3$-dimensional lattices.\cite{frontera00,machta03,machta05,dahmen07} The ground-state solution $\varphi_0(h+J)$ should rather be written as $\sum_i \mathcal C_{0,i}(h+J)  \varphi_{0,i}(h+J)$ with $\mathcal C_{0,i}$ the characteristic of the $i$th interval along the axis $h+J$ in which the ground state  $\varphi_{0,i}$ is a continuous function of $h+J$ (see also Appendix~\ref{appendix:toy}). More generally, including a single solution in the generating functional $\mathcal Z_h(\hat J,J, \bar K, K)$, provided this solution is piecewise defined for \textit{all} values of $h+J$, also leads to the ``ultralocal property'' in the Grassmann subspace for $\mathcal W_h[\mathcal J]$, Eq.~(\ref{eq_ultralocal_1}).

On the other hand, one easily realizes that this property cannot be true when several different solutions are included in the generating functional, no matter how one chooses to weigh these solutions. The weighting of the solutions that corresponds to the above Parisi-Sourlas  supersymmetric construction, Eq.~(\ref{eq_generating_func2}), is given by the sign of the determinant of the Hessian $S_B^{(2)}$.\cite{parisi82} Then, setting to zero the fermionic sources for simplicity, one has
\begin{equation}
\label{eq_generating4_app}
\mathcal Z_h(\hat J,J)= \sum_{\alpha,i}  (-1)^{n(\alpha)} \mathcal C_{\alpha,i}(h+J) e^{\hat J \varphi_{\alpha,i}(h+J)},
\end{equation}
where $n(\alpha)$ is the index of the $\alpha$th solution, \textit{i.e.} the associated number of negative eigenvalues of the Hessian $S_B^{(2)}$, and $\mathcal C_{\alpha,i}(h+J)$ is the characteristic function of the $i$th interval over which $\varphi_{\alpha}=\varphi_{\alpha,i}$ is a continuous function of $h+J$ [at the boundaries of $\mathcal C_{\alpha,i}(h+J)$ the solution $\varphi_{\alpha}$ has a discontinous jump or possibly stops to exist]. Obviously,
\begin{equation}
\label{eq_ultralocal_breakdown1_app}
\mathcal W_h(\hat J,J)= \log \sum_{\alpha} (-1)^{n(\alpha)} e^{\hat J \varphi_{\alpha}(h+J)}, 
\end{equation}
where $\varphi_{\alpha}$ denotes $\sum_i \mathcal C_{\alpha,i}(h+J) \varphi_{\alpha,i}(h+J)$, is not ``ultralocal'' in the Grassmann coordinates, \textit{i.e.}  cannot be put in the form of Eqs.~(\ref{eq_generating3_app}-\ref{eq_ultralocal_app}) [observe in particular that the form in Eq.~(\ref{eq_generating3_app}) implies a linear dependence on $\hat J$, which is not satisfied by the above expression]. When $\hat J =0$, one finds that $\mathcal Z_h=1$ and $\mathcal W_h=0$, just as in the case where a unique solution is taken into account.\cite{footnote01} However, the result now follows from the property that the sum of all solutions weighted by the sign of the determinant of the Hessian is a topological invariant which is equal to $1$ in the present case.\cite{kurchan02}

Note that contrary to the toy model discussed in Ref.~[\onlinecite{parisi82}], in which the breakdown of the supersymmetric formalism is related to the fact that the stochastic equation has no solutions for a certain range of the random field, the stochastic equation associated with the present scalar field theory in a Gaussian distributed random field has always at least one solution (the ground state always exists). 

\subsection{Selecting the ground state}

A natural procedure to select the ground state is to add in the generating functional, Eq.(\ref{eq_generating_func1}), a weighting factor that strongly favors the solution with the smallest action. This can be done through a Bolzmann-like factor, namely,
\begin{equation}
  \label{eq_generating_func1beta}
\begin{aligned}
&\mathcal Z_h^{(\beta)}[\hat{J},J]=\int \mathcal D\varphi \; \delta\left[\dfrac{\delta S_B[\varphi]}{\delta \varphi}-h-J\right] \; \det\left[ \dfrac{\delta^2 S_B[\varphi]}{\delta \varphi \delta \varphi}\right]\; \times  \\& \exp \big [-\beta \left ( S_B[\varphi]- \int_{x} [J(x) +h(x)] \varphi(x) \right) + \int_{x}  \hat{J}(x)  \varphi(x)\big ]
\end{aligned}
\end{equation}
where $\beta$ is the inverse of an auxiliary temperature (the actual temperature is equal to zero). Note that $\mathcal Z_h^{(\beta)}$ is \textit{not} the same as the partition function obtained from the equilibrium Boltzmann-Gibbs measure at a temperature $T=1/\beta$. Even if $\beta$ is large enough that only minima contribute to Eq.~(\ref{eq_generating_func1beta}), the latter expression only includes the contribution of the minima whereas the Boltzmann-Gibbs measure also takes into account the contribution of the basins of attraction of the minima (roughly speaking, the thermal fluctuations around the minima).

With the above generating functional, one finds that the average of the field $\varphi$ is given by
\begin{equation}
  \label{eq_generating_func2beta}
\begin{aligned}
\langle\varphi(x)\rangle=\frac{\delta \log \mathcal Z_h^{(\beta)}}{\delta \hat J(x)}\bigg \vert_{\hat J=0}=\frac{1}{\mathcal Z_h^{(\beta)}[\hat J=0,J]}\frac{\delta \mathcal Z_h^{(\beta)}}{\delta \hat J(x)}\bigg \vert_{\hat J=0}.
\end{aligned}
\end{equation}
The simplicity of the Parisi-Sourlas formalism is however lost as $\mathcal Z_h^{(\beta)}[\hat J=0,J]\neq 1$. In consequence, simply considering the average over the disorder of $\mathcal Z_h^{(\beta)}$ is no longer sufficient to generate the $\varphi$-field correlation functions of the original problem.

Interestingly, the superfield formalim can still prove useful. After having introduced one bosonic and two fermionic auxiliary fields as before, added two fermionic sources, grouped all the sources in a supersource as  in Eq.~(\ref{eq_supersource}), and similarly grouped all the fields in a superfield as in Eq.~(\ref{eq_superfield}), the generating functional can be rewritten as
\begin{equation}
\begin{aligned}
\label{eq_part_funcbeta}
\mathcal Z_h^{(\beta)}[\mathcal J]&=\int\mathcal D\Phi  \exp \bigg (- \int \int d\theta d\bar\theta  [1+\beta \bar \theta \theta] S_{B}[\Phi(\underline \theta)] \\&+   \int_{x} \int \int d\theta d\bar\theta [1+\beta \bar \theta \theta] \left [h(x) + \mathcal J(x,\underline \theta) \right ] \Phi(x, \underline \theta)\bigg ) \\&=\exp (\mathcal W_h^{(\beta)}[\mathcal J]).
\end{aligned}
\end{equation}
The construction may appear rather formal and untractable but it turns out that Eq.~(\ref{eq_part_funcbeta}) can be expressed in a way which will prove efficient to study the symmetries of the theory and investigate its long-distance properties. It is convenient to introduce a superspace combining the $d$ Euclidean  and the $2$ Grassmannian dimensions in which the Grassmann subspace is now curved. To be more specific, we replace the metric tensor of the Parisi-Sourlas formalism (see section II-A) by
\begin{equation}
\begin{aligned}
\label{eq_metric_curved}
&g_{\theta \bar\theta}=- g_{\bar\theta \theta}=- (1-\beta \bar \theta \theta)\,,\\&
g_{\theta \theta}= g_{\bar\theta \bar\theta}=0\,,
\end{aligned}
\end{equation}
keeping $g_{mn}=\delta_{\mu \nu}$ for the Euclidean sector and all cross-components between Euclidean and Grassmannian coordinates equal to zero. So long as we are not interested in mixing bosonic and fermionic directions (mixing occurs when considering superrotations, see section II-A), there is no need to introduce the factor $2/\Delta_B$ in the supermetric as was done in the Parisi-Sourlas formalism above.

The properties of the curved superspace are discussed in Ref.~[\onlinecite{wschebor09}]. In a nutshell, one has the usual prescriptions of Riemannian geometry that in order to satisfy the isometries of the curved Grassmann subspace, one should (i) contract Grassmann indices with either the metric tensor or its inverse, (ii) integrate over the Grassmann coordinates with a measure $\sqrt{\mathrm{ sdet}g}\,d\theta d\bar \theta=(1+ \beta\bar \theta \theta)\, d\theta d\bar \theta$, where $\mathrm{sdet}g=(1+ \beta\bar \theta \theta)^2$ is the superdeterminant of the metric tensor in the Grassmann sector, (iii) use, if necessary, covariant derivatives as well as the proper Laplacian operator $\Delta_{\underline \theta}=g^{mn}(\partial_m\partial_n -\Gamma^p_{mn}\partial_p)$, where $m,n,p=\theta, \bar \theta$, summation over repeated indices is implied, and the $\Gamma^p_{mn}$'s are the Christoffel symbols with nonzero components $\Gamma^{\theta}_{\theta \bar \theta}=-\Gamma^{\theta}_{\bar\theta \theta}=\beta \theta$ and 
$\Gamma^{\bar \theta}_{\theta \bar \theta}=-\Gamma^{\bar \theta}_{\bar\theta \theta}=\beta \bar \theta$; the Laplacian for the Grassmann subspace is then explicitly given by
\begin{equation}
\begin{aligned}
\label{eq_laplacian_curved}
\Delta_{\underline \theta}=2(1+ \beta\bar \theta \theta)(\partial_{\theta}\partial_{\bar \theta}-\beta \bar \theta \partial_{\bar \theta}-\beta  \theta \partial_{\theta}).
\end{aligned}
\end{equation}
It is easy to show that the parameter $\beta$ (inverse of an auxiliary temperature) is up to a factor  $1/6$ equal to the Ricci scalar curvature of the Grassmann subspace.\cite{wschebor09}

We now come back to the discussion of the previous subsection concerning ``ultralocality'' in the Grassmann subspace. For ease of notation, we consider again a $d=0$ Euclidean subspace and we momentarily drop the two fermionic sources. The random generating functional $\mathcal W_h^{(\beta)}[\mathcal J]$ corresponding to the superfield construction in Eqs.~(\ref{eq_generating_func2beta}) and (\ref{eq_part_funcbeta}) is expressed as
\begin{equation}
\begin{aligned}
\label{eq_ultralocal_breakdown_beta}
&\mathcal W_h^{(\beta)}(\hat J,J) =\\& - \beta \big [ S_{B}[\varphi_0(h+J)]-(h+J)\varphi_0(h+J) \big ] +\hat J\, \varphi_{0}(h+J) +\\&  \log \big ( 1+\sum_{\alpha\neq 0} (-1)^{n(\alpha)} e^{- \beta \Delta S_{\alpha,0}(h+J) +\hat J \left [\varphi_{\alpha}(h+J)-\varphi_{0}(h+J)\right ]}\big ), 
\end{aligned}
\end{equation}
where $\Delta S_{\alpha,0}= (S_{B}[\varphi_{\alpha}] - S_{B}[\varphi_0])-(h+J)(\varphi_{\alpha}-\varphi_0)$.

Assume first that the parameter $\beta$ can be taken sufficiently large for ensuring $\Delta S_{\alpha,0}\ll 1/\beta$ for all solutions $\alpha\neq 0$, so that all contributions but that of the ground state can be neglected in Eq.~(\ref{eq_ultralocal_breakdown_beta}). It is then easy to show that the random functional can be put in an ``ultralocal'' form appropriate for the curved Grassmann subspace,
\begin{equation}
\begin{aligned}
\label{eq_ultralocal_beta1}
&\mathcal W_h^{(\beta)}[\mathcal J] =\int_{\underline \theta}W_h^{(\beta)}[\mathcal J(\underline \theta)], 
\end{aligned}
\end{equation}
where
\begin{equation}
\begin{aligned}
\label{eq_ultralocal_beta2}
W_h^{(\beta)}[J]= S_{B}[\varphi_0(h+J)]-(h+J)\varphi_0(h+J)
\end{aligned}
\end{equation}
is actually independent of $\beta$ and where the integral over $\underline \theta$ now involves the metric factor $(1+\beta \bar \theta \theta)$, \textit{i.e.} $\int_{\underline \theta}\equiv \int \int (1+\beta \bar\theta \theta) d\theta d\bar\theta$.

In fact, no matter how large (but finite) $\beta$, the above assumption may not be valid for all realizations of the random field. Excited states with $\Delta S_{\alpha,0}\lesssim 1/\beta$ may be present. The vast majority of them involve only local changes of configuration with respect to the ground state.\cite{frontera01,zumsande08} As a result, the term $\hat J (\varphi_{\alpha}-\varphi_{0})$ remains small in $d$ Euclidean dimensions, provided $\hat J$ of course stays small or finite (here, we are ultimately interested in the limit $\hat J=0$). In addition, rare excitations, also known as ``droplets'',\cite{droplets} may involve a large-scale reorganization of the ground-state configuration while satisfying $\Delta S_{\alpha,0}\lesssim 1/\beta$ and therefore provide a significant contribution to $\mathcal W_h^{(\beta)}$. In such a situation, the ``ultralocality'' in the curved Grassmann subspace is broken and the random generating function has the general form
\begin{equation}
\begin{aligned}
\label{eq_ultralocal_broken_beta}
&\mathcal W_h^{(\beta)}[\mathcal J]=\\& \int_{\underline \theta}W_h^{(\beta)}[\mathcal J(\underline{\theta}),(1+\frac{\beta}{2} \bar{\theta} \theta)\partial_{\underline{\theta}}\mathcal J(\underline{\theta}), \Delta_{\underline{\theta}}\mathcal J(\underline{\theta})],
\end{aligned}
\end{equation}
where $\partial_{\underline{\theta}}$ is a short-hand notation for designating either $\partial_{\bar \theta}$ or $\partial_{\theta}$, and we recall that $\Delta_{\underline{\theta}}=2 (1+\beta \bar{\theta} \theta)( \partial_{\theta}\partial_{\bar{\theta}}-\beta \bar \theta \partial_{\bar{\theta}}-\beta \theta \partial_{\theta})$ is the Laplacian in the curved Grassmann subspace (see above). Note that the functional dependence on the Euclidean coordinates is completely general at this point.

The droplets being nonetheless rare events, and the other excitations being essentially local, we expect that the deviation from ``ultralocality'' is small when $\beta$ is large, \textit{i.e.},
\begin{equation}
\label{eq_ultralocal_broken2_beta}
W_h^{(\beta)}\simeq W_h^{(\beta)UL}[\mathcal J(\underline{\theta})] + {\rm corrections},
\end{equation}
with the corrections going to zero (and $W_h^{(\beta)UL}$ going to a well defined limit) when $\beta \rightarrow \infty$. In the following, we will check the correctness of this behavior and show how in the FRG flow the contributions coming from errors in selecting the ground state actually become subdominant as one approaches the critical fixed point.

\section{Cumulants of the renormalized disorder and the need for multiple copies}
\label{sec:cumulants}

\subsection{Why the need for multiple copies ?}

A central quantity is the random (``free energy'') functional $\mathcal W_h^{(\beta)}[ \mathcal J]$, introduced above. This random functional is characterized by its (functional) probability distribution or, alternatively, by the infinite set of its cumulants (if of course the cumulants exist). Dealing with cumulants has the advantage of involving an average over the bare disorder: as a result, one recovers the translational (and rotational) invariance in Euclidean space which is otherwise broken by the space-dependent random field. In the following, we will therefore consider a formalism based on cumulants. However, a crucial point when working with such disorder-averaged quantities is that one does not want to lose track of the rare events that characterize systems with quenched disorder. For random-field models, these rare events are expected to take the form of ``avalanches'' or ``shocks'' that are seen in the dependence of the ground state on the applied source at zero temperature and of low-energy excitations known as ``droplets'' at nonzero temperature (see above). As shown in previous work,\cite{balents-fisher93,BLbalents93, balents96,BLchauve00,CUSPledoussal,BLbalents04,BLbalents05,CUSPledoussal09,BLledoussal10,tarjus04,tissier06,tarjus08,tissier08} these phenomena show up in the cumulants of the renormalized disorder as a singular dependence on the arguments (for an illustration in a simple zero-dimensional toy model, see Appendix~\ref{appendix:toy}). Describing such features therefore requires the functional dependence of the cumulants for \textit{generic arguments}. For instance, a complete characterization of the random functional $\mathcal W_h^{(\beta)}[ \mathcal J]$ implies the knowledge of all its cumulants, $\mathcal W_1^{(\beta)}[ \mathcal J_1]$, $\mathcal W_2^{(\beta)}[ \mathcal J_1,  \mathcal J_2]$, $\mathcal W_3^{(\beta)}[ \mathcal J_1,  \mathcal J_2,  \mathcal J_3]$, ..., which are defined as
\begin{equation}
  \label{eq_cumW1}
\mathcal W_1^{(\beta)}[ \mathcal J_1]= \overline{\mathcal W_h^{(\beta)}[ \mathcal J_1]}
\end{equation}
\begin{equation}
\begin{aligned}
  \label{eq_cumW2}
\mathcal W_2^{(\beta)}[ \mathcal J_1,  \mathcal J_2]= &\overline{\mathcal W_h^{(\beta)}[ \mathcal J_1]\mathcal W_h^{(\beta)}[ \mathcal J_2]}\\&-\overline{\mathcal W_h^{(\beta)}[ \mathcal J_1]}\,\, \overline{\mathcal W_h^{(\beta)}[ \mathcal J_2]},
\end{aligned}
\end{equation}
etc.  \textit{Generic}, \textit{i.e.} independently tunable, arguments require the introduction of several copies or replicas of the original system, each with the same bare disorder (random field) but coupled to \textit{different} external supersources. It is worth stressing that this is \textit{not} what is done in the Parisi-Sourlas supersymmetric approach nor in the conventional replica trick. In the former, a single copy of the system is considered (see section \ref{sec:model}) and in the latter the sources acting on the replicas are all taken equal. As a consequence, in both cases, one has only access to cumulants in the specific configuration with all arguments equal. Quite differently in the present formalism, we consider multiple copies or replicas and supersources that explicitly break the (permutational) symmetry among these replicas.\cite{mouhanna10} We note in passing that, by construction, this takes care of the problem coming from having to average the logarithm of the partition function over disorder when $\beta\neq 0$.

\subsection{Multi-copy superfield formalism}
\label{sec:multi-copy}

The cumulants of $\mathcal{W}_h^{(\beta)}$ for generic arguments can be obtained from the average over the bare disorder of the extension of Eq.~(\ref{eq_part_funcbeta}) to an arbitrary large number $n$ of copies submitted to independently controllable (super)sources, namely,
\begin{equation}
\begin{aligned}
  \label{eq_part_func_multicopy0}
\exp (\mathcal W^{(\beta)}[\{\mathcal J_a\}]) &=\overline{\prod_{a=1}^n \mathcal{Z}_h^{(\beta)}[\mathcal J_a]}\\&=\overline{\exp(\sum_{a=1}^n\mathcal W_h^{(\beta)}[ \mathcal J_a])}.
\end{aligned}
\end{equation}
With the help of the curved superspace introduced above and by combining Eq.~(\ref{eq_part_funcbeta}) and Eq.~(\ref{eq_part_func_multicopy0}), we end up with a superfield theory associated with the following partition function:
\begin{equation}
\begin{aligned}
  \label{eq_part_func_multicopy}
&\mathcal Z^{(\beta)}[\{\mathcal J_a\}]=\exp (\mathcal W^{(\beta)}[\{\mathcal J_a\}]) \\&= \int \prod_{a=1}^{n}\mathcal D\Phi_a  \exp \bigg(-S^{(\beta)}[\{\Phi_a\}] +  \sum_{a=1}^{n} \int_{\underline x}  \mathcal J_a(\underline x)  \Phi_a(\underline x)\bigg),
\end{aligned}
\end{equation}
where the multicopy action is given by
\begin{equation}
\begin{aligned}
\label{eq_superaction_multicopy}
S^{(\beta)}&[\{\Phi_a\}] = \sum_{a=1}^{n} \int_{\underline{x}} [ \frac 12 (\partial_{\mu}\Phi_a(\underline{x}))^2+U_{B}(\Phi_a(\underline{x}))] \\&- \frac{\Delta_B}{2}\sum_{a_1=1}^{n}\sum_{a_2=1}^{n} \int_{x}\int_{\underline{\theta}_1\underline{\theta}_2} \Phi_{a_1}(x,\underline{\theta}_1)\Phi_{a_2}(x,\underline{\theta}_2)
\end{aligned}
\end{equation}
and it should be kept in mind that the integral over the Grassmann coordinates include a metric factor due to the curvature $\beta$ (as a result, $\int_{\underline x}\equiv \int_x \int \int (1+\beta \bar\theta \theta)d\theta d\bar\theta$). Note that for $n\geq 2$ the first (kinetic) and last (disorder-induced) terms in the above expression of the multicopy action can no longer be combined and simply expressed with the super-Laplacian, as it was the case for the $1$-copy action $S_{SS}[\Phi]$ in the absence of curvature ($\beta=0$) which is defined in Eq.~(\ref{eq_susyaction2}).

The cumulants of $\mathcal{W}_h^{(\beta)}$ can now be generated by the expansion in increasing number of sums over copies,
\begin{equation}
\begin{aligned}
\label{eq_gen_func_multicopy}
\mathcal W^{(\beta)}[\{\mathcal J_a\}] = \sum_{p\geq1}\sum_{a_1=1}^n...\sum_{a_p =1}^n
\frac {1}{p!} \mathcal{W}_{p}^{(\beta)}[\mathcal{J}_{a_1},...,\mathcal{J}_{a_p}],
\end{aligned}
\end{equation}
where the $p$th cumulant $\mathcal{W}_{p}^{(\beta)}$ is fully symmetric under any permutation of its arguments (and independent of $n$). As is obvious from the above equation, at least $p$ distinct copies must be considered to describe the $p$th cumulant with generic arguments, so that an arbitrary large number of copies (or replicas) are needed to generate all cumulants.

By using Eq.~(\ref{eq_ultralocal_broken_beta}), the first cumulant can be formally rewritten as
\begin{equation}
\begin{aligned}
  \label{eq_cumW1_formal}
\mathcal W_1^{(\beta)}[ \mathcal J_1]= \int_{\underline{\theta}_1}W_1^{(\beta)}[1]
\end{aligned}
\end{equation}
where $W_1^{(\beta)}[1]$ is a short-hand notation for indicating a functional of $\mathcal J_1(\underline{\theta}_1)$,  $(1+\frac{\beta}{2} \bar{\theta}_1 \theta_1) \partial_{\underline{\theta}_1}\mathcal J_1(\underline{\theta}_1)$, and $\Delta_{\underline{\theta}_1}\mathcal J_1(\underline{\theta}_1)$ (the functional character comes from the Euclidean dependence that remains completely general at this point); there is no additional explicit dependence on Grassmann coordinates in $W_1^{(\beta)}$. The second cumulants reads
\begin{equation}
\begin{aligned}
  \label{eq_cumW2_formal}
\mathcal W_2^{(\beta)}[ \mathcal J_1,  \mathcal J_2]=\int_{\underline{\theta}_1}\int_{\underline{\theta}_2} W_2^{(\beta)}[1,2],
\end{aligned}
\end{equation}
with an obvious extension of the above notation, and similar expressions hold for the higher-order cumulants. A physical interpretation of the cumulants when the random functional  $\mathcal W_h^{(\beta)}$ is ``ultralocal'' in the Grassmann coordinates will be given in the next section.

\subsection{Legendre transform and effective action in a curved superspace}

Due to the specific form of the source term in the presence of curvature, \textit{i.e.}, explicitly, 
$\sum_a  \int_{x}\int \int d\theta d\bar{\theta} (1+\beta \bar{\theta}\theta) \Phi_a(x,\underline\theta) \mathcal J_a(x,\underline\theta)$, the functional $\mathcal W^{(\beta)}[\{\mathcal J_a\}]$ does not exactly generate the average of the superfields; rather, one has an extra metric factor,
\begin{equation}
  \label{eq_magn}
\frac{\delta \mathcal W^{(\beta)}[\{\mathcal J_f\}]}{\delta \mathcal J_a(x,\underline\theta)}=(1+\beta\bar \theta \theta)\Phi_a(x,\underline\theta).
\end{equation}
The Legendre transform defining the effective action is then expressed as
\begin{equation}
  \label{eq_legendre_multicopy}
\Gamma^{(\beta)}[\{\Phi_a\}]=-\mathcal W^{(\beta)}[\{\mathcal J_a\}]+\sum_a \int_x \int_{\underline\theta} \Phi_a(\underline\theta,x) \mathcal J_a(\underline\theta,x).
\end{equation}
From this equation, one can determine the relation between the second functional derivatives $\Gamma^{(2)}$ and $\mathcal W^{(2)}$ (where we have omitted the superscript $(\beta)$ to alleviate the notation, as we shall do each time superscripts indicating functional derivatives are involved):
\begin{equation}
\label{eq_legendre_invert0}
\begin{aligned}
&\frac{\delta\Phi_a(x_1,\underline\theta_1)}{\delta\Phi_b(x_2,\underline\theta_2)}=\delta_{ab}\delta^{(d)}(x_1-x_2)\delta_{\underline\theta_1 \underline\theta_2}\\&=\sum_c\int d\bar\theta_3 d\theta_3\int_{x_3} \frac{\delta\Phi_a(x_1,\underline\theta_1)}{\delta
 \mathcal  J_c(x_3,\underline\theta_3)}\, \frac{\delta \mathcal J_c(x_3,\underline\theta_3)}{\delta\Phi_b(x_2,\underline\theta_2)}\\
  &=\sum_c\int_{\underline\theta_3}(1-\beta \bar \theta_3\theta_3)\int_{x_3}
  \frac{\delta\Phi_a(x_1,\underline\theta_1)}{\delta \mathcal J_c(x_3,\underline\theta_3)}\, \frac{\delta
    \mathcal J_c(x_3,\underline\theta_3)}{\delta\Phi_b(x_2,\underline\theta_2)}
    \end{aligned}
\end{equation}
so that
\begin{equation}
\label{eq_legendre_invert}
\begin{aligned}
&(1+\beta \bar\theta_1 \theta_1) \delta_{ab}\delta^{(d)}(x_1-x_2)\delta_{\underline\theta_1 \underline\theta_2}=\sum_c\int_{\underline\theta_3}\int_{x_3}(1-2\beta\bar\theta_3\theta_3)\\&
\times  \frac{\delta^2\mathcal W^{(\beta)}}{\delta \mathcal J_a(x_1,\underline\theta_1) \delta \mathcal J_c(x_3,\underline\theta_3)}
\,  \frac{\delta^2 \Gamma^{(\beta)}}{\delta \Phi_c(x_3,\underline\theta_3)\delta\Phi_b(x_2,\underline\theta_2)},
\end{aligned}
\end{equation}
where $\delta_{\underline{\theta}_1 \underline{\theta}_2}=\delta_{\bar{\theta}_1 \bar{\theta}_2} \delta_{{\theta}_1 {\theta}_2}= (\bar{\theta}_1 - \bar{\theta}_2) ({\theta}_1 - {\theta}_2)$ (with this definition, $\int_{\underline{\theta}_2}\delta_{\underline{\theta}_1 \underline{\theta}_2}= 1+ \beta \bar \theta_1 \theta_1$). $\Gamma^{(2)}$ and $\mathcal W^{(2)}$ are thus essentially inverse operators, provided that the effect of the curvature of the Grassmann subspace is appropriately taken into account.

Like the generating functional $\mathcal W^{(\beta)}[\{ \mathcal J_a\}]$, the effective action $\Gamma^{(\beta)}[\{\Phi_a\}]$ can be expanded in increasing number of unrestricted sums over copies (considering now the superfields $\{\Phi_a\}$ as fundamental variables in place of the supersources $\{ \mathcal J_a\}$):
\begin{equation}
\begin{aligned}
\label{eq_multicopy1_app}
\Gamma^{(\beta)}[\{\Phi_{a}\}] = \sum_{p\geq1} \sum_{a_1=1}^n...\sum_{a_p =1}^n \frac{(-1)^{p-1}}{p!}
\mathsf \Gamma_p^{(\beta)}[\Phi_{a_1},...,\Phi_{a_p}],
\end{aligned}
\end{equation}
where $\mathsf \Gamma_p^{(\beta)}$  is a fully symmetric functional of its $p$ arguments whose functional form is independent of the number $n$ of copies. The sign $(-1)^{p-1}$ is chosen for further convenience.

The above expansion is similar to that in number of unrestricted (free) replica sums developed in Ref.~[\onlinecite{tarjus08}]. In consequence, one can apply, \textit{mutatis mutandis}, the results of 
Ref.~[\onlinecite{tarjus08}] and relate the $\mathsf{\Gamma}_p^{(\beta)}$'s to the cumulants $\mathcal W_p^{(\beta)}$ as follows [we drop for simplicity the superscript $(\beta)$ in the expressions]. The first-order term $\mathsf{\Gamma}_1^{(\beta)}[\Phi]$ is the Legendre transform of $\mathcal W_1^{(\beta)}[\mathcal J]$, namely,
\begin{equation}
\label{legendre_Gamma_1}
\mathsf{\Gamma}_1[\Phi] = - \mathcal W_1[\mathcal J] +  \int_{\underline x} \mathcal J(\underline x) \Phi(\underline x),
\end{equation}
with
\begin{equation}
\label{legendre_Phi_curved}
(1+\beta \bar\theta \theta) \Phi(\underline x)=\frac{\delta \mathcal W_1 [\mathcal J ] }{\delta \mathcal J(\underline x)}= \mathcal W_{1;\underline x}^{(1)}[\mathcal J],
\end{equation}
whereas the second-order term $\mathsf{\Gamma}_2^{(\beta)}[\Phi_1,  \Phi_2]$ is given by
\begin{equation}
  \label{eq_cumG2}
\mathsf{\Gamma}_2[\Phi_1,  \Phi_2] =\mathcal W_2[\mathcal J[ \Phi_1],\mathcal J[\Phi_2]],
\end{equation}
where  $\mathcal J[ \Phi]$ is the \textit{nonrandom} source defined via the inverse of the Legendre transform relation in Eq. (\ref{legendre_Phi_curved}), \textit{i.e.}, 
\begin{equation}
\label{eq_nonrandom_source_curved}
(1+\beta \bar \theta \theta) \mathcal J[\Phi](\underline x)=\mathsf{\Gamma}_{1;\underline x}^{(1)} [\Phi].
\end{equation}
The above expression of $\mathsf{\Gamma}_2$ motivates our choice of signs for the terms of the expansion of the effective action $\Gamma$, Eq.~(\ref{eq_multicopy1_app}): $\mathsf{\Gamma}_2[\Phi_1,  \Phi_2]$ is directly the second cumulant of $\mathcal W_h[\mathcal J]$ (with the proper choice of $\mathcal J[ \Phi]$). 

For the higher-order terms, one finds after some lenghty but straightforward manipulations (see Appendix~\ref{appendix:expansion_copies})
\begin{equation}
\begin{split}
  \label{eq_cumG3}
&\mathsf{\Gamma}_3 [ \Phi_1,  \Phi_2, \Phi_3]  =  - \mathcal W_3[\mathcal J[ \Phi_1],\mathcal J[ \Phi_2],\mathcal J[ \Phi_3]]  +   \int_{\underline x}\int_{ \underline x'}\\&(1-\beta \bar \theta \theta)(1-\beta \bar \theta' \theta') \bigg \{\mathcal W_{2;\underline x,.}^{(10)}[\mathcal J[ \Phi_1], \mathcal J[ \Phi_2]] \mathsf{\Gamma}_{1;\underline x \underline x'}^{(2)}[\mathcal J[ \Phi_1]    \\& \times \mathcal W_{2;\underline x',.}^{(10)}[\mathcal J[ \Phi_1],\mathcal J[ \Phi_3]] + perm (123)\bigg \} ,
\end{split}
\end{equation}
where $perm (123)$ denotes the two additional terms obtained by circular permutations of the superfields $ \Phi_1,  \Phi_2,  \Phi_3$ and $\mathsf{\Gamma}_{1}^{(2)}$ is the inverse of $\mathcal W_1^{(2)}$ through an equation similar to Eq.~(\ref{eq_legendre_invert}), and so on. This procedure leads to a unique functional form for $\mathsf \Gamma_p^{(\beta)}$ which is expressed in terms of cumulants of $\mathcal W_h^{(\beta)}$ of order $p$ or less. (Note that this guarantees that the functional form is independent of the number $n$ of copies.) 

As illustrated by Eq.~(\ref{eq_cumG3}), $\mathsf{\Gamma}_p^{(\beta)}[ \Phi_1, ...,  \Phi_p]$  for $p\geq 3$ cannot be directly taken as the $p$th  cumulant of a physically accessible random functional, in particular not of the disorder-dependent Legendre transform of $\mathcal W_h^{(\beta)}[\mathcal J[\Phi]]$. However, as it can be expressed in terms of such cumulants of order equal to or lower than $p$, we will call for simplicity the $\mathsf{\Gamma}_p^{(\beta)}$'s ``cumulants of the renormalized disorder'' (which is true for $p=2$) in what follows. See also section V-B below.

\subsection{Symmetries and WT identities for multiple copies}
\label{sec:symmetries}

The presence of curvature in the Grassmann subspace and of multiple copies still allows invariance of the theory under a large group of transformations. The multi-copy action, which is given in Eq.~(\ref{eq_superaction_multicopy}), is indeed invariant under the following symmetries:

(i) The permutations $S_n$ among copies and a global $Z_2$ symmetry;

(ii) the global rotations and translations in the $d$-dimensional Euclidean space;

(iii) the symplectic transformations with generators $\bar t=\bar{\theta}\partial_{\theta},t= \theta \partial_{\bar{\theta}}$  and $N=\bar{\theta}\partial_{\bar{\theta}}-\theta \partial_{\theta}$ acting on the  $2$-dimensional curved Grassmann subspace, independently for each copy;

(iv) the two isometries of the curved Grassmann subspace that generalize the translations of flat space, independently for each copy; their generators are $(1-\beta \bar \theta \theta)\partial_{\theta}$ and $(1-\beta \bar \theta \theta)\partial_{\bar \theta}$.

(v)  It was shown in Ref.~[\onlinecite{wschebor09}] that the above isometries (iii) and (iv) are the only possible ones in the presence of a nonzero curvature $\beta$. However, when the curvature is set to zero and, in addition, when restricting the supersources such that in all copies except a given copy $a$, the components $\hat{J}_b = \bar{K}_b = K_b = 0$ ($b\neq a$), the partition function in Eq.~(\ref{eq_part_func_multicopy})  is invariant under the superrotations considered in section III-B. In this case, one is effectively back to a 1-copy system since it is found that
\begin{equation}
\begin{aligned}
  \label{eq_part_func_1copy}
\mathcal Z^{(\beta=0)}&[\mathcal J_a, \{\mathcal J_b=J_b\}]=\\& \int\mathcal D\Phi_a  \exp \big[-S_{ss}[\Phi_a] +   \int_{\underline x}  \mathcal J_a(\underline x)  \Phi_a(\underline x)\big],
\end{aligned}
\end{equation}
where $S_{ss}$ is given in Eq.~(\ref{eq_susyaction2}) and where we have used that the $1$-copy partition function $\mathcal Z^{(\beta=0)}[\mathcal J_b=J_b,h]=1$ for all copies $b\neq a$. Invariance under the superrotations follows directly.

The above continuous (super) symmetries imply a set of WT identities satisfied by the generating functionals $\mathcal W^{(\beta)}[\{\mathcal J_a\}]$ and $\Gamma^{(\beta)}[\{\Phi_a\}]$. Denoting by $\mathcal D_{\underline{\theta}}$ any one of the generators acting on the Grassmann coordinates $\underline{\theta}$ in a chosen copy $a$, one finds that
\begin{equation}
\label{eq_ward_multicopycurved_W}
\int_{\underline{x}}\mathcal J_a(\underline{x})\mathcal D_{\underline{\theta}}\big (\left [1-\beta \bar\theta \theta \right]\mathcal W_{a \underline{x}}^{(1)}[\{ \mathcal J_f\}]\big )=0,
\end{equation}
where we have again dropped the superscript $(\beta)$ to avoid confusion with the superscripts indicating functional derivation. This WT identity carries over to the effective action:
\begin{equation}
\label{eq_ward_multicopycurved0_gamma}
\int_{\underline{x}} \left (1-\beta \bar\theta \theta \right) \Gamma_{a \underline{x}}^{(1)}[\{ \Phi_f\}]\mathcal D_{\underline{\theta}} \Phi_a(\underline{x})=0,
\end{equation}
where we recall that the integral over the Grassmann coordinates comes with a factor $(1+\beta \bar\theta \theta)$.

Through differentiation with respect to the superfield $\Phi_a$, the above equation leads to WT identities for the 1PI vertices. For the symplectic transformations, by using the fact that Eq.~(\ref{eq_ward_multicopycurved0_gamma}) can be rewritten as
\begin{equation}
\label{eq_ward_multicopycurved1_Gamma}
\int_{\underline{x}} \left (1-\beta \bar\theta \theta \right)  \Phi_a(\underline{x})\, \bar t \, \Gamma_{a \underline{x}}^{(1)}[\{ \Phi_f\}]=0,
\end{equation}
one finds for $p\geq 1$
\begin{equation}
\begin{aligned}
\label{eq_wardcurved1_p}
&\big(\sum_{q=1}^p \delta_{a a_q}\, \bar t_q\big) \Gamma_{(a_1\underline{x}_1)...(a_p \underline{x}_{p})}^{(p)}[\{\Phi_f\}] \\&+ \int_{\underline{x}_{p+1}}\Phi_a(\underline{x}_{p+1})\, \bar t_{p+1}\Gamma_{(a_1\underline{x}_1)...(a_{p+1}\underline{x}_{p+1})}^{(p+1)}[\{\Phi_f\}]=0,
\end{aligned}
\end{equation}
where $\bar t_q=\bar \theta_q \partial_{\theta_q}$, and similarly for
the transformations $t$ and $N$. 

For the generalized translations $(1-\beta \bar \theta \theta)\partial_{\theta}$ and $(1-\beta \bar \theta \theta)\partial_{\bar \theta}$, a little more care is needed; Eq.~(\ref{eq_ward_multicopycurved0_gamma}) can now be reexpressed as
\begin{equation}
\label{eq_ward_multicopycurved2_Gamma}
\int_{\underline{x}} \left (1-\beta \bar\theta \theta \right)  \Phi_a(\underline{x})\,\partial_{\theta}\big([1-\beta \bar \theta \theta] \Gamma_{a \underline{x}}^{(1)}[\{ \Phi_f\}]\big)=0,
\end{equation}
so that the WT identities for the 1PI vertices become
\begin{equation}
\begin{aligned}
\label{eq_wardcurved2_p}
&\big(\sum_{q=1}^p \delta_{a a_q}\, \partial_{\theta_q}[1-\beta \bar \theta_q \theta_q]\big) \Gamma_{(a_1\underline{x}_1)...(a_p \underline{x}_{p})}^{(p)}[\{\Phi_f\}] + \int_{\underline{x}_{p+1}}\\&\Phi_a(\underline{x}_{p+1})\partial_{\theta_{p+1}}[1-\beta \bar \theta_{p+1} \theta_{p+1}]\Gamma_{(a_1\underline{x}_1)...(a_{p+1}\underline{x}_{p+1})}^{(p+1)}[\{\Phi_f\}]=0
\end{aligned}
\end{equation}
and similarly with the other generator. These WT identities generalize those already encountered at the $1$-copy level for a flat Grassmann subspace in Sec. II-B.

In addition, when the curvature $\beta$ is set to zero and the supersources are restricted as discussed above, the superrotational invariance of the partition function also leads to WT identities. After making use of the invariance under translations and symplectic transformations in the (flat) Grassmann subspace, one finds that the solution of the Legendre relation for a supersource $\mathcal J_b(\underline{x})\equiv J_b(x)$, \textit{i.e.},
\begin{equation}
  \label{eq_legendre_reduction1copy}
\Gamma_{(b,\underline{x})}^{(1)}[\Phi_b,\{\Phi_f\}']\big \vert_{\beta=0}= J_b(x),
\end{equation}
where $\left\lbrace  \Phi_f\right\rbrace'$ denotes the set of all copy superfields but $\Phi_b$, satisfies $\Phi_b(\underline{x})\equiv \phi_b(x)$ (with $\psi_b(x) = \bar{\psi}_b(x) = \hat{\phi}_b(x)=0$); one then derives the following WT identities for the superrotation invariance:
\begin{equation}
\label{eq_ward_0_1copy}
\int_{\underline{x}}\Phi_a(\underline{x})\mathcal Q_{\underline{x}}\Gamma_{\underline{x}}^{(1)}[\Phi_a,\{\Phi_b=\phi_b\}]\big \vert_{\beta=0}=0,
\end{equation}
where the field components $\psi_b(x), \bar{\psi}_b(x), \hat{\phi}_b(x)$ have been set to zero in all copies $b\neq a$ and $\mathcal Q_{\underline{x}}$ is defined in section II-B. A similar expression holds with the generators $\bar{\mathcal Q}_{\underline{x}}$. By functional differentiation, WT identites are also obtained for the higher-order 1PI vertices. This will be further exploited in the next section.

\section{Exploring the formal consequences of ``Grassmannian ultralocality''}
\label{sec:ultralocal}

\subsection{Expansion in ``ultralocal'' cumulants}

Assume now that the random functional $\mathcal{W}_h^{(\beta)}$ is ``ultra-local'' in the Grassmann coordinates, which, as discussed before, implies that a unique configuration is included in the computation of the random generating functional for each realization of the random field. In this case, the expansion of the generating functional $\mathcal W^{(\beta)}[\{\mathcal J_a\}]$ in increasing number of sums over copies, Eq.~(\ref{eq_gen_func_multicopy}), coincides with a ``multilocal'' expansion in Grassmann coordinates, \textit{i.e.},
\begin{equation}
\begin{aligned}
\label{eq_multilocal_multicopy}
\mathcal W_p^{(\beta)}[\mathcal J_1,..., \mathcal J_p] = \int_{\underline{\theta}_1}...\int_{\underline{\theta}_p}
W_{p}^{(\beta)}[\mathcal{J}_{1}(\underline{\theta}_1),...,\mathcal{J}_{p}(\underline{\theta}_p)],
\end{aligned}
\end{equation}
where $W_{p}^{(\beta)}$ no longer depends on the derivatives of the supersources in the Grassmann directions and is the $p$th cumulant of the ``ultralocal'' functional $W_h^{(\beta)}[\mathcal{J}(\underline{\theta})]$ defined in Eq.~(\ref{eq_ultralocal_beta1}). Note that due to the assumed uniqueness of the solution included in the computation of the random generating functional,  $W_h[\mathcal{J}(\underline{\theta})]$ as well as its cumulants,
\begin{equation}
  \label{eq_cumW1_1copy}
W_1[ \mathcal J_{1}(\underline{\theta}_1)]= \overline{W_h[ \mathcal J_{1}(\underline{\theta}_1)]},
\end{equation}
\begin{equation}
  \label{eq_cumW2_1copy}
\begin{aligned}
W_2[ \mathcal J_{1}(\underline{\theta}_1), \mathcal J_{2}(\underline{\theta}_2)]= &\overline{W_h[ \mathcal J_{1}(\underline{\theta}_1)]W_h[ \mathcal J_{2}(\underline{\theta}_2)]}\\&-\overline{W_h[ \mathcal J_{1}(\underline{\theta}_1)]}\, \overline{W_h[ \mathcal J_{2}(\underline{\theta}_2)]},
\end{aligned}
\end{equation}
etc, are independent of the curvature $\beta$ (see section III-B). By an abuse of language, we will characterize the cumulants  obtained from an ``ultralocal'' random functional as ``ultralocal'' in the following.

A physical interpretation of $\mathcal W_h^{(\beta)}$ and its cumulants is next obtained by restricting the supersources to their physical component, $\mathcal J(\underline{x})\equiv J(x)$, which plays the role of an applied magnetic field. [Said otherwise, this amounts to considering supersources that are uniform in the Grassmann subspace, \textit{i.e.} to set $\theta = \bar \theta =0$ in the defining expression in Eq.~(\ref{eq_supersource}).]  $W_h$ is then given by Eq.~(\ref{eq_ultralocal_beta2}) (properly generalized to $d$-dimensional Euclidean space) and, as stated above, is independent of the auxiliary parameter $\beta$.  Its first derivative is by construction equal to the ground-state configuration $\varphi_0[h+J]$. The first cumulant $W_1[J]$ then gives access to the thermodynamics and its first derivative is the average of the physical field (the ``magnetization''), which corresponds at zero temperature to the ground-state configuration. Its higher-order derivatives
\begin{equation}
 \label{eq_notation_partderiv_W1}
W_{1; x_1 ...  x_p}^{(p)}[ J]=\frac{\delta^p W_1[ J]}{\delta J( x_1)... \delta J( x_p)}
\end{equation}
correspond to Green's functions that are related to what is known in the literature on disordered systems as  ``connected'' correlation functions of the $\varphi$ field. For instance, the second derivative is related to the linear response of the magnetization to a change of the applied magnetic field. 

The higher-order cumulants describe the distribution of the renormalized disorder. They generate through differentiation, 
and after setting all sources equal ($J_1=J_2=\cdots=J$), the so-called ``disconnected'' correlation functions of the original system. We introduce the notation
\begin{equation}
 \label{eq_notation_partderiv_W2}
\begin{split}
W&_{p;x_{11}..x_{1q_1},...,x_{p 1}..x_{pq_p}}^{(q_1...q_p)}[J_1... J_p]=\\& \frac{\delta^{q_1+...+q_p} W_p[J_1... J_p]}{\delta J_1(x_{11})..\delta J_1(x_{1q_1})...\delta J_p (x_{p 1})..\delta J_p (x_{pq_p})}.
\end{split}
\end{equation}
Then, for instance, $W_{2; x_1,x_2}^{(11)}[J, J]$ is equal to the standard two-point ``disconnected'' correlation function defined in Eq.~(\ref{eq_physical_disconn}), which, in magnetic systems, is directly accessible to experimental measurements.\cite{footnote011}

\subsection{Expansion of the effective action}

From the above equations and the definition of the Legendre transform, one derives that the effective action has also a ``multilocal'' expansion in Grassmann coordinates, which corresponds to having an ``ultralocal'' form for each term of the expansion of $\Gamma^{(\beta)}$ in number of copies [Eq.~(\ref{eq_multicopy1_app})], \textit{i.e.},
\begin{equation}
\begin{aligned}
\label{eq_cumulant_p_ultralocal}          
\mathsf \Gamma_p^{(\beta)}[\Phi_{a_1},...,\Phi_{a_p}] = \int_{\underline{\theta}_1}...\int_{\underline{\theta}_p}
\Gamma_{p}^{(\beta)}[\Phi(\underline{\theta}_1),...,\Phi(\underline{\theta}_p)],
\end{aligned}
\end{equation}
where $\Gamma_{p}^{(\beta)}$ does not contain any additional explicit dependence on Grassmann coordinates (nor any dependence on the derivatives of the superfields in the Grassmann directions). As before, the dependence on the Euclidean coordinates, which actually still makes the $\Gamma_p^{(\beta)}$'s nonlocal functionals of the superfields, is left implicit. The proof is easily derived from the expression of the $\mathsf \Gamma_p^{(\beta)}$'s in terms the $\mathcal W_q^{(\beta)}$'s with $q \leq p$ and the ``ultralocal'' property of the latter. A more specific discussion of the relation between the  $\Gamma_p^{(\beta)}$'s and the $W_p^{(\beta)}$'s is provided below.

From the above property one can derive expressions for the proper (1PI) vertices, defined as
\begin{equation}
\Gamma_{(a_1,\underline{x}_1), ... ,(a_p,\underline{x}_p)}^{(p)}[\left\lbrace \Phi_a \right\rbrace ]=\frac{\delta^p \Gamma^{(\beta)} [\left\lbrace \Phi_a \right\rbrace ]}{\delta \Phi_{a_1}(\underline{x}_1)... \delta \Phi_{a_p}(\underline{x}_p)},
\end{equation}
which will prove useful in the following. After specializing to fields in the physical subspace, $\Phi_a(\underline{x})\equiv \phi_a(x)$, which are relevant for the equilibrium behavior of the RFIM, and introducing the notation
\begin{equation}
\begin{aligned}
\label{eq_deriv_Gamma_p}
\Gamma&_{p;x_{11}..x_{1q_1},...,x_{p_1}..x_{pq_p}}^{(q_1...q_p)}[\phi_1... \phi_p]=\\& \frac{\delta^{q_1+...+q_p} \Gamma_p[\phi_1... \phi_p]}{\delta \phi_1(x_{11})..\delta \phi_1(x_{1q_1})...\delta \phi_p (x_{p_1})..\delta \phi_p (x_{pq_p})},
\end{aligned}
\end{equation}
one obtains the following expression for the $1$-point  proper vertex,
\begin{equation}
\label{eq_expand_Gamma_1}
\begin{aligned}
&\Gamma_{(a_1\underline{x}_1)}^{(1)}[\left\lbrace \phi_a \right\rbrace ]=(1+\beta\overline \theta_1\theta_1)\bigg \{\Gamma_{1;x_1}^{(1)}[\phi_{a_1}]-  \beta \, \times \\& \sum_{a_2}\Gamma_{2;x_1, .}^{(10)}[\phi_{a_1},\phi_{a_2}]+ \beta^2\sum_{a_2,a_3}\Gamma_{3;x_1,.,.}^{(100)}[\phi_{a_1},\phi_{a_2},\phi_{a_3}] + \cdots \bigg \},
\end{aligned}
\end{equation}
and for the $2$-point proper vertex,
\begin{equation}
\begin{aligned}
\label{eq_expand_Gamma_2}
\Gamma^{(2)}_{(a_1\underline{x}_1),(a_2\underline{x}_2)}&[\left\lbrace \phi_a \right\rbrace ] \\&= \delta_{a_1a_2}\widehat{\Gamma}_{a_1;\underline{x}_1\underline{x}_2}^{(2)}[\left\lbrace \phi_a \right\rbrace ] + 
\widetilde{\Gamma}_{(a_1\underline{x}_1),(a_2\underline{x}_2)}^{(2)}[\left\lbrace \phi_a \right\rbrace ]
\end{aligned}
\end{equation}
with
\begin{equation}
\begin{aligned}
\label{eq_expand_Gamma_2hat}
\widehat{\Gamma}_{a_1;\underline{x}_1\underline{x}_2}^{(2)}[\left\lbrace \phi_a \right\rbrace ] =&(1+\beta\overline \theta_1\theta_1)\delta_{\underline{\theta}_1\underline{\theta}_2} \bigg \{\Gamma_{1;x_1x_2}^{(2)}[\phi_{a_1}]\\& - \beta \sum_{a_2}\Gamma_{2;x_1x_2, .}^{(20)}[\phi_{a_1},\phi_{a_2}]+ \cdots \bigg \},
\end{aligned}
\end{equation}
where $\delta_{\underline{\theta}_1 \underline{\theta}_2}$ is defined below Eq.~(\ref{eq_legendre_invert}), and
\begin{equation}
\begin{aligned}
\label{eq_expand_Gamma_2tilde}
&\widetilde{\Gamma}_{(a_1\underline{x}_1)(a_2\underline{x}_2)}^{(2)}[\left\lbrace \phi_a \right\rbrace ]=-(1+\beta\overline \theta_1\theta_1)(1+\beta\overline \theta_2\theta_2)\times \\&\bigg \{\Gamma_{2;x_1,x_2}^{(11)}[\phi_{a_1},\phi_{a_2}] - \beta \sum_{a_3}\Gamma_{3;x_1,x_2, .}^{(110)}[\phi_{a_1},\phi_{a_2},\phi_{a_3}]+ \cdots \bigg \}.
\end{aligned}
\end{equation}
As the order increases, the formulas go along the same lines but become more involved; for instance, one finds
\begin{equation}
\begin{aligned}
\label{eq_expand_Gamma_3}
&\Gamma_{(a_1\underline{x}_1),(a_2\underline{x}_2),(a_3\underline{x}_3)}^{(3)}[\left\lbrace \phi_a \right\rbrace ] = \delta_{a_1a_2a_3}(1+\beta\overline \theta_1\theta_1) \delta_{\underline{\theta}_1\underline{\theta}_2\underline{\theta}_3}\\& \bigg \{\Gamma_{1;x_1x_2x_3}^{(3)}[\phi_{a_1}] - \beta \sum_{a_4}\Gamma_{2;x_1x_2x_3,.}^{(30)}[\phi_{a_1},\phi_{a_4}]+ \cdots \bigg \}- \\&\bigg(  \delta_{a_1a_2}(1+\beta\overline \theta_1\theta_1)(1+\beta\overline \theta_2\theta_2)\delta_{\underline{\theta}_1\underline{\theta}_2} \bigg \{ \Gamma_{2;x_1x_2,x_3}^{(21)}[\phi_{a_1},\phi_{a_3}]- \\&\beta \sum_{a_4}\Gamma_{3;x_1x_2,x_3,. }^{(210)}[\phi_{a_1},\phi_{a_3},\phi_{a_4}]+ \cdots \bigg \}+ perm(123) \bigg) +\\&(1+\beta\overline \theta_1\theta_1)(1+\beta\overline \theta_2\theta_2)(1+\beta\overline \theta_3\theta_3)\bigg \{ \Gamma_{3;x_1,x_2,x_3}^{(111)}[\phi_{a_1},\phi_{a_2},\phi_{a_3}]\\& - \beta \sum_{a_4}\Gamma_{4;x_1,x_2,x_3,.}^{(1110)}[\phi_{a_1},\phi_{a_2},\phi_{a_3},\phi_{a_4}]+ \cdots \bigg \},
\end{aligned}
\end{equation}
etc, where $\delta_{\underline{\theta}_1 \underline{\theta}_2 \underline{\theta}_3}=\delta_{\underline{\theta}_1 \underline{\theta}_2} \delta_{\underline{\theta}_2 \underline{\theta}_3}$ and $perm(123)$ denotes the two additional terms obtained by circular permutations of the indices $1, 2, 3$.

\subsection{Interpretation of the $\Gamma_p$'s}

The relations between the $\Gamma_p$'s and the $W_p$'s follow from Eqs.~(\ref{legendre_Gamma_1}-\ref{eq_cumG3}) and Eq.~(\ref{eq_multilocal_multicopy}). They are straighforward for the first terms, but get more involved as the order increases.  To obtain full information, it is sufficient to consider configurations of the superfields that are uniform in the Grassmann subspace. (Note that the $W_p$'s being independent of $\beta$, so are the $\Gamma_p$'s.)

More precisely, one obtains that $\Gamma_1[ \phi]$ is the Legendre transform of $W_1[J]$, namely,
\begin{equation}
\label{legendre_gamma_1}
\Gamma_1[ \phi ] = - W_1[J] +  \int_{x} J(x) \phi( x),
\end{equation}
with
\begin{equation}
\label{legendre_phi}
\phi(x)=\frac{\delta W_1 [J ] }{\delta J(x)}= W_{1;x}^{(1)}[J] .
\end{equation}
The second-order term is given by
\begin{equation}
  \label{eq_cumg2}
\Gamma_2[\phi_1,  \phi_2] = W_2[J[ \phi_1], J[\phi_2]],
\end{equation}
where  $J[ \phi]$ is the (physical) \textit{nonrandom} source defined via the inverse of the Legendre transform relation in Eq.~(\ref{legendre_phi}), \textit{i.e.}, 
\begin{equation}
  \label{eq_nonrandom_source}
J[\phi]( x)= \Gamma_{1;x}^{(1)} [\phi].
\end{equation}
and the third-order one by
\begin{equation}
\begin{split}
  \label{eq_cumg3}
\Gamma_3 [ \phi_1,  \phi_2,&  \phi_3]  =  - W_3[J[ \phi_1], J[ \phi_2], J[ \phi_3]]  +  \\& \int_{x  y} \bigg \{ W_{2; x,.}^{(10)}[J[ \phi_1],  J[ \phi_2]] \left( W_{1}^{(2)}[J[ \phi_1]]\right)^{-1} _{ x\,  y}  \\& \times W_{2; y,.}^{(10)}[J[ \phi_1], J[ \phi_3]] + perm (123)\bigg \} ,
\end{split}
\end{equation}
etc..., where $perm (123)$ denotes the two additional terms obtained by circular permutations of the fields $ \phi_1,  \phi_2,  \phi_3$. As stated above, by an abuse of language, we will generically call the $\Gamma_p$'s ``cumulants of the renormalized disorder''.

As we did in Ref.~[\onlinecite{tarjus08}], it is also instructive to introduce a ``renormalized random field''  $\breve{h}[ \phi]( x)$ as the derivative of a random free-energy functional (it can equivalently be defined at the level of superfields),
\begin{equation}
\label{def_ren_randomfield}
 \breve{h}[ \phi](x)=- \frac{\delta }{\delta \phi(x)}\left(W_h[J[ \phi]] - \overline{W_h[J[ \phi]]} \right) ,
\end{equation}
where $J[\phi]$ is the nonrandom source given by Eq.~(\ref{eq_nonrandom_source}). The first moment of $\breve{h}[ \phi]$ is equal to zero by construction, and it is easy to derive that the $p$th cumulant ($p\geq 2$) is given by the derivative with respect to $ \phi_1, ...,  \phi_p$ of $W_p[J[ \phi_1], ..., J[ \phi_p]]$; this derivative in turn can  be  related to derivatives of the $\Gamma_q$'s with $q\leq p$; for instance,
\begin{equation}
  \label{eq_cumhren2}
\overline{\breve{h}[ \phi_1](x_1) \breve{h}[ \phi_2](x_2)}= \Gamma_{2; x_1,x_2}^{(11)}[ \phi_1, \phi_2].
\end{equation}
The cumulants of order $p\geq 3$  are  given by $\Gamma_{p; x_1,.. ,x_P}^{(1..1)}[\phi_1,...,\phi_p]$  plus additional terms involving higher derivatives of $\Gamma_q$'s with $q<p$; again, by an abuse of language, we will simply refer to $\Gamma_{p}^{(1...1)}$ as the ``$p$th cumulant of the renormalized random field''.\cite{tarjus08}

\subsection{WT identities for the superrotational invariance}

Finally, we make use of the above developments to derive explicitly the WT identities associated with superrotational invariance when $\beta=0$ and the multi-copy theory is reduced to a one-copy problem by an appropriate choice of the supersources (see above). We start with Eq.~(\ref{eq_ward_0_1copy}), which we functionally differentiate to introduce higher-order 1PI vertices, and we consider configurations of the superfields that are uniform in the Grassmann subspace but nonuniform in the Euclidean subspace. Thanks to the expressions in section V-B, we decompose each identity into components associated with different polynomials in the Grassmann coordinates. We then find several types of relations: first, relations merely expressing the translational invariance in Euclidean space, namely,
\begin{equation}
\begin{aligned}
\partial_{1\mu}\Gamma_{1;x_1}^{(1)}[\phi] = - \int_{x_2}\phi(x_2)
\partial_{2\mu}\Gamma_{1;x_1x_2}^{(2)}[\phi],
\end{aligned}
\end{equation}
\begin{equation}
\begin{aligned}
\left( \partial_{1\mu} +  \partial_{2\mu}\right) \Gamma_{1;x_1x_2}^{(2)}[\phi] = - \int_{x_3}\phi(x_3)
\partial_{3\mu}\Gamma_{1;x_1x_2x_3}^{(3)}[\phi],
\end{aligned}
\end{equation}
\begin{equation}
\begin{aligned}
&\left(  \partial_{1\mu} + \partial_{2\mu}\right)  \Gamma_{2;x_1,x_2}^{(11)}[\phi,\phi] = \\&- \int_{x_3}\phi(x_3)
\partial_{3\mu}\left( \Gamma_{2;x_1x_3, x_2}^{(21)}[\phi,\phi] + \Gamma_{2;x_1,x_2x_3}^{(12)}[\phi,\phi]\right) ,
\end{aligned}
\end{equation}
etc.

Secondly, we also obtain more specific and interesting identities that relate $\Gamma_p$ and $\Gamma_{p+1}$, \textit{e.g.},
\begin{equation}
\label{eq_ward_susy_nonunif_2}
\begin{aligned}
\partial_{1\mu}\Gamma_{2;x_1,x_2}^{(11)}[\phi,\phi] &- \frac{\Delta_B}{2} (x_1^\mu-x_2^\mu)\Gamma_{1;x_1x_2}^{(2)}[\phi]  = \\&- \int_{x_3}\phi(x_3)
\partial_{3\mu}\Gamma_{2;x_1x_3,x_2}^{(21)}[\phi,\phi],
\end{aligned}
\end{equation}
\begin{equation}
\begin{aligned}
\label{eq_ward_susy_nonunif_3}
&\partial_{1\mu}\Gamma_{3;x_1,x_2,x_3}^{(111)}[\phi,\phi,\phi] - \frac{\Delta_B}{2}\times \\&\big[ (x_1^\mu-x_2^\mu)\Gamma_{2;x_1x_2,x_3}^{(21)}[\phi,\phi]  + (x_1^\mu-x_3^\mu)\Gamma_{2;x_1x_3,x_2}^{(21)}[\phi,\phi] \big] \\& = - \int_{x_4}\phi(x_4)
\partial_{4\mu}\Gamma_{3;x_1x_4,x_2,x_3}^{(211)}[\phi,\phi,\phi],
\end{aligned}
\end{equation}
etc. For a uniform physical field $\phi(x) = \phi$, Eq.~(\ref{eq_ward_susy_nonunif_2}) becomes
\begin{equation}
\label{eq_ward_susy_unif_2}
\begin{aligned}
\partial_{1\mu}\Gamma_{2;x_1,x_2}^{(11)}(\phi,\phi) - \frac{\Delta_B}{2} (x_1^\mu-x_2^\mu)\Gamma_{1;x_1x_2}^{(2)}(\phi)  = 0,
\end{aligned}
\end{equation}
which after Fourier transforming and using the translational and rotational invariance in Euclidean space leads to
\begin{equation}
\label{eq_ward_susy_momentum_2}
\Gamma_{2}^{(11)}(q^2;\phi,\phi) = \Delta_B \partial_{q^2}\Gamma_{1}^{(2)}(q^2;\phi),
\end{equation}
with the obvious notation: $\Gamma_{2;q_1,q_2}^{(11)}(\phi,\phi)=(2\pi)^d \delta^{(d)}(q_1+q_2)\Gamma_{2}^{(11)}(q_1^2;\phi,\phi)$, etc. Similar identities are derived for the higher orders.

\section{NP-FRG in the superfield formalism}
\label{sec:NPFRG}

\subsection{Nonperturbative FRG}

The main purpose of the present work is to develop a formalism that allows one to study the supersymmetry, more specifically the invariance under superrotations, and its spontaneous breaking in the RFIM. To this end we upgrade our NP-FRG approach\cite{tarjus04,tissier06,tarjus08,tissier08} to a superfield formulation and use the tools developed in the previous sections to extend the Parisi-Sourlas formalism to the relevant situations in which multiple minima of the bare action may be present. The key points involve:

(1) adding an infrared regulator that enforces a progressive account of the fluctuations while ensuring that the initial condition of the RG flow satisfies the (super) symmetries of the action in Eq.~(\ref{eq_superaction_multicopy}) and corresponds to a unique solution of the stochastic field equation,

(2) considering copies of the original disordered system that give access to the full functional field dependence of the renormalized cumulants of the disorder,

(3) selecting the ground state through the introduction of a proper weighting factor and use of a curved superspace,

(4) using the WT identities associated with the (super) symmetries to ensure that neither the regulator nor the approximations explicitly break the latter.

Extending our previous work to the superfield theory, we introduce a generating functional of the correlation functions at the running scale $k$ for an arbitrary number $n$ of copies of the system (coupled to the same random field but submitted to different external sources) and a weighting factor involving the auxiliary parameter $\beta$,
\begin{equation}
 \begin{aligned}
 \label{eq_part_func}
\mathcal Z_k^{(\beta)}[\big \{ \mathcal J_a \big \}&]= \int \prod_{a=1}^n\mathcal D\Phi_a  \exp \big \lbrace - \Delta S_k^{(\beta)}[\left\lbrace \Phi_a \right\rbrace] \\&- S^{(\beta)}[\left\lbrace \Phi_a \right\rbrace]+ \sum_{a=1}^n \int_{\underline{x}}\mathcal J_a(\underline x) \Phi_a(\underline x))\big \rbrace,
 \end{aligned}
\end{equation}
where $S^{(\beta)}[\{\Phi_a \}]$ is defined in Eq.~(\ref{eq_superaction_multicopy}) and $\int_{\underline{x}}$ involves the curved superspace measure as in the preceding sections. As previously discussed, the $n$-copy action  is invariant under the $S_n$ permutational symmetry, the $Z_2$ symmetry, the translations and rotations in the Euclidean space, and, separately for each copy, under the isometries of the curved Grassmann subspace. The regulator is as usual taken quadratic in the superfields. Demanding that it satisfies the above symmetries and requiring in addition that it keeps the multilocal form of the bare action in Eq.~(\ref{eq_superaction_multicopy}) imply the following form:
\begin{equation}
\label{eq_regulator}
\Delta S_k^{(\beta)}=\frac 12  \sum_{a_1,a_2=1}^n\int_{\underline{x}_1}\int_{\underline{x}_2} \Phi_{a_1}(\underline{x}_1)\mathcal R_{k,a_1a_2}(\underline{x}_1,\underline{x}_2)\Phi_{a_2}(\underline{x}_2),
\end{equation}
where $\mathcal R_{k,a_1a_2}$ denotes infrared cutoff functions satisfying
\begin{equation}
 \begin{aligned}
 \label{eq_replicatedregulator}
\mathcal R_{k,a_1a_2}(\underline{x}_1,\underline{x}_2) = &\delta_{a_1a_2}\delta_{\underline{\theta}_1 \underline{\theta}_2}(1-\beta \bar \theta_1 \theta_1) \widehat{R}_k(|x_1-x_2|)\\&+\widetilde{R}_k(|x_1-x_2|),
 \end{aligned}
 \end{equation}
where $\delta_{\underline{\theta}_1 \underline{\theta}_2}$ is defined below Eq.~(\ref{eq_legendre_invert}). The infrared cutoff functions are chosen such that the integration over modes with momentum $\vert q \vert \ll k$ is suppressed (see below); these functions must go to zero when $k\rightarrow 0$ so that full integration is recovered in this limit.\cite{berges02,tarjus04,tarjus08}
[More precisely, we shall see below that $\widetilde{R}_k(q^2)$ goes to zero except for $q=0$, which nonetheless does not alter the property that the regularized theory converges to the full theory when $k \rightarrow 0$.]

The central quantity of our NP-FRG approach is the so-called ``effective average action'' $\Gamma_k^{(\beta)}$,\cite{wetterich93,berges02} which is the generating functional of the 1PI (proper) vertices\cite{zinnjustin89} at scale $k$ and is obtained from $\mathcal W_k^{(\beta)}\equiv \log \mathcal Z_k^{(\beta)}$ by a modified Legendre transform [compare with Eq.~(\ref{eq_legendre_multicopy})],
\begin{equation}
\begin{aligned}
\label{eq_legendre_effective_average_action}
&\Gamma_k^{(\beta)}[\left\lbrace \Phi_a \right\rbrace] = \\&-\mathcal W_k^{(\beta)}[\left\lbrace \mathcal J_a \right\rbrace] +\sum_{a=1}^n  \int_{\underline x}  \mathcal J_a(\underline x) \Phi_a(\underline x)-\Delta S_k^{(\beta)}[\left\lbrace \Phi_a \right\rbrace].
\end{aligned}
\end{equation}
Its flow with the infrared scale $k$ is described by an exact RG equation (ERGE),\cite{wetterich93,berges02} which, after accounting for the curvature, reads
\begin{equation}
 \begin{split}
\label{eq_ERGE_functional}
\partial_t \Gamma_k^{(\beta)}[\left\lbrace \Phi_a \right\rbrace]&=\frac 12 \sum_{a_1,a_2=1}^n\int_{\underline{x}_1}\int_{\underline{x}_2} (1-\beta \bar{\theta}_1 \theta_1)  (1-\beta \bar{\theta}_2 \theta_2) \\&\times \partial_t \mathcal R_{k;a_1 a_2}(\underline{x}_1,\underline{x}_2)  \mathcal P_{k;(a_1,\underline{x}_1)(a_2,\underline{x}_2)}^{(\beta)}[\left\lbrace \Phi_a \right\rbrace] , 
 \end{split}
 \end{equation}
where $t=\log(k/\Lambda)$; the (modified) propagator $\mathcal P_{k}^{(\beta)}$ is defined as the inverse of $\Gamma_k^{(2)} +  \mathcal R_k$ in the sense defined in Eq.~(\ref{eq_legendre_invert}), \textit{i.e.},
\begin{equation}
\label{eq_full_propagator}
\begin{aligned}
&\sum_{a_3}\int_{\underline x_3}(1-2\beta\bar\theta_3\theta_3)\,  \mathcal P_{k;(a_1,\underline{x}_1)(a_3,\underline{x}_3)}^{(\beta)} \bigg (\Gamma^{(2)}_{k;(a_3,\underline{x}_3)(a_2,\underline{x}_2)} \\&+  \mathcal R_{k;(a_3,\underline{x}_3)(a_2,\underline{x}_2)}\bigg )  =(1+\beta\bar\theta_1\theta_1) \delta_{a_1 a_2} \delta_{\underline x_1 \underline x_2} \, ,
\end{aligned}
\end{equation}
where $\Gamma_k^{(2)}[\left\lbrace \Phi_a \right\rbrace] $ is the second functional derivative of the effective average action with respect to the superfields and $\delta_{\underline x_1 \underline x_2}\equiv \delta^{(d)}(x_1-x_2) \delta_{\underline\theta_1 \underline\theta_2}$ [with, again, $\int_{\underline \theta_2}\delta_{\underline \theta_1 \underline\theta_2}=(1+\beta\bar\theta_1\theta_1)$]. Here and in the rest of this section, we omit the superscript $(\beta)$ in the functional derivatives of $\Gamma_k$ in order to avoid the awkward proliferation of superscripts. (Note that conventional  translational invariance being explicitly broken in the curved Grassmann space, Fourier transforming is of no use when $\beta \neq 0$.)

Due to the properties of the infrared cutoff functions, the effective average action reduces to the standard effective action $\Gamma^{(\beta)}[\left\lbrace \Phi_a \right\rbrace]$ when $k\rightarrow 0$. The initial condition at the microscopic (UV) scale, when $k\rightarrow \Lambda$, is more subtle. By using the definition of the effective average action, Eq.~(\ref{eq_legendre_effective_average_action}), and of the regulator,  Eqs.~(\ref{eq_regulator},\ref{eq_replicatedregulator}), and after a change of integration variables, one obtains the expression:
\begin{equation}
 \begin{aligned}
\label{eq_UVscale}
\exp&(-\Gamma_k^{(\beta)}[\left\lbrace \Phi_a \right\rbrace])=\int \prod_{a=1}^n\mathcal D\chi_a  \exp \big \{- S^{(\beta)}[\left\lbrace \Phi_a +\chi_a\right\rbrace]\\&  + \sum_{a=1}^n \int_{\underline{x}} \Gamma_{k;\underline{x}}^{(1)}[\left\lbrace \Phi_a +\chi_a\right\rbrace] \chi_a(\underline{x}) - \frac{1}{2}\int_{x_1 x_2} \widehat{R}_k(|x_1-x_2|) \\& \times \sum_{a=1}^n \int_{\underline{\theta}}\chi_a(x_1,\underline{\theta})\chi_a(x_2,\underline{\theta}) - \frac{1}{2}\int_{x_1 x_2} \widetilde{R}_k(|x_1-x_2|) \\& \times \big( \sum_{a=1}^n \int_{\underline{\theta}} \chi_a(x_1,\underline{\theta})\big) \big( \sum_{a=1}^n \int_{\underline{\theta}} \chi_a(x_2,\underline{\theta})\big)\big \}.
 \end{aligned}
\end{equation}
If one requires that $\widehat{R}_k(|x|)$ diverges for all $x$ when $k \rightarrow \Lambda$ while $\widetilde{R}_k(|x|)$ stays bounded, the term in $\widehat{R}_k$ in the above expression acts as a delta functional for all the superfield variables $\chi_a$. As a consequence,
\begin{equation}
\Gamma_\Lambda^{(\beta)} [\left\lbrace \Phi_a \right\rbrace]= S^{(\beta)}[\left\lbrace \Phi_a \right\rbrace], 
\end{equation}
\textit{i.e.}, the effective average action reduces to the bare action defined in Eq.~(\ref{eq_superaction_multicopy}).

\subsection{Properties and role of the cutoff functions}

We first go back from Eq.~(\ref{eq_part_func}) to the original formulation with the $\varphi_a$ fields in the presence of disorder (setting the fermionic sources to zero). The generating functional can then be expressed as
\begin{equation}
 \begin{aligned}
 \label{eq_part_func_regulator}
&\mathcal Z_k^{(\beta)}[\{ \hat J_a, J_a\}]\propto \int \mathcal D h\; \frac{\exp \big[\frac{-\vert h_q\vert^2}{2(\Delta_B-\widetilde{R}_k(q^2))}\big]}{\sqrt{\det(\Delta_B-\widetilde R_k)}}\prod_{a=1}^n \int \mathcal D\varphi_a \\&\delta\left [\dfrac{\delta S_B[\varphi_a]}{\delta \varphi_a(x)}+\int_y \widehat{R}_k(x-y)\varphi_a(y) -h(x)-J_a(x)\right ] \times \\& \det \left(\dfrac{\delta^2 S_B[\varphi_a]}{\delta \varphi_a \delta \varphi_a}+\widehat{R}_k\right)\;  \exp  \int_{x}  \hat{J}_a(x)  \varphi_a(x).
 \end{aligned}
\end{equation}
One can see that $\widehat{R}_k$ is added to the second functional derivative $S_B^{(2)}[\varphi]$ of the bare action and plays the role of an infrared cutoff at scale $k$ for the fluctuations of the $\varphi$ field, as in the NP-FRG of pure systems.\cite{berges02} More specifically in the present case, a large enough cutoff function $\widehat{R}_k$ ensures that the operator $S_B^{(2)}[\varphi]+\widehat{R}_k$ is definite positive which guarantees that the stochastic field equation has a unique solution. This is certainly true at the UV scale $\Lambda$ where $\widehat{R}_{k\rightarrow \Lambda}$ diverges. At the beginning of the flow, the regularized theory is therefore ``ultralocal'' in the Grassmann subspace. Since as discussed in section V the ``ultralocal'' cumulants $W_p$ and $\Gamma_p$ are independent of the curvature $\beta$ when uniqueness of the solution is enforced, the regularized theory at the start of the RG flow is also invariant under the superrotations.

The other cutoff function $\widetilde{R}_k$ on the other hand (with the requirement that it is always positive)  reduces the variance of the random magnetic source, \textit{i.e.} reduces the fluctuations of the bare disorder. From this, one gets the additional constraint at the beginning of the flow, when the random field distribution is not yet renormalized, that $\Delta_B \geq \widetilde{R}_{\Lambda}(q^2)$ for all $q$'s. Note that as one follows the RG flow by reducing the IR scale $k$, Eq.~(\ref{eq_part_func_regulator}) is still valid but rather useless:  it is indeed more convenient to work with renormalized quantities, namely a renormalized disorder and a renormalized action or effective action. When $k$ decreases, so does the cutoff function $\widehat{R}_k$, which presumably leads at some point to a breaking of the ``ultralocal'' property when $\beta$ is finite. One should however keep in mind that the relevant solutions are associated with a renormalized random functional and no longer with the bare action, so that the deviation from ``Grassmannian ultralocality'' may be small and, even further, vanish along the flow (see below).

Finally, the superfield formalism offers a way to constrain the cutoff functions by relating them. As noted before, the action in Eq.~(\ref{eq_superaction_multicopy}) is invariant under the superrotations when the supersources are taken as uniform in the Grassmann subspace (\textit{i.e.}, $\hat{J}_b= \bar{K}_b = K_b = 0$) for all copies but one and when the curvature $\beta$ is set to zero. The regulator $\Delta S_k$ can be made explicitly invariant under the same conditions by choosing $\mathcal R_{k,aa}$ to be a function of the super-Laplacian $\Delta_{SS}$ only. As a result,\cite{footnote1}
\begin{equation}
\label{eq_susy_regulator}
\widetilde{R}_k(q^2)=-\Delta_B\, \partial_{q^2}\widehat{R}_k(q^2),
 \end{equation}
where $q$ denotes the momentum in Euclidean $d$-dimensional space. Note that the above expression involves the strength of the bare disorder whereas, as also discussed above, a more general relation should allow for a proportionality factor expressed in terms of renormalized quantities rather than bare ones. This will be discussed in section VII-B.

\subsection{ERGE for the cumulants}

Aiming at deriving exact FRG flow equations for the ``cumulants'' $\mathsf \Gamma_{kp}^{(\beta)}$, we first rewrite more explicitly the ERGE for the effective average action,
\begin{equation}
 \begin{split}
\label{eq_ERGE_functional_explicit}
&\partial_t \Gamma_k^{(\beta)}[\left\lbrace \Phi_a \right\rbrace]=\frac 12 \int_q \bigg \{ \partial_t \widehat R_{k}(q^2)  \int_{\underline{\theta}_1}(1-2 \beta \bar{\theta}_1 \theta_1) \sum_{a_1} \\& \mathcal P_{k;(a_1,-q \underline{\theta}_1)(a_1,q \underline{\theta}_1)}^{(\beta)}[\left\lbrace \Phi_a \right\rbrace]  + \partial_t \widetilde R_{k}(q^2) \int_{\underline{\theta}_1}\int_{\underline{\theta}_2} (1-\beta \bar{\theta}_1 \theta_1) \\& (1-\beta \bar{\theta}_2 \theta_2)  \sum_{a_1 a_2} \mathcal P_{k;(a_1,-q \underline{\theta}_1)(a_2,q \underline{\theta}_2)}^{(\beta)}[\left\lbrace \Phi_a \right\rbrace] \bigg \},
 \end{split}
 \end{equation}
and we use the expansion in increasing number of sums over copies. For the expansion of the full propagator  $\mathcal P_{k}^{(\beta)}$ appearing in the right-hand side, it is convenient to introduce notations as follows. A generic matrix $\mathcal A_{(a_1 \underline x_1)(a_2 \underline x_2)}[\left\lbrace \Phi_a\right\rbrace ]$ can be decomposed as
\begin{equation}
\label{eq_genericdecompos}
\mathcal A_{a_1a_2}[\left\lbrace \Phi_a\right\rbrace ]=\delta_{a_1a_2} \widehat{\mathcal A}_{a_1}[\left\lbrace \Phi_a\right\rbrace ] + \widetilde{\mathcal A}_{a_1a_2}[\left\lbrace  \Phi_a\right\rbrace ],
\end{equation}
where we have dropped the explicit dependence on the coordinates. In the above expression,  it is understood that the second term $\widetilde{\mathcal A}_{a_1a_2}$ no longer contains any explicit Kronecker symbol. Each component can now be expanded in increasing number of sums over copies,
\begin{equation}
\label{eq_widehatA}
\widehat{\mathcal A}_{a_1}[\left\lbrace  \Phi_a\right\rbrace ]=\widehat{\mathcal A}^{[0]}[ \Phi_{a_1} ]+\sum_{b} \widehat{\mathcal A}^{[1]}[ \Phi_{a_1} \vert  \Phi_b  ]+\cdots
\end{equation}
\begin{equation}
\label{eq_widetildeA}
\widetilde{\mathcal A}_{a_1a_2}[\left\lbrace  \Phi_a\right\rbrace ]=\widetilde{\mathcal A}^{[0]}[ \Phi_{a_1},  \Phi_{a_2}]+\sum_{b}\widetilde{\mathcal A}^{[1]}[ \Phi_{a_1},  \Phi_{a_2} \vert   \Phi_{b}]+\cdots,
\end{equation}
where the superscripts in square brackets denote the order in the expansion (and should not be confused with superscripts in parentheses indicating functional derivatives). The $\widehat{\mathcal A}^{[p]}$'s and $\widetilde{\mathcal A}^{[p]}$'s are symmetric under any permutation of their arguments, and their functional form is independent of the number $n$ of copies (taken as arbitrarily large).

After inserting Eq.~(\ref{eq_multicopy1_app}) for $\Gamma_{k}^{(\beta)}$ and the above expressions applied to the  propagator $\mathcal{P}_k^{(\beta)}$, we obtain a hierarchy of ERGE's for the cumulants  $\mathsf \Gamma_{kp}^{(\beta)}$. The first equations read:
\begin{equation}
\label{eq_flow_Gamma1}
\begin{split}
&\partial_t \mathsf \Gamma_{k 1}^{(\beta)}\left[ \Phi_1\right ]=
\dfrac{1}{2} \int_{ q} \bigg \{ \partial_t \widehat{R}_k(q^2) \int_{\underline{\theta}_1}(1-2 \beta \bar{\theta}_1 \theta_1) \times \\& \big (\widehat{\mathcal P}_{k;(-q \underline{\theta}_1)(q \underline{\theta}_1)}^{[0]}[ \Phi_1] +\widetilde{\mathcal P}_{k;(-q \underline{\theta}_1)(q \underline{\theta}_1)}^{[0]}[ \Phi_1, \Phi_1] \big ) + \partial_t \widetilde R_{k}(q^2) \times \\& \int_{\underline{\theta}_1}\int_{\underline{\theta}_2} (1-\beta \bar{\theta}_1 \theta_1)(1-\beta \bar{\theta}_2 \theta_2)\widehat{\mathcal P}_{k;(-q \underline{\theta}_1)(q \underline{\theta}_2)}^{[0]}[\Phi_1] \bigg \} ,
\end{split}
\end{equation}
\begin{equation}
\label{eq_flow_Gamma20}
\begin{split}
\partial_t &\mathsf \Gamma_{k 2}^{(\beta)}\left[ \Phi_1 , \Phi_2\right ]= - 
\dfrac{1}{2} \int_{ q} \bigg \{ \partial_t \widehat{R}_k(q^2) \int_{\underline{\theta}_1}(1-2 \beta \bar{\theta}_1 \theta_1) \times \\&\big (\widehat{\mathcal P}_{k;(-q \underline{\theta}_1)(q \underline{\theta}_1)}^{[1]}[ \Phi_1\vert \Phi_2] +\widetilde{\mathcal P}_{k;(-q \underline{\theta}_1)(q \underline{\theta}_1)}^{[1]}[ \Phi_1, \Phi_1 \vert \Phi_2] \big ) + \\& \partial_t \widetilde R_{k}(q^2)  \int_{\underline{\theta}_1}\int_{\underline{\theta}_2} (1-\beta \bar{\theta}_1 \theta_1)(1-\beta \bar{\theta}_2 \theta_2)  \times \\& \big (\widehat{\mathcal P}_{k;(-q \underline{\theta}_1)(q \underline{\theta}_2)}^{[1]}[ \Phi_1\vert \Phi_2] + \widetilde{\mathcal P}_{k;(-q \underline{\theta}_1)(q \underline{\theta}_2)}^{[0]}[ \Phi_1, \Phi_2] \big )\\&+ perm (12)\bigg \},
\end{split}
\end{equation}
and so on,  where $perm (12)$ denotes the expression obtained by permuting $ \Phi_1$ and $ \Phi_2$; for clarity we have omitted the superscript $(\beta)$ in the right-hand side.

The components of the propagator, $\widehat{\mathcal P}_{k}^{[p]}$ and $\widetilde{\mathcal P}_{k}^{[p]}$, can be expressed in terms of second derivatives of the cumulants $\mathsf \Gamma_{kq}$ by inserting into Eq.~(\ref{eq_full_propagator}) the expressions (\ref{eq_widehatA},\ref{eq_widetildeA}) with $\mathcal A$ equal to the second functional derivative of Eq.~(\ref{eq_multicopy1_app}). The algebraic manipulations are given in Appendix~\ref{appendix:expansion_copies}. One finds for instance that the propagator $\widehat{\mathcal P}_{k}^{[0]}$ can be symbolically expressed as
\begin{equation}
\label{eq_hatpropagator}
\widehat {\mathcal P}_{k}^{[0]}[  \Phi_1 ]=\left( \mathsf{\Gamma} _{k 1}^{(2)}[  \Phi_1 ]+\widehat R_k  U\right) ^{-1},
\end{equation}
which means that
\begin{equation}
\label{eq_hatpropagator_definition}
\begin{aligned}
&\int_{\underline{x}_3}(1-2\beta \bar \theta_3 \theta_3)\, \widehat{\mathcal P}_{k;\underline{x}_1 \underline{x}_3}^{[0]} \bigg (\mathsf \Gamma^{(2)}_{k1;\underline{x}_3 \underline{x}_2}+  \widehat R_{k;x_3 x_2} U_{\underline{\theta}_3 \underline{\theta}_2}\bigg ) \\& =U_{\underline{\theta}_1 \underline{\theta}_2} \delta^{(d)}( x_1 - x_2) \,  ,
\end{aligned}
\end{equation}
where we have introduced $U_{\underline{\theta}_1 \underline{\theta}_2}=(1 + \beta \bar \theta_1 \theta_1) \delta_{\underline{\theta}_1 \underline{\theta}_2}$, and that $\widetilde{\mathcal P}_{k}^{[0]}$ is given by
\begin{equation}
\begin{aligned}
\label{eq_tildepropagator}
&\widetilde {\mathcal P}_{k;\underline{x}_1 \underline{x}_2}^{[0]}[  \Phi_1,   \Phi_2 ]=\int_{\underline{x}_3}\int_{ \underline{x}_4}(1 -2 \beta \bar \theta_3 \theta_3)(1 -2 \beta \bar \theta_4 \theta_4)\\& \widehat {\mathcal P}_{k;\underline{x}_1 \underline{x}_3}^{[0]}[  \Phi_1 ] \big( \mathsf{\Gamma} _{k2;\underline{x}_3, \underline{x}_4}^{(11)}[  \Phi_1,   \Phi_2 ]-(1 + \beta \bar \theta_3 \theta_3)(1 + \beta \bar \theta_4 \theta_4)\\&
\times \widetilde R_{k;x_3 x_4} \big ) \widehat {\mathcal P}_{k;\underline{x}_4 \underline{x}_2}^{[0]}[  \Phi_2 ].
\end{aligned}
\end{equation}
Eqs.~(\ref{eq_hatpropagator}) and (\ref{eq_tildepropagator}) can be inserted in Eq.~(\ref{eq_flow_Gamma1}), which provides an ERGE for $\mathsf \Gamma_{k 1}^{(\beta)}$ only expressed in terms of cumulants associated with the effective average action.

After introducing the short-hand notation $\widetilde{\partial}_t$ to indicate a derivative acting only on the cutoff functions (\textit{i.e.},  $\widetilde{\partial}_t \equiv \partial_t \widehat{R}_k\, \delta/\delta \widehat{R}_k + \partial_t \widetilde{R}_k \, \delta/\delta \widetilde{R}_k$), Eq. (\ref{eq_flow_Gamma20}) can now be rewritten as
\begin{equation}
\label{eq_flow_Gamma2_final}
\begin{split}
&\partial_t \mathsf \Gamma_{k2}^{(\beta)}\left[ \Phi_1 , \Phi_2\right ]= \dfrac{1}{2} \widetilde{\partial}_t \bigg \{ \int_{\underline{x}_3}\int_{ \underline{x}_4}(1 -2 \beta \bar \theta_3 \theta_3)(1 -2 \beta \bar \theta_4 \theta_4) \times \\&\bigg [ \widehat{\mathcal P}_{k;\underline{x}_3 \underline{x}_4}^{[0]}\left[ \Phi_1 \right ] ( \mathsf \Gamma _{k2;\underline{x}_4 \underline{x}_3,.}^{(20)}\left[ \Phi_1,  \Phi_2 \right ] - \mathsf \Gamma _{k3;\underline{x}_4, \underline{x}_3,.}^{(110)}\left[ \Phi_1,  \Phi_1,  \Phi_2 \right ])  \\&+ \widetilde{\mathcal P}_{k;\underline{x}_3 \underline{x}_4}^{[0]}\left[ \Phi_1, \Phi_1 \right ] \mathsf \Gamma _{k2;\underline{x}_4 \underline{x}_3,.}^{(20)}\left[ \Phi_1, \Phi_2 \right ]  +\dfrac{1}{2} \widetilde{ \mathcal P}_{k;\underline{x}_3 \underline{x}_4}^{[0]}\left[ \Phi_1, \Phi_2 \right ]\times \\&(\mathsf \Gamma _{k2;\underline{x}_4, \underline{x}_3}^{(11)}\left[ \Phi_1, \Phi_2 \right ] - (1 + \beta \bar \theta_4 \theta_4)(1 + \beta \bar \theta_3 \theta_3)\widetilde{R}_{k;x_4x_3} ) \\& + perm (12) \bigg ]\bigg \},
\end{split}
\end{equation}
where, again, $perm (12)$ denotes the expression obtained by permuting $ \Phi_1$ and $ \Phi_2$. Similar ERGE's are obtained for the higher-order cumulants. A graphical representation of the hierachy of ERGE's is provided is Appendix~\ref{appendix:graphical}.

This provides a hierachy of exact RG equations for the cumulants of the renormalized disorder (including the first one which leads to a description of the thermodynamics). One should note that (i) the cumulants are functionals of the superfields and contain full information on the complete set of 1PI correlation functions, (ii) the flow equations are coupled, the $(p+1)$th cumulant appearing in the right-hand side of the equation for the $p$th cumulant, and (iii)  to obtain the flow equation for $\mathsf \Gamma_{kp}[\Phi_1,...,\Phi_p]$ with its full functional dependence on the $p$ field arguments, one needs to consider at least $p$ copies  of the original system. Formally, the whole hierarchy of flow equations for the cumulants can thus be obtained by considering an arbitrary large number of copies.

\subsection{ERGE for the cumulants under the hypothesis of  ``Grassmannian ultralocality''}

As such the hierarchy of ERGE's obtained above is  expressed in terms of superfields and superspace coordinates and is untractable in general. It is instructive to consider the simplification that arises under the hypothesis of  ``Grassmannian ultralocality''. This is achieved by considering the flow equations when the cumulants are evaluated for ``physical'' configurations $\Phi_a(\underline x)=\phi_a(x)$ of the superfields, \textit{i.e.} configurations that are uniform in the Grassmann subspace.  Then, from Eq.~(\ref{eq_cumulant_p_ultralocal}), $\mathsf \Gamma_{kp}^{(\beta)}\left[ \Phi_1 ,..., \Phi_p\right ]=\beta^p\, \Gamma_{kp}^{(\beta)}\left[ \phi_1 ,...,\phi_p\right ]$. We next insert  the results of section V in the ERGE's and we introduce the following ``hat'' and ``tilde'' propagators,
\begin{equation}
\label{eq_hatP_zero}
\widehat {P}_{k}^{[0]}[\phi ]=\left(\Gamma _{k1}^{(2)}[ \phi ]+\widehat R_k\right) ^{-1}
\end{equation}
and
\begin{equation}
\label{eq_tildeP_zero}
\widetilde {P}_{k}^{[0]}[\phi_1, \phi_2 ]= \widehat {P}_{k}^{[0]}[ \phi_1 ](\Gamma _{k2}^{(11)}[\phi_1, \phi_2 ]-\widetilde R_k ) \widehat {P}_{k}^{[0]}[ \phi_2 ],
\end{equation}
which are obtained from Eqs.~(\ref{eq_hatpropagator}-\ref{eq_tildepropagator}) with, as in Eqs.~(\ref{eq_expand_Gamma_2hat},\ref{eq_expand_Gamma_2tilde}), $\widehat {\mathcal P}_{k;\underline \theta_1 \underline \theta_2}^{[0]}[\phi]=(1+\beta \bar\theta_1 \theta_1) \delta_{\underline \theta_1 \underline \theta_2}\widehat {P}_{k}^{[0]}[\phi]$ and $\widetilde {\mathcal P}_{k;\underline \theta_1 \underline \theta_2}^{[0]}[\phi_1,\phi_2]=(1+\beta \bar\theta_1 \theta_1)(1+\beta \bar\theta_2 \theta_2) \widetilde {P}_{k}^{[0]}[\phi_1,\phi_2]$.

The exact flow equation for the first cumulant of the renormalized disorder is finally obtained as
\begin{equation}
\label{eq_flow_Gamma1_ULapp}
\begin{split}
&\partial_t\Gamma_{k1}\left[\phi_1\right ]= \\&-
\dfrac{1}{2} \tilde{\partial}_t \int_{x_2x_3}\widehat{P}_{k;x_2x_3}^{[0]}[\phi_1] \big(\Gamma_{k2;x_2,x_3}^{(11)}\left[\phi_1,\phi_1\right ] - \widetilde{R}_{k;x_2x_3}\big),
\end{split}
\end{equation}
where the explicit dependence on the Euclidean coordinates has been momentarily reinstalled. One similarly derives an ERGE for the second cumulant,
\begin{equation}
\label{eq_flow_Gamma2_ULapp}
\begin{split}
&\partial_t\Gamma_{k2}\left[\phi_1,\phi_2\right ]=\\&
\dfrac{1}{2} \tilde{\partial}_t \int_{x_3x_4}\big \{- \widehat{P}_{k;x_3x_4}^{[0]}\left[\phi_1\right ] \Gamma_{k3;x_3,.,x_4}^{(101)}\left[\phi_1,\phi_2,\phi_1\right ]+\\& \widetilde{P}_{k;x_3x_4}^{[0]}\left[\phi_1,\phi_1\right ] \Gamma_{k2;x_3x_4,.}^{(20)}\left[\phi_1,\phi_2\right ]+ \frac{1}{2}\widetilde{P}_{k;x_3x_4}^{[0]}\left[\phi_1,\phi_2\right ] \\& \times \left( \Gamma_{k2;x_3,x_4}^{(11)}\left[\phi_1,\phi_2\right ] - \widetilde{R}_{k;x_3x_4}\right) + perm(12)\big \},
\end{split}
\end{equation}
where $perm (12)$ denotes the expression obtained by permuting $\phi_1$ and $\phi_2$.

Generically, the flow of $\Gamma_{kp}\left[\phi_1,...,\phi_p\right ]$ involves three types of quantities: the propagators  $\widehat{P}_{k}^{[0]}$ and  $\widetilde{P}_{k}^{[0]}$, second functional derivatives of $\Gamma_{kp}$ in which all the arguments are different, and second functional derivatives of $\Gamma_{k(p+1)}$ with two of their arguments equal to each other (for a graphical representation of the hierachy of ERGE's, see Appendix~\ref{appendix:graphical}). We will come back in more detail to the structure of these flow equations in the following paper, but we note for future reference that these equations are independent of the auxialiary parameter $\beta$, and so is their solution if the initial condition is itself ``ultralocal''.

Finally, we point out that the above ERGE's coincide with those previously derived without the superfield formalism by means of an expansion in number of free replica sums, when evaluated at $T=0$.\cite{tarjus04, tarjus08} The same is true for the ERGE for all higher-order cumulants. It is however important to stress that our previous replica approach provides no obvious way to make the superrotational invariance (or lack of it) explicit and, therefore, neither to relate the two infrared cutoff functions $\widetilde R_k$ and  $\widehat R_k$ nor to provide guidance for devising nonperturbative approximations to the ERGE's that do not explicitly break the underlying supersymmetry (see the companion paper).

\section{Ground-state dominance in the NP-FRG}
\label{sec:ground-state_dominance}

\subsection{Taking the limit of infinite curvature (zero auxiliary temperature)}

We now come to a central step of our approach. Introducing the auxiliary temperature $\beta^{-1}$ has allowed us to place the superfield formalism on a firm ground. However, being ultimately interested in studying the ground-state properties of the system, we would like to take the limit $\beta^{-1} \rightarrow 0$ in the exact flow equations derived above. To make the dependence on  $\beta$ explicit in the ERGE's, we first rescale the Grassmann coordinates and, accordingly, the auxiliary fields: $(\bar \theta, \theta) = \beta^{-1/2}(\bar \omega, \omega)$, $(\bar \psi, \psi) = \beta^{1/2}(\widetilde{\bar \psi}, \widetilde \psi)$, and $\hat \phi=\beta \widetilde{\hat \phi}$, so that $\Phi(\underline \theta)\equiv \widetilde \Phi(\underline \omega)$ with $\widetilde \Phi(\underline \omega)=\phi +\bar \omega \widetilde \psi+\widetilde{\bar \psi} \omega+\bar \omega \omega \widetilde{\hat \phi}$.

The cumulants associated with the effective average action can be formally written as
\begin{equation}
\label{eq_}
\mathsf \Gamma_{kp}^{(\beta)}[\Phi_{a_1},...,\Phi_{a_p}]=\int_{\underline \theta_1}...\int_{\underline \theta_p} \Gamma_{kp}^{(\beta)}[1,...,p],
\end{equation}
where in the right-hand side, as in Eqs.~(\ref{eq_cumW1_formal},\ref{eq_cumW2_formal}), $q \in \{1,...,p\}$ denotes
\begin{equation}
\{\Phi_{a_q}(\underline{\theta}_q),  (1+\frac{\beta}{2} \bar{\theta}_q \theta_q)\, \partial_{\underline \theta_q}\Phi_{a_q}(\underline{\theta}_q),  \Delta_{\underline{\theta}_q}\Phi_{a_q}(\underline{\theta}_q)\}. 
\end{equation}
When changing Grassmann coordinates from $\underline\theta$ to $\underline\omega$, we assume that
$\Gamma_{kp}^{(\beta)}[1,...,p]=\Gamma_{kp}^{(\beta)}[\tilde 1,...,\tilde p]$, where $\tilde 1$ denotes $\{\widetilde \Phi_{a_1}(\underline{\omega}_1)$,  $(1+\frac{1}{2} \bar{\omega}_1\omega_1 ) \partial_{\underline{\omega}_1}\widetilde \Phi_{a_1}(\underline{\omega}_1), \Delta_{\underline{\omega}_1}\widetilde \Phi_{a_1}(\underline{\omega}_1) \}$, and so on, with the same functional form for $\Gamma_{kp}^{(\beta)}$. This guarantees that the derivatives of the ``non-ultralocal'' contributions to the $\Gamma_{kp}^{(\beta)}$'s do not come with increasing factors of $\beta$ which would completely spoil the limit of infinite $\beta$. We shall support this argument by studying the first ``non-ultralocal'' corrections (see below).

Then, taking into account that an integral over a Grassmann variable acts like a derivative leads to
\begin{equation}
\label{eq_betadependence_cumulant}
\mathsf \Gamma_{kp}^{(\beta)}[\Phi_{a_1},...,\Phi_{a_p}]=\beta^p \int_{\underline \omega_1}...\int_{\underline \omega_p} \Gamma_{kp}^{(\beta)}[\tilde 1,...,\tilde p].
\end{equation}
In addition, by using the identity $\delta /\delta \Phi_a(\underline\theta) \equiv \delta /\delta \widetilde \Phi_a(\underline\omega)$, which can be proven by considering the components of the superfield in the old and the new coordinate system, we arrive at
\begin{equation}
\begin{aligned}
\label{eq_betadependence_derivativecumulant}
&\frac{\delta ^q}{\delta \Phi_{b_1}(\underline\theta_{b_1})...\delta \Phi_{b_q}(\underline\theta_{b_q})}\mathsf \Gamma_{kp}^{(\beta)}[\Phi_{a_1},...,\Phi_{a_p}]\\&=\beta^{p-q}\int_{\underline \omega_1}...\int_{\underline \omega_p} \frac{\delta ^q}{\delta\widetilde \Phi_{b_1}(\underline\omega_{b_1})...\delta\widetilde \Phi_{b_q}(\underline\omega_{b_q})}\Gamma_{kp}^{(\beta)}[\tilde 1,...,\tilde p],
\end{aligned}
\end{equation}
where the functional dependence on the Euclidean coordinates is left implicit.

After the above preliminaries, we consider the ERGE's for the cumulants, \textit{e.g.} Eqs.~(\ref{eq_flow_Gamma1},\ref{eq_flow_Gamma2_final}) for the first two cumulants. We have to study the explicit $\beta$ dependence of the propagators $\widehat{\mathcal P}_{k}^{[0]}$ and $\widetilde{\mathcal P}_{k}^{[0]}$. After changing the Grassmann coordinates in Eqs.~(\ref{eq_hatpropagator_definition},\ref{eq_tildepropagator}) and using Eq.~(\ref{eq_betadependence_derivativecumulant}), we easily obtain
\begin{equation}
\label{eq_betadependence_hatP0}
\widehat {\mathcal P}_{k;\underline{\theta}_1 \underline{\theta}_2}^{[0]}[  \Phi_1 ] =\frac{1}{\beta} \widehat {\mathcal P}_{k;\underline{\omega}_1 \underline{\omega}_2}^{[0]}[ \widetilde \Phi_1 ] 
\end{equation}
and
\begin{equation}
\label{eq_betadependence_tildeP0}
\widetilde {\mathcal P}_{k;\underline{\theta}_1 \underline{\theta}_2}^{[0]}[  \Phi_1,   \Phi_2 ]=\widetilde {\mathcal P}_{k;\underline{\omega}_1 \underline{\omega}_2}^{[0]}[ \widetilde \Phi_1,  \widetilde \Phi_2 ],
\end{equation}
where $\widetilde \Phi$ is the superfield defined with the new coordinate system and the rescaling of the auxiliary fields (see above).

The ERGE's for the cumulants can then be reexpressed as
\begin{equation}
\label{eq_flow_Gamma1_scaled}
\begin{split}
&\partial_t \int_{\underline{\omega}_1} \Gamma_{k 1}^{(\beta)}\left[\tilde1\right ]=
\dfrac{1}{2} \mathrm{Tr} \bigg \{ \partial_t \widehat{R}_k \int_{\underline{\omega}_1}(1-2 \bar{\omega}_1 \omega_1) \times \\& \big (\frac{1}{\beta} \widehat{\mathcal P}_{k;\underline{\omega}_1 \underline{\omega}_1}^{[0]}[\widetilde \Phi_1] +\widetilde{\mathcal P}_{k;\underline{\omega}_1 \underline{\omega}_1}^{[0]}[\widetilde \Phi_1,\widetilde \Phi_1] \big ) + \partial_t \widetilde R_{k} \times \\& \int_{\underline{\omega}_1}\int_{\underline{\omega}_2}(1- \bar{\omega}_1 \omega_1)(1- \bar{\omega}_2 \omega_2)\widehat{\mathcal P}_{k;\underline{\omega}_1 \underline{\omega}_2}^{[0]}[\widetilde \Phi_1] \bigg \} ,
\end{split}
\end{equation}
\begin{equation}
\label{eq_flow_Gamma2_final_scaled}
\begin{split}
&\partial_t \int_{\underline{\omega}_1}\int_{ \underline{\omega}_2} \Gamma_{k2}^{(\beta)}\left[ \tilde1 , \tilde2\right ]= \dfrac{1}{2} \widetilde{\partial}_t   \mathrm{Tr} \bigg \{ \int_{\underline{\omega}_3}\int_{ \underline{\omega}_4}(1-2 \bar{\omega}_3 \omega_3)\times \\&(1-2 \bar{\omega}_4 \omega_4)\bigg [ \frac{1}{\beta}\, \widehat{\mathcal P}_{k;\underline{\omega}_3 \underline{\omega}_4}^{[0]}\left[ \widetilde \Phi_1 \right ]  \mathsf \Gamma _{k2;\underline{\omega}_4 \underline{\omega}_3,.}^{(20)}\left[ \widetilde  \Phi_1,  \widetilde  \Phi_2 \right ] -\\& \widehat{\mathcal P}_{k;\underline{\omega}_3 \underline{\omega}_4}^{[0]}\left[ \widetilde \Phi_1 \right ] \mathsf \Gamma _{k3;\underline{\omega}_4, \underline{\omega}_3,.}^{(110)}\left[  \widetilde \Phi_1,   \widetilde \Phi_1,   \widetilde \Phi_2 \right ]+ \widetilde{\mathcal P}_{k;\underline{\omega}_3 \underline{\omega}_4}^{[0]}\left[ \widetilde \Phi_1,  \widetilde \Phi_1 \right ]\times \\& \mathsf \Gamma _{k2;\underline{\omega}_4 \underline{\omega}_3,.}^{(20)}\left[  \widetilde \Phi_1,  \widetilde \Phi_2 \right ]  +\dfrac{1}{2} \widetilde{ \mathcal P}_{k;\underline{\omega}_3 \underline{\omega}_4}^{[0]}\left[  \widetilde \Phi_1,  \widetilde \Phi_2 \right ]\bigg (\mathsf \Gamma _{k2;\underline{\omega}_4, \underline{\omega}_3}^{(11)}\left[  \widetilde \Phi_1,  \widetilde \Phi_2 \right ] \\& - (1+ \bar{\omega}_4 \omega_4)(1+ \bar{\omega}_3 \omega_3)\widetilde{R}_{k} \bigg ) + perm (12) \bigg ]\bigg \},
\end{split}
\end{equation}
where $\mathrm{Tr}$ indicates a trace over the Euclidean momenta (which are not shown explicitly). Similar expressions are obtained for the higher-order cumulants. One actually finds a structure that is analogous to that derived in our previous replica approach\cite{tarjus08} when considering the (bath) temperature $T$: the cumulant of order $p$ comes with a factor $T^{-p}$, its $q$th derivative with a factor $T^{-(p-q)}$, the ``hat'' propagator has a factor $1/T$ and the ``tilde'' propagator no explicit factor of $T$. The limit $\beta \rightarrow \infty$ allows one to drop the terms in the above flow equations that have an explicit factor of $1/\beta$, namely the first term of the right-hand sides. It should however be kept in mind that as in the ``thermal'' case when $T\rightarrow 0$, the limit $1/\beta \rightarrow 0$ is expected to be nonuniform in the (super)field dependence. This will be discussed further down and in the following paper.

Having introduced the change of Grassmann coordinates and the formal way to study the limit $\beta \rightarrow \infty$, we can go one step beyond in the analysis of the ERGE's. To do so, we consider superfields that are uniform in the Grassmann subspace, \textit{i.e.} $\Phi_a=\widetilde \Phi_a=\phi_a$, and we generalize the results of sections V-B and VI-B to the case where the cumulants $\mathsf \Gamma_{kp}^{(\beta)}$ have a generic  ``non-ultralocal'' component,
\begin{equation}
\begin{aligned}
\label{eq_cumulant_p_nonultralocal}          
\mathsf \Gamma_{kp}^{(\beta)}[\Phi_{1},...,\Phi_{p}] = \beta^p\int_{\underline{\omega}_1}...\int_{\underline{\omega}_p}&
\bigg (\Gamma_{kp}^{(\beta)UL}[\widetilde \Phi_1(\underline{\omega}_1),...,\widetilde \Phi_p(\underline{\omega}_p)]\\&+ \Gamma_{kp}^{(\beta)NUL}[\widetilde 1,...,\widetilde p] \bigg ),
\end{aligned}
\end{equation}
where $\Gamma_{kp}^{(\beta)NUL}$ involves derivatives of the superfields in the Grassmann directions and is equal to zero when the superfields are uniform in the Grassmann subspace: as a result, $\mathsf{\Gamma}_{kp}^{(\beta)}[\phi_{1},...,\phi_{p}]=\beta^p\Gamma_{kp}^{(\beta)UL}[\phi_1,...,\phi_p]$ . The second functional derivative of the effective average action can be decomposed as in Eq.~(\ref{eq_genericdecompos}) and, by applying the WT identities associated with invariance under the isometries of the curved Grassmann subspace (see section IV-D and Appendix~\ref{appendix:nonultralocal}), one finds the following general structure for the ``hat'' and ``tilde'' components:
\begin{equation}
\begin{split}
\label{eq_hatGamma2_unifNUL}
\widehat {\Gamma}_{k;a_1\underline{\omega}_1 \underline{\omega}_2}^{(2)}[ \{ \phi_a\}]& =(1+ \bar{\omega}_1 \omega_1)\delta_{\underline{\omega}_1 \underline{\omega}_2}\widehat {A}_{k;a_1}[ \{ \phi_a\}] \\&+ (1+\bar \omega_1 \omega_2+\bar \omega_2 \omega_1)\widehat {B}_{k;a_1}[ \{ \phi_a\}]
\end{split}
\end{equation}
and
\begin{equation}
\label{eq_tildeGamma2_unifNUL}
\widetilde {\Gamma}_{k;(a_1\underline{\omega}_1)(a_2 \underline{\omega}_2)}^{(2)}[ \{\phi_a\}]=(1+ \bar{\omega}_1 \omega_1) (1+ \bar{\omega}_2 \omega_2)\widetilde {C}_{k;a_1a_2}[ \{\phi_a\}],
\end{equation}
where we recall that a factor $1/\beta$ is present when changing variables from $\underline \theta$ to $\underline \omega$ in $\widehat {\Gamma}_{k}^{(2)}$.

In addition, it can be proven that $\widehat{A}_{k;a_1}$ and $\widetilde{C}_{k;a_1a_2}$ are obtained only from the ``ultralocal'' part of the cumulants, namely (leaving again implicit the dependence on the Euclidean coordinates),
\begin{equation}
\begin{aligned}
\label{eq_expand_Gamma_2hatNUL1}
\widehat{A}_{k;a_1}[\left\lbrace \phi_a \right\rbrace ] =\Gamma_{k1}^{UL(2)}[\phi_{a_1}] - \beta \sum_{a_2}\Gamma_{k2}^{UL(20)}[\phi_{a_1},\phi_{a_2}]+ \cdots,
\end{aligned}
\end{equation}
\begin{equation}
\begin{aligned}
\label{eq_expand_Gamma_2tildeNUL}
\widetilde{C}_{k;a_1a_2}[\left\lbrace \phi_a \right\rbrace ]=-&\Gamma_{k2}^{UL(11)}[\phi_{a_1},\phi_{a_2}]\\& + \beta \sum_{a_3}\Gamma_{k3}^{UL(110)}[\phi_{a_1},\phi_{a_2},\phi_{a_3}]+ \cdots,
\end{aligned}
\end{equation}
whereas $\widehat{B}_{k;a_1}$, whose precise expression is not particularly illuminating at this point, involves the ``non-ultralocal'' part of the cumulants. As a result, $\widehat{B}_{k;a_1}$ is equal to zero when the effective average action is purely ``ultralocal'' in the Grassmann subspace. The demonstration of the above property and of Eqs.~(\ref{eq_hatGamma2_unifNUL}-\ref{eq_expand_Gamma_2tildeNUL}) is provided in Appendix~\ref{appendix:nonultralocal}.

The full propagator $\mathcal P_{k;(a_1\underline{\omega}_1)(a_2 \underline{\omega}_2)}$, which is the inverse of $\Gamma_k^{(2)} +\mathcal R_k$, has the same structure as in Eqs.~(\ref{eq_hatGamma2_unifNUL},\ref{eq_tildeGamma2_unifNUL}) with
\begin{equation}
\begin{split}
\label{eq_hatP0_unifNUL}
\widehat {\mathcal P}_{k;a_1\underline{\omega}_1 \underline{\omega}_2}[ \{ \phi_a\}]& =(1+ \bar{\omega}_1 \omega_1)\delta_{\underline{\omega}_1 \underline{\omega}_2}\widehat {P}_{k;a_1}[ \{ \phi_a\}] \\&+ (1+\bar \omega_1 \omega_2+\bar \omega_2 \omega_1)\widehat {Q}_{k;a_1}[ \{ \phi_a\}]
\end{split}
\end{equation}
and
\begin{equation}
\label{eq_tildeP0_unif}
\widetilde {\mathcal P}_{k;(a_1\underline{\omega}_1)(a_2 \underline{\omega}_2)}[ \{\phi_a\}]=(1+ \bar{\omega}_1 \omega_1) (1+ \bar{\omega}_2 \omega_2)\widetilde {P}_{k;a_1a_2}[ \{\phi_a\}],
\end{equation}
where $\widehat {\mathcal P}_{k;a}$ and $\widetilde {\mathcal P}_{k;a_1a_2}$ are expressed only in terms of  ``ultralocal'' cumulants whereas $\widehat {Q}_{k;a_1}$ also involves ``non-ultralocal'' terms and therefore vanishes when the latter go to zero. All these quantities can be expanded in increasing number of sums over copies. The zeroth-order components of the propagator, $\widehat {\mathcal P}_{k}^{[0]}$, $\widetilde{P}_{k}^{[0]}$ and $\widehat{Q}_{k}^{[0]}$, are then expressed as
\begin{equation}
\label{eq_hatP_zero_NUL}
\widehat {P}_{k}^{[0]}[\phi ]=\left(\Gamma _{k1}^{UL(2)}[ \phi ]+\widehat R_k\right) ^{-1},
\end{equation}
which denotes an inversion in the sense of operators in Euclidean space,
\begin{equation}
\label{eq_tildeP_zero_NUL}
\widetilde {P}_{k}^{[0]}[\phi_1, \phi_2 ]= \widehat {P}_{k}^{[0]}[ \phi_1 ](\Gamma _{k2}^{UL(11)}[\phi_1, \phi_2 ]-\widetilde R_k ) \widehat {P}_{k}^{[0]}[ \phi_2 ],
\end{equation}
and
\begin{equation}
\begin{split}
\label{eq_hatQ_zero_NUL}
&\widehat{Q}_{k}^{[0]}[\phi ]=\\&-\left(\Gamma _{k1}^{UL(2)}[ \phi ]+\widehat R_k\right) ^{-1}\widehat B_k^{[0]}[\phi]\left(\Gamma _{k1}^{UL(2)}[ \phi ]-\widehat B_k^{[0]}[\phi]+\widehat R_k\right) ^{-1}
\end{split}
\end{equation}
where the Euclidean indices are omitted for simplicity. Note the (expected) correspondence between the above expressions for $\widehat {P}_{k}^{[0]}$ and $\widetilde {P}_{k}^{[0]}$ and those derived under the hypothesis of ``Grassmannian ultralocality'', Eqs.~(\ref{eq_hatP_zero}) and (\ref{eq_tildeP_zero}).

Reinstalling the explicit dependence on the Euclidean momenta, we find the following \textit{exact} RG flow equations for the ``ultralocal'' parts of the cumulants,
\begin{equation}
\label{eq_flow_Gamma1_scaledUL}
\begin{split}
&\partial_t \Gamma_{k 1}^{UL(\beta)}\left[\phi_1\right ]=-
\dfrac{1}{2} \tilde{\partial}_t \int_{qq'} \widehat{P}_{k;qq'}^{[0]}[\phi_1]  \bigg( \Gamma_{k2;q,q'}^{UL(11)}\left[\phi_1,\phi_1\right ] \\&- \delta^{(d)}(q-q') \widetilde{R}_{k}(q^2)\bigg)+\dfrac{1}{2\beta}\int_{q}\partial_t \widehat{R}_k(q^2)\widehat{Q}_{k;-qq}^{[0]}[\phi_1],
\end{split}
\end{equation}
\begin{equation}
\label{eq_flow_Gamma2_final_scaledUL}
\begin{split}
&\partial_t\Gamma_{k2}^{UL(\beta)}\left[\phi_1,\phi_2\right ]=\dfrac{1}{2} \tilde{\partial}_t \int_{qq'}\bigg \{\widetilde{P}_{k;qq'}^{[0]}\left[\phi_1,\phi_1\right ] \Gamma_{k2;qq',.}^{UL(20)}\left[\phi_1,\phi_2\right ]\\&- \widehat{P}_{k;qq'}^{[0]}\left[\phi_1\right ] \Gamma_{k3;q,.,q'}^{UL(101)}\left[\phi_1,\phi_2,\phi_1\right ] + \frac{1}{2}\widetilde{P}_{k;qq'}^{[0]}\left[\phi_1,\phi_2\right ] \times \\& \left( \Gamma_{k2;q,q'}^{UL(11)}\left[\phi_1,\phi_2\right ] - (2\pi)^{d}\delta^{(d)}(q-q') \widetilde{R}_{k}(q^2)\right) + perm(12)\bigg \}\\&+\dfrac{1}{2\beta} \tilde{\partial}_t \int_{qq'}\bigg \{\widehat{Q}_{k;qq'}^{[0]}\left[\phi_1\right ] \Gamma_{k2;q'q,.}^{UL(20)}\left[\phi_1,\phi_2\right] + perm(12)\bigg \},
\end{split}
\end{equation}
and similarly for the higher orders.

Setting $1/\beta=0$ in the above ERGE's allows one to get rid of the terms that involve the ``non-ultralocal'' contributions to the effective average action. The  $\beta \rightarrow \infty$ limit therefore coincides with the RG equations for the cumulants obtained under the hypothesis of ``Grassmannian ultralocality'' (compare with section VI-D). This shows that the ERGE's derived under the assumption of ``Grassmannian ultralocality'' describe the renormalization of the RFIM at equilibrium with a proper selection of the ground state. This is an important piece in the resolution of the problem associated with the long-distance physics of the model.

\subsection{Illustration of the RG flow for ``non-ultralocal'' contributions}

We now illustrate the structure of the FRG flow for the ``non-ultralocal'' components of the cumulants by looking at the lowest-order correction to the first cumulant, assuming that all other cumulants are purely ``ultralocal''. This allows us to verify that this flow is well-behaved, thereby giving a direct confirmation to the arguments used in the previous subsection.
More specifically, we consider 
\begin{equation}
\begin{aligned}
\label{eq_cumulant_1_nonultralocal_correction}          
&\mathsf \Gamma_{k1}[\Phi_{1}] = \beta \int_{\underline{\omega}_1}
\bigg (\Gamma_{k1}^{UL}[\widetilde \Phi_1(\underline{\omega}_1)]\\&+\frac{1}{2} Y_{k}[\widetilde \Phi_1(\underline{\omega}_1)](1+ \bar \omega_1 \omega_1)\partial_{ \omega_1}\widetilde \Phi_1(\underline{\omega}_1)\partial_{\bar \omega_1}\widetilde \Phi_1(\underline{\omega}_1) \bigg ),
\end{aligned}
\end{equation}
and for $p \geq 2$,
\begin{equation}
\begin{aligned}
\label{eq_cumulant_p_ultralocal_correction}          
\mathsf \Gamma_{kp}[\Phi_{1},...,\Phi_{p}] = \beta^p\int_{\underline{\omega}_1}...\int_{\underline{\omega}_p}
\Gamma_{kp}^{UL}[\widetilde \Phi_1(\underline{\omega}_1),...,\widetilde \Phi_p(\underline{\omega}_p)],
\end{aligned}
\end{equation}
where we have omitted the superscript $(\beta)$.

The FRG equations for the ``ultralocal'' components are the same as in Eqs.~(\ref{eq_flow_Gamma1_scaledUL},\ref{eq_flow_Gamma2_final_scaledUL}) and the propagators $\widehat P_k^{[0]}[\phi]$, $\widetilde P_k^{[0]}[\phi_1,\phi_2]$ and $\widehat Q_k^{[0]}[\phi]$ are given in eqs.~(\ref{eq_hatP_zero_NUL}-\ref{eq_hatQ_zero_NUL}) with $\widehat B_k^{[0]}[\phi]$ now simply equal to
\begin{equation}
\label{eq_tildeB_Y}
\widehat B_k^{[0]}[\phi]=-Y_k[\phi].
\end{equation}
The most convenient way to obtain  the flow of $Y_k[\phi]$ is to differentiate twice the ERGE for $\mathsf \Gamma_{k1}$ in Eq.~(\ref{eq_flow_Gamma1_scaled})  and evalutate the resulting expression for a superfield configuration that is uniform in the Grassmann subspace (\textit{i.e.} $\Phi=\phi$). 
The final RG equation for $Y_k[\phi]$ reads
\begin{equation}
\begin{aligned}
\label{eq_flow_correction}
&\partial_t Y_k\left[\phi \right]=
\tilde \partial_t \mathrm{Tr} \bigg\{\widetilde{P}_{k}^{[0]}\left[\phi,\phi \right] \bigg (\frac{1}{2} Y_k^{(2)}\left[\phi \right]+Y_k^{(1)}\left[\phi \right]^2 \times \\&
\big [\widehat Q_k^{[0]}[\phi]-\widehat{P}_{k}^{[0]}\left[\phi \right]\big ] \bigg )
+2 \big [\widehat Q_k^{[0]}[\phi]-\widehat{P}_{k}^{[0]}\left[\phi \right]\big ]\bigg (\widetilde{P}_{k}^{[0]}\left[\phi,\phi \right] \times \\&
\Gamma_{k1}^{UL(3)}\left[\phi \right ]- \widehat{P}_{k}^{[0]}\left[\phi \right]  \Gamma_{k2}^{UL(21)}\left[\phi,\phi \right]\bigg )Y_k^{(1)}\left[\phi \right]
+\widehat Q_k^{[0]}[\phi] \times \\&
\bigg (\widetilde{P}_{k}^{[0]}\left[\phi,\phi \right ]\Gamma_{k1}^{UL(3)}\left[\phi \right ]^2+\Gamma_{k2}^{UL(22)}\left[\phi,\phi \right]-2\widehat{P}_{k}^{[0]}\left[\phi \right]\times \\&
\Gamma_{k1}^{UL(3)}\left[\phi \right ] \Gamma_{k2}^{UL(21)}\left[\phi,\phi \right]\bigg )\bigg \}
+\frac{1}{\beta} \tilde \partial_t \mathrm{Tr} \bigg\{ \frac{1}{2} Y_k^{(2)}\left[\phi \right]\widehat Q_k^{[0]}[\phi]\\&
+ \frac{1}{4} Y_k^{(1)}\left[\phi \right]^2\big [\widehat Q_k^{[0]}[\phi]-\widehat{P}_{k}^{[0]}\left[\phi \right]\big ]\big [7\widehat Q_k^{[0]}[\phi]-3\widehat{P}_{k}^{[0]}\left[\phi \right]\big ] \\&
+2 Y_k^{(1)}\left[\phi \right]\widehat Q_k^{[0]}[\phi]\big [\widehat Q_k^{[0]}[\phi]-\widehat{P}_{k}^{[0]}\left[\phi \right]\big ]\Gamma_{k1}^{UL(3)}\left[\phi \right]
+\\&
\frac{1}{2} \widehat Q_k^{[0]}\left[\phi \right]^2\Gamma_{k1}^{UL(3)}\left[\phi \right]^2 \bigg \},
\end{aligned}
\end{equation} 
where $\mathrm{Tr}$ denotes a trace over Euclidean momenta (which are left implicit in the right-hand side of the equation). 

Provided that $Y_k[\phi]$ converges to a finite (nondiverging) fixed-point value, its contribution to the flow of the ``ultralocal'' components of the cumulants, which appears through the propagator $\widehat Q_k^{[0]}[\phi]$ in Eqs.~(\ref{eq_flow_Gamma1_scaledUL}) and (\ref{eq_flow_Gamma2_final_scaledUL}),  can indeed be neglected when $\beta \rightarrow \infty$. This will be shown in the companion paper. (More generally, it can be shown that the contribution becomes subdominant at long distance, \textit{i.e.} when $k \rightarrow 0$, even when $\beta$ is large but finite.)

\subsection{Scaling dimensions near a zero-temperature fixed point and asymptotic dominance of the ground state}

To search for the fixed point that controls the critical behavior of the RFIM associated with the spontaneous breaking of the (global) $Z_2$ symmetry, the NP-FRG flow equations must first be recast in a scaled form. This can be done by introducing appropriate scaling dimensions.\cite{tarjus04,tarjus08} Near a zero-temperature fixed point, it is convenient to introduce a ``renormalized temperature''. As shown in Refs.~[\onlinecite{tarjus04},\onlinecite{tarjus08}], this can be done by considering (i) the strength of the renormalized random field $\Delta_k$, which can be defined from the vertex $\Gamma_{k2}^{(11)}$ evaluated for a specific configuration of the fields and reduces to the bare value $\Delta_B$ at the UV scale $\Lambda$, and (ii) the amplitude of the field renormalization constant $Z_k$, which is as usual obtained from $\Gamma_{k1}^{(2)}$ for a specific field configuration and is equal to $1$ at the UV scale. Specifically,
\begin{equation}
\label{eq_running_temperature}
T_k\propto \frac{Z_{k}  k^2}{ \Delta_{k}}.
\end{equation}
An associated running exponent $\theta_k$ is defined from
\begin{equation}
\label{eq_running_theta}
\theta_k = \partial_t \log T_k
\end{equation}
whereas the (running) anomalous dimension $\eta_k$ is obtained as
\begin{equation}
\label{eq_running_eta}
\eta_k= -\partial_t \log Z_{k}.
\end{equation}
One may also introduce a running exponent $\bar{\eta}_k=2-\theta_k+\eta_k$ from
\begin{equation}
\label{eq_running_etabar}
\bar{\eta}_k-2\eta_k= \partial_t \log \Delta_{k}.
\end{equation}
Note that the two anomalous dimensions $\eta$ and $\bar \eta$ describe the spatial decay of the pair ``connected'' and ``disconnected'' correlations functions (see section \ref{sec:ultralocal}) at
criticality; this translates into 
\begin{equation}
\begin{aligned}
\widehat P_k^{[0]}(q=0) \sim k^{-(2-\eta)}\, , \; \widetilde P_k^{[0]}(q=0) \sim k^{-(4-\bar \eta)}\, ,
\end{aligned}
\end{equation}
with $\eta \leq \bar \eta \leq 2 \eta$\cite{nattermann98,soffer85}.

A systematic way to proceed is to introduce, on top of the canonical scaling dimensions for the effective average action and the related cumulants and of a rescaling of the Euclidean momenta and coordinates ($\hat q=q/k$, $\hat x=kx$), a rescaling of the Grassmann coordinates via a renormalized curvature $\beta_k \propto 1/T_k$, with $T_k$ defined in Eq.~(\ref{eq_running_temperature}):
\begin{equation}
  \label{eq_grassmann_rescaling}
\widehat\theta=\left (\frac{\beta}{\beta_k}\right )^{\frac{1}{2}} \theta \; ,  \widehat{\bar\theta}=\left (\frac{\beta}{\beta_k}\right )^{\frac{1}{2}} \bar \theta 
\end{equation}
where
\begin{equation}
\label{eq_running_curvature}
\beta_k=\beta \frac{ \Delta_{k}/\Delta_B}{Z_{k}  (k/\Lambda)^2}.
\end{equation}
The renormalized curvature reduces to the (bare) curvature $\beta$ at the UV scale $\Lambda$ and diverges as $k^{-\theta}$ in the IR. It is worth pointing out that a symmetry between the rescaling of the Euclidean and of the Grassmann coordinates exists in the case where the dimensional reduction applies: then, $\theta=2$ so that all coordinates are rescaled by the same factor $k=k^{\theta/2}$.

We can now introduce dimensionless quantities. The dimensionless superfield is defined through
\begin{equation}
  \label{eq_superfield_dimension}
 \Phi(x,\underline \theta)=\left( \frac{\beta_k k^{d-2}}{\beta Z_{k}}\right)^{1/2}\Phi_{ren}(\hat x, \widehat{\underline \theta}),
\end{equation}
which implies for the components:
\begin{equation}
  \label{eq_field_dimension}
 \phi(x)=\left( \frac{\beta_k k^{d-2}}{\beta Z_{k}}\right)^{1/2}\varphi(\hat x),
\end{equation}
\begin{equation}
  \label{eq_response_field_dimension}
 \hat \phi(x)=\left( \frac{k^{d-2}}{Z_{k} \beta_k}\right)^{1/2}\hat \varphi(\hat x),
\end{equation}
etc.

The ``ultralocal'' component of the $p$th cumulant is rescaled as
\begin{equation}
\label{eq_cumulant_dimension}
\Gamma_{kp}^{UL}[\phi_1,...,\phi_p]= k^d(\beta_k)^p\, \gamma_{kp}^{UL}[\varphi_1,...,\varphi_p] ,
\end{equation}
and  the ``non-ultralocal'' cumulant can similarly be put in a dimensionless form . For instance, the term $\widehat B_k^{[0]}[\phi]$ appearing in the definition of the propagator $\widehat Q_k^{[0]}[\phi]$ is expressed as
\begin{equation}
\label{eq_NUL_dimension}
\widehat B_k^{[0]}[\phi]=\beta_k Z_k k^2 \hat b_k[\varphi].
\end{equation}
The dimensionless quantities will be systematically denoted by lower-case letters (except for the superfield and for the coordinates, for obvious reasons).

In addition, and following the discussion in section VI-B, the cutoff functions are chosen according to
\begin{equation}
\label{eq_scaled_hatR}
\widehat{R}_k(q^2)=Z_k k^2 s(\hat q^2),
\end{equation}
with $s(x)$ such that $s(0)>0$ and $s(x\rightarrow \infty)=0$ (see the companion paper and Refs.~[\onlinecite{tarjus04},\onlinecite{tarjus08}]), and
\begin{equation}
\label{eq_scaled_wideR}
\widetilde{R}_k(q^2)=-\Delta_k s'(\hat q^2),
\end{equation}
with $s'(x)$ the derivative of $s(x)$; the above form of the regulator then satisfies the superrotational invariance whenever $\Delta_k/Z_k=\Delta_B$.

We focus here on the ``ultralocal'' components of the cumulants. The ERGE's for the latter can then be put in a scaled form which has the following generic structure:
\begin{equation}
 \label{eq_ERGE_dimensionlessform_p}
\begin{split}
\partial_t \gamma_{kp}^{UL}&[\varphi_1,...,\varphi_p] =- p (2+ \eta_k -\bar \eta_k) \gamma_{kp}^{UL}[\varphi_1,...,\varphi_p]+ \\& \frac{1}{2}\sum_{a=1}^p(d-4+\bar\eta_k)\int_{\hat q_a} \varphi_a(\hat q_a) \gamma_{kp;\hat q_a}^{UL(1)}[\varphi_1,...,\varphi_p] \\&+ \mathcal F_{\gamma_p}^{UL}[\varphi_1,...,\varphi_p]+\frac{1}{\beta_k}\mathcal F_{\gamma_p}^{NUL}[\varphi_1,...,\varphi_p],
\end{split}
\end{equation}
where $\mathcal F_{\gamma_p}^{UL}$ is a dimensionless beta-functional expressed only in terms of (dimensionless) ``ultralocal'' components and $\mathcal F_{\gamma_p}^{NUL}$ is a dimensionless beta-functional that contains  (dimensionless) ``non-ultralocal'' components. Explicit expressions for the first cumulants are easily derived from Eqs.~(\ref{eq_flow_Gamma1_scaledUL},\ref{eq_flow_Gamma2_final_scaledUL}) and will be given in the following paper. Here, only the general structure is needed. 

Fixed points are then found by setting the left-hand side of Eq.~(\ref{eq_ERGE_dimensionlessform_p}) to zero. This will be studied in detail in the companion paper. At this point we would only like to stress that the ``non-ultralocal'' contributions to the flow of the dimensionless renormalized cumulants come with a factor $1/\beta_k$. Even if $\beta^{-1}$ is not taken equal to zero, and provided the ``non-ultralocal'' contributions remain bounded (see following paper), the flow leads to a fixed point characterized by the property of ``Grassmannian ultralocality'': one indeed expects $1/\beta_k\propto T_k$ to flow to zero, \textit{i.e.} $\theta_{k\rightarrow 0 }>0$, as already shown in computer studies\cite{nattermann98} and in Refs.~[\onlinecite{tissier06},\onlinecite{tissier08}]. Ground-state dominance is thus found asymptotically as the flow goes to the zero-temperature fixed point. Guided by our previous work\cite{tissier06,tissier08} and by that of Balents, Ledoussal and coworkers on the random manifold model,\cite{BLchauve00,BLbalents04,BLbalents05,BLledoussal10} we anticipate that when the fixed point is characterized by a nonanalytic, cusp-like, dependence of the cumulants of the renormalized random field, the approach to the $k\rightarrow 0$ limit (which coincides to the fixed point obtained by setting rightaway $\beta^{-1}= 0$ in the ERGE's) involves a boundary layer, generically in $\vert\varphi_a - \varphi_b \vert /T_k$.\cite{footnote2} This boundary layer is physically associated with the presence of  rare, power-law distributed, ``droplet'' excitations above the ground state at the running scale $k$.\cite{footnote3}

\section{Conclusion}
\label{sec:conclusion}

In this paper we have extended our nonperturbative FRG approach of disordered systems, which was presented in the previous articles of this series.\cite{tarjus08,tissier08} The objective was to discuss the property of dimensional reduction and its breakdown in the RFIM from the Parisi-Sourlas\cite{parisi79} perspective of an underlying supersymmetry and its breaking. To this end, we have reformulated the FRG in a superfield formalism.

We have identified two sources of problems in the Parisi-Sourlas supersymmetric formalism. One stems from the presence of metastable states which, due to the resulting multiplicity of solutions to the stochastic field equation that describes the long-distance physics of the RFIM, prevents the selection of the ground state in the random generating functional. The other one was already discussed in our previous papers and comes from the fact that a single copy of the original system is considered. As such, the formalism is therefore unable to describe rare events, such as avalanches and droplets, that manifest themselves as nonanalyticities in the field dependence of the cumulants of the renormalized disorder.

We have provided ways to cure both problems, through the introduction of a weighting factor with an auxiliary temperature and through the use of multiple copies. The resulting theory involves superfields in a curved superspace whose curvature is related to the inverse of the auxiliary temperature. The presence of metastable states leads to a breakdown of a formal property which we have called ``Grassmannian ultralocality'' and is associated with the fact that a unique solution is incorporated in the random generating functional. On the other hand, as will be discussed in detail in the companion paper, nonanalyticities originating from the presence of avalanches in the ground state as one varies the applied source trigger a spontaneous breaking of supersymmetry, more precisely of ``superrotational invariance''.

Through the introduction of an appropriate infrared regulator which guarantees that the stochastic field equation has a unique solution at the initial condition of the flow and which can be chosen to satisfy all (super) symmetries of the theory, we have derived exact RG flow equations for the cumulants of the renormalized disorder in the effective average action formalism. We have shown that in the limit of infinite curvature, \textit{i.e.} of zero auxiliary temperature, the hierarchy of exact FRG equations coincides with that obtained under the hypothesis of ``Grassmannian ultralocality''. Through this procedure, the ground state is properly selected and is the only solution that contributes to the random generating functional. We have moreover found that the corrections to ``ultralocality'' that are present for a finite curvature become subdominant at long distance as one approaches the expected zero-temperature fixed point.

In the following paper, we investigate the exact FRG equations derived for the cumulants in the limit of infinite curvature and show that nonanalytic behavior of the effective average action in its field arguments is related to the appearance of a spontaneous breaking of the superrotational invariance. We also devise and study a nonperturbative approximation scheme that does not break explicitly the superrotational invariance and allows us to study the critical behavior of the RFIM and the breakdown of dimensional reduction as a function of the spatial dimension $d$.

\appendix

\section{``Metastable'' states and nonanalyticities in the zero-dimensional RFIM}
\label{appendix:toy}

In section~\ref{sec:ultralocality}, we have introduced the property of ``Grassmannian ultralocality'' by considering the simple case of the $d=0$ RFIM at zero temperature. We use this same toy model here to illustrate on the one hand the different proposed ways to construct generating functionals when ``metastable'' states are present and on the other hand the nonanalyticities that may be present in the field (or source) dependence of these functionals and of the associated Green's functions.
In the $d=0$ RFIM, a variable $\varphi$ is submitted
to an interaction characterized by the action:
\begin{equation}
S(\varphi)=U_B(\varphi)-(J+h)\varphi,
\end{equation}
with, as in Eq.~(\ref{eq_ham_dis}),
\begin{equation}
U_B(\varphi)=\frac{\tau}{2}\varphi^2+\frac u{4!}\varphi^4
\end{equation}
and the random field is Gaussian distributed with variance $\Delta_B$.

In the case where $\tau$ is negative, the ``stochastic field equation'' which reduces to  $U_B'(\varphi)= J+h$ has  three branches of solutions:

(1) $\phi_-(J+h)$, which exists for $J+h<J_c$ with $J_c=\frac{2\vert \tau \vert}3\sqrt{\frac{2\vert \tau \vert}{u}}$,

(2) $\phi_+(J+h)$, which exists for $J+h>-J_c$, and 

(3) $\phi_\ddag(J+h)$, which exists for $-J_c<J+h<J_c$.

In the region $-J_c<J+h<J_c$, the equation has then three solutions, two minima, $\phi_-$ and $\phi_+$, and a maximum, $\phi_\ddag(J+h)$. This is a situation where one encounters multiple solutions or ``metastable states''.

We focus on a ``Green's function'' which, when a single solution is present and all constructions lead to the same results, is obtained from the second derivative of the second cumulant of generating function $\mathcal W(\hat J, J)$, namely [compare with Eq.~(\ref{eq_physical_disconn})],
\begin{equation}
\begin{split}
&\overline{\langle\varphi(J_1+h)\rangle\langle \varphi(J_2+h)\rangle}\\&-\overline{\langle\varphi(J_1+h)\rangle}\;\, \overline{\langle\varphi(J_2+h)\rangle}.
\end{split}
\end{equation}
At $T=0$, the thermal average $\langle\varphi\rangle$ reduces to the ground state $\phi_0$, \textit{i.e.} the absolute minimum of the action. In the Parisi-Sourlas supersymmetric construction, $\langle\varphi\rangle$ is replaced by $\phi_++\phi_--\phi_\ddag$ (in the range where the three solutions coexit) and in the construction proposed in this work (see section III-B), $\langle\varphi\rangle$ is replaced by a  Boltzmann-like weighted average with an auxiliary temperature $\beta^{-1}$.

As we want to study the possible nonanalyticities in the dependence on the sources $J_1,J_2$ as $J_2\rightarrow J_1$, we rewrite $J_1=J-\delta J$, $J_2=J+\delta J$ and consider the limit $\delta J\to 0$. Moreover, we consider the difference between the Green's functions
\begin{equation}
\label{eq_green_cusp}
\overline{\phi_*(J+\delta J+h)\phi_*(J-\delta J+h)}-\overline{\phi_*(J+h)^2},
\end{equation}
where we have dropped the difference in the disconnected pieces that is easily shown to be at least of $\mathcal O(\delta J^2)$ and $\phi_*$ is given by any of the three above definitions (ground state, Parisi-Sourlas weighting, Boltzmann-like weighting). The  expression in Eq.~(\ref{eq_green_cusp}) is symmetric in the inversion $\delta J\rightarrow -\delta J$ and so, without loss of generality, we choose $\delta J >0$ in the following.

\subsubsection{Ground state}

If $\phi_*$ is taken as the gound state, it then  presents a discontinuity at $J+h=0$ (being equal to $\phi_-$ when $J+h<0$ and to $\phi_+$ when $J+h >0$). It is 
easy to show that $\phi_*(J+\delta J+h)\phi_*(J-\delta
J+h)-\phi_*(J+h)^2$ is of order $\mathcal O(\delta J^2)$
everywhere except in a region centered around $J+h=0$ and of width $2\delta J$, where the function
is of order 1 (actually $-12\vert \tau \vert/u$). 
Averaging over the disorder therefore generates a  linear cusp in
$\delta J$,
\begin{equation}
\begin{split}
&\overline{\phi_*(J+\delta J+h)\phi_*(J-\delta J+h)}-\overline{\phi_*(J+h)^2}=\\&-\frac{24}{\sqrt{2\pi}u\Delta_B}\exp(-\frac{J^2}{2\Delta_B})\vert \delta J \vert +\mathcal O(\delta J^2).
\end{split}
\end{equation}

\subsubsection{Parisi-Sourlas weighting}

If we use the prescription that $\phi_*=\phi_++\phi_--\phi_\ddag$ when the three solutions coexist (\textit{i.e.} for $-J_c<J+h<J_c$) and $\phi_*$ is given by the unique solution otherwise, the
function $\phi_*(J+h)$ is now continuous but presents a nonanalytic
behavior around $\pm J_c$. When approaching $J_c$ from below, one has 
$\phi_*(J)\simeq \phi_+(J_c)-A\sqrt{J_c-J}+\mathcal O(J-J_c)$ with
$A$ a positive expression that is not worth to be given here. A similar behavior is observed when $J$ approaches $-J_c$ from above. It is now easy to
show that $\phi_++\phi_--\phi_\ddag$ goes to zero as $\mathcal
O(\delta J^2)$ everywhere except in regions of width $\delta J$
around $J+h=\pm J_c$. In these regions, the function behaves as $\mathcal
O(\sqrt{\delta J})$. After averaging over disorder, one therefore
obtains a singular behavior
\begin{equation}
\begin{split}
\overline{\phi_*(J+\delta J+h)\phi_*(J-\delta J+h)}-\overline{\phi_*(J+h)^2}\propto \vert \delta J\vert^{3/2},
\end{split}
\end{equation}
as $\delta J \rightarrow 0$. However, this nonanalytic behavior originates from branches that are  not
absolute minima of the action and should therefore be discarded in studying equilibrium behavior. It is an artifact of the Parisi-Sourlas construction.

\subsubsection{Boltzmann-like weighting}

In this case, one has $\phi_*=(\phi_+ e^{-\beta S_+}+\phi_- e^{-\beta
  S_-}-\phi_\ddag e^{-\beta S_\ddag})/(e^{-\beta S_+}+ e^{-\beta S_-}-
e^{-\beta S_\ddag})$ where $S_+=S(\phi_+)$, etc, when the three solutions coexist and $\phi_*$ is given by the unique solution otherwise. As in the Parisi-Sourlas weighting, $\phi_*(J+h)$ is continuous with a nonanalytic behavior around $\pm J_c$. In this case however, the resulting nonanalyticity in 
$\vert \delta J\vert^{3/2}$ in the difference of Green's functions, Eq.~(\ref{eq_green_cusp}), is exponentially damped by a factor
$\exp(-\beta\Delta S)$ where $\Delta S$ is the difference of action
between the ground state and the first excited state. 

The behavior of $\phi_*$ around $J+h=0$ is continuous and regular, but with a rapid change of $\mathcal O(1)$ over a narrow interval of width $\mathcal O(\beta^{-1})$. After averaging over disorder, this gives rise to a ``boundary layer'' in $\beta \vert \delta J\vert$, \textit{i.e.},
\begin{equation}
\begin{split}
&\overline{\phi_*(J+\delta J+h)\phi_*(J-\delta J+h)}-\overline{\phi_*(J+h)^2} = \\& \beta^{-1}f(J,\beta \vert \delta J\vert)+ \mathcal O(e^{-\beta \Delta S}\vert \delta J\vert^{3/2}, \delta J^2),
\end{split}
\end{equation}
with $f(J,y)= f_2(J) y^2+ \cdots$ when $y \rightarrow 0$ and $f(J,y) \sim f_c(J) y$ when $y \rightarrow \infty$.

When $\beta \rightarrow \infty$, the spurious nonanalyticity in $\vert \delta J\vert^{3/2}$ vanishes exponentially fast whereas the boundary layer leads to a linear cusp in $\vert \delta J\vert$. One exactly recovers ground-state dominance [and it is easy to check that $f_c(J)=-\frac{24}{(\sqrt{2\pi}u\Delta_B)}\exp(-J^2/(2\Delta_B))$].

\section{Algebraic manipulations for the expansion in number of sums over copies}
\label{appendix:expansion_copies}

In this appendix we provide a brief survey of some of the algebraic manipulations that are associated with the expansion in increasing number of sums over copies. We recall that the $p$th term of such an expansion involves exactly $p$ copies and is a symmetric and continuous functional of either $p$ copy superfields or $p$ copy supersources (depending on whether one works with superfields or supersources as fundamental variables), the functional form being independent of the total number of copies (which however must be larger than $p$ to give access to the full functional dependence).

For instance, to relate the terms of the expansion of the effective action $\Gamma[\{\Phi_a\}]$ [we omit the superscript $(\beta)$ throughout this appendix] to those of the expansion of $W[\{\mathcal J_a\}]$, \textit{i.e.} the cumulant of $\mathcal W_h$, one can use a pedestrian method that starts with the Legendre transform in Eq.~(\ref{eq_legendre_multicopy}). Keeping the superfields $\{\Phi_{a}\}$ as variables, we also expand $-\mathcal W[\{\mathcal{J}_a\}]+ \sum_a\ \int_{\underline{x}}  \mathcal{J}_a(\underline{x})  \Phi_{a}(\underline{x})$ in number of sums over copies as
\begin{equation}
\begin{aligned}
\label{eq_multicopy3_app}
&\sum_a \bigg(-\mathcal W_1[\mathcal{J}_a[\{\Phi_{f}\}]] + \int_{\underline{\theta}} \mathcal{J}_a(\underline{\theta};\{\Phi_{f}\}) \Phi_{a}(\underline{\theta})  \bigg)
-  \\& \sum_{p\geq2} \frac {1}{p!} \sum_{a_1=1}^n..\sum_{a_p =1}^n
\mathcal W_{p}[\mathcal{J}_{a_1}[\{\Phi_{f}\}],..,\mathcal{J}_{a_p}[\{\Phi_{f}\}]],
\end{aligned}
\end{equation}
where the supersource has the following expansion:
\begin{equation}
\begin{aligned}
\label{eq_multicopy2_app}
&(1+\beta \bar \theta \theta) \mathcal J_{a}(\underline{\theta};\{\Phi_f\}) = \mathsf \Gamma_{1,\underline{\theta}}^{(1)}[\Phi_{a}] + \sum_{p\geq1} \sum_{a_1=1}^n...\sum_{a_p =1}^n \frac{(-1)^{p}}{(p+1)!} \\&  \bigg( 
\mathsf \Gamma_{p+1,\underline{\theta}..}^{(10..0)}[\Phi_{a},\Phi_{a_1},..,\Phi_{a_p}]+ .. +\mathsf \Gamma_{p+1,..\underline{\theta}}^{(0..01)}[\Phi_{a_1},..,\Phi_{a_p},\Phi_{a}]\bigg).
\end{aligned}
\end{equation}
Here, we have dropped the explicit mention of the dependence on the Euclidean coordinates to keep only that on the Grassmann ones.

We next proceed to a term-by-term identification of both sides of Eq.~(\ref{eq_legendre_multicopy}) and we symmetrize the output if necessary. The first orders are easily derived and are given in Eqs.~(\ref{legendre_Gamma_1}-\ref{eq_cumG3}). Higher-order terms are similarly derived. One finds that the $p$th term $\mathsf \Gamma_p[\Phi_{a_1},..,\Phi_{a_p}]$ is equal to  $(-1)^{p}W_p[ \mathcal J[\Phi_{a_1}],..,\mathcal J[\Phi_{a_p}]]$, where $\mathcal J[\Phi_{a}]$ is the nonrandom supersource defined in Eq.~(\ref{eq_nonrandom_source_curved}),  plus terms involving cumulants $W_q$ with $q<p$ that are also functionals of the nonrandom supersource.

We also have to consider the expansion in increasing number of sums over copies of matrices, as in section VI-C. A generic matrix $A_{(a_1 \underline x_1)(a_2 \underline x_2)}[\left\lbrace \Phi_a\right\rbrace ]$ can then be expressed as in Eqs.~(\ref{eq_genericdecompos},\ref{eq_widehatA},\ref{eq_widetildeA}). Algebraic manipulations on such matrices can again be performed through a term-by-term identification of the orders of the expansions. For instance, the inverse $ \mathcal B=`` {\mathcal A}^{-1}"$ of the matrix $ \mathcal A$ in the sense of Eq.~(\ref{eq_legendre_invert}), \textit{i.e.}
\begin{equation}
\label{eq_invert_A}
\begin{aligned}
&(1+\beta \bar\theta_1 \theta_1) \delta_{a_1a_2}\, \delta_{\underline x_1\underline x_2}=\sum_{a_3}\int_{\underline x_3}(1-2\beta\bar\theta_3\theta_3)\\&
\times \mathcal B_{(a_1 \underline x_1)(a_3 \underline x_3)}[\left\lbrace \Phi_a\right\rbrace ]\mathcal A_{(a_3 \underline x_3)(a_2 \underline x_2)}[\left\lbrace \Phi_a\right\rbrace ]  ,
\end{aligned}
\end{equation}
can also be put in the form of Eq.~(\ref{eq_genericdecompos}) and its components, $\widehat{\mathcal B}_{a_1}$ and $\widetilde{\mathcal B}_{a_1a_2}$, can be expanded in number of sums over copies. The term-by-term identification of both sides of Eq.~(\ref{eq_invert_A}) leads to a unique expression of the various orders,  $\widehat{\mathcal B}^{[p]}$and $\widetilde{\mathcal B}^{[p]}$, of the expansion of $\mathcal B$  in terms of the $\widehat{\mathcal A}^{[q]}$'s and $\widetilde{\mathcal A}^{[q]}$'s with $q\leq p$. The algebra becomes rapidly tedious, but the first few terms are easily derived. $\widehat{\mathcal B}^{[0]}$ is the inverse of $\widehat{\mathcal A}^{[0]}$, \textit{i.e.},
\begin{equation}
\label{eq_widehatB_0}
\int_{\underline{x}_3}(1-2\beta \bar{\theta}_3 \theta_3) \widehat{\mathcal B}_{\underline x_1\underline x_3}^{[0]}[\Phi_1] \widehat{\mathcal A}_{\underline x_3\underline x_2}^{[0]}[  \Phi_1]=(1+\beta \bar{\theta}_1 \theta_1)\delta_{\underline x_1\underline x_2},
\end{equation}
whereas
\begin{equation}
\begin{split}
\label{eq_widetildeB_0}
\widetilde{\mathcal B}^{[0]}_{\underline x_1\underline x_2}[  \Phi_1,  \Phi_2] =& - \int_{\underline{x}_3}\int_{\underline{x}_4} (1-2\beta \bar{\theta}_3 \theta_3) (1-2\beta \bar{\theta}_4 \theta_4)\\& \widehat{\mathcal B}_{\underline x_1\underline x_3}^{[0]}[ \Phi_1] \widetilde{\mathcal A}_{\underline x_3\underline x_4}^{[0]}[  \Phi_1, \Phi_2] \widehat{\mathcal B}_{\underline x_4\underline x_2}^{[0]}[  \Phi_2]
\end{split}
\end{equation}
and
\begin{equation}
\begin{split}
\label{eq_widehatB_1}
\widehat{\mathcal B}^{[1]}[  \Phi_1 \vert  \Phi_2] =&  - \int_{\underline{x}_3}\int_{\underline{x}_4} (1-2\beta \bar{\theta}_3 \theta_3) (1-2\beta \bar{\theta}_4 \theta_4)\\&\widehat{\mathcal B}_{\underline x_1\underline x_3}^{[0]}[  \Phi_1]\widehat{\mathcal A}_{\underline x_3\underline x_4}^{[1]}[  \Phi_1\vert  \Phi_2] \widehat{\mathcal B}_{\underline x_4\underline x_2}^{[0]}[  \Phi_1],
\end{split}
\end{equation}
\begin{equation}
\label{eq_widetildeB_1}
\begin{split}
\widetilde{\mathcal B} ^{[1]}[  \Phi_1 &,  \Phi_2 \vert  \Phi_3]  =  - \int_{\underline{x}_3}\int_{\underline{x}_4} (1-2\beta \bar{\theta}_3 \theta_3) (1-2\beta \bar{\theta}_4 \theta_4)\\& \bigg\{\widehat{\mathcal B}_{\underline x_1\underline x_3}^{[0]}[  \Phi_1] \widetilde{\mathcal A}_{\underline x_3\underline x_4}^{[1]}[  \Phi_1, \Phi_2 \vert  \Phi_3] \\& + \widetilde{\mathcal A}_{\underline x_1\underline x_3}^{[0]}[  \Phi_1, \Phi_3]\widetilde{\mathcal B}_{\underline x_3\underline x_4}^{[0]}[  \Phi_3, \Phi_2] \\&+ \widehat{\mathcal A}_{\underline x_1\underline x_3}^{[1]}[  \Phi_1\vert  \Phi_3] \widetilde{\mathcal B}_{\underline x_3\underline x_4}^{[0]}[  \Phi_1, \Phi_2] \\& + \widetilde{\mathcal A}_{\underline x_1\underline x_3}^{[0]}[  \Phi_1, \Phi_2] \widehat{\mathcal B}_{\underline x_3\underline x_4}^{[1]}[  \Phi_2\vert  \Phi_3]\bigg\} \widehat{\mathcal B}_{\underline x_4\underline x_2}^{[0]}[  \Phi_2],
\end{split}
\end{equation}
etc. These equations allow us to relate the expansion of the full propagator to that of the effective average action and to derive the ERGE's for the cumulants in section VI-C.

\section{A graphical representation of the hierarchy of ERGE's for
  the cumulants }
\label{appendix:graphical}

In this appendix we give a graphical representation of the flow equations
for the cumulants $\Gamma_{kp}$. We concentrate on the replica
structure and remain general on the way the momentum and Grassmann
coordinates must be integrated over in the loops. Similarly, we use a notation with ``simple'' fields but superfields could be used as well. In doing so, we obtain
results which apply to the curved space, to the ``ultralocal'' case (when curvature drops out), 
but also in the replica formalism of papers I and II.\cite{tarjus08,tissier08}

It is convenient to represent the first cumulant by a filled circle
\begin{equation}
\raisebox{-1mm}{\begin{picture}(0,0)%
\includegraphics{gamma_1.pstex}%
\end{picture}%
\setlength{\unitlength}{3947sp}%
\begingroup\makeatletter\ifx\SetFigFont\undefined%
\gdef\SetFigFont#1#2#3#4#5{%
  \reset@font\fontsize{#1}{#2pt}%
  \fontfamily{#3}\fontseries{#4}\fontshape{#5}%
  \selectfont}%
\fi\endgroup%
\begin{picture}(296,165)(3511,-2401)
\put(3526,-2386){\makebox(0,0)[lb]{\smash{{\SetFigFont{12}{14.4}{\rmdefault}{\mddefault}{\updefault}{\color[rgb]{0,0,0}1}%
}}}}
\end{picture}%
} \qquad \to \qquad \Gamma_{k1}[\phi_1],
\end{equation}
the second cumulant by two filled circles joined by a dashed line
\begin{equation}
\raisebox{-3mm}{\input gamma_2.pstex_t} \qquad \to \qquad \Gamma_{k2}[\phi_1,\phi_2]
\end{equation}
the third cumulant by three filled circles joined by dashed lines
\begin{equation}
\raisebox{-3mm}{ \input gamma_3.pstex_t} \qquad \to \qquad \Gamma_{k3}[\phi_1,\phi_2,\phi_3]
\end{equation}
etc. The rule is that each circle corresponds to one copy (or replica), \textit{i.e.}, to one field $\phi_a$. We represent the zeroth-order ``hat'' propagator by a line:
\begin{equation}
 \begin{picture}(0,0)%
\includegraphics{prop.pstex}%
\end{picture}%
\setlength{\unitlength}{3947sp}%
\begingroup\makeatletter\ifx\SetFigFont\undefined%
\gdef\SetFigFont#1#2#3#4#5{%
  \reset@font\fontsize{#1}{#2pt}%
  \fontfamily{#3}\fontseries{#4}\fontshape{#5}%
  \selectfont}%
\fi\endgroup%
\begin{picture}(549,24)(2389,-1423)
\end{picture}%
 \qquad \to \qquad \widehat P_{\underline{x}_1,\underline{x}_2}[\phi].
\end{equation}

The vertices are represented by extracting legs from the cirles. For
the first cumulant, the graphical rule is as usual
\begin{equation}
\raisebox{-8mm}{ \input gamma_1q.pstex_t }\qquad \to \qquad \Gamma_{k1,\underline{x}_1\cdots \underline{x}_q}^{(q)}[\phi_1]
\end{equation}
For the second cumulant, we must specify on which circle we extract
the leg. We represent the vertex $\Gamma_{k2,\underline{x}_1\cdots
  \underline{x}_{q_1},\underline{y}_1\cdots \underline{y}_{q_2}}^{(q_1,q_2)}[\phi_1,\phi_2]$ with $q_1$ legs
in the circle associated with the copy field $\phi_1$ and $q_2$ legs in
the circle associated with the copy field $\phi_2$. Graphically:
\begin{equation}
  \raisebox{-13mm}{ \input gamma_2q1q2.pstex_t} \qquad \to \qquad \Gamma_{k2,\underline{x}_1\cdots
    \underline{x}_{q_1},\underline{y}_1\cdots \underline{y}_{q_2}}^{(q1,q2)}[\phi_1,\phi_2]
\end{equation}
Some caution must be taken in the particular case of 
$\Gamma_{k2}^{(1,1)}$ as it always comes with the cutoff function
$\widetilde R_k$. We therefore use a special graphical rule in this case:
\begin{equation}
\begin{split}
\raisebox{0mm}{ \begin{picture}(0,0)%
\includegraphics{gamma_211.pstex}%
\end{picture}%
\setlength{\unitlength}{3947sp}%
\begingroup\makeatletter\ifx\SetFigFont\undefined%
\gdef\SetFigFont#1#2#3#4#5{%
  \reset@font\fontsize{#1}{#2pt}%
  \fontfamily{#3}\fontseries{#4}\fontshape{#5}%
  \selectfont}%
\fi\endgroup%
\begin{picture}(624,110)(3439,-2216)
\end{picture}%
} \quad \to & \qquad
\Gamma_{k2,\underline{x}_1,\underline{x}_2}^{(1,1)}[\phi_1,\phi_2]\\&-(1+\beta\bar\theta_1\theta_1)(1+\beta\bar\theta_2\theta_2)\widetilde R_k(x-y).
\end{split}
\end{equation}

Finally, when needed, we encode the time derivative of the regulating
functions in the following way:
\begin{equation}
  \raisebox{-0mm}{ \begin{picture}(0,0)%
\includegraphics{dtrh.pstex}%
\end{picture}%
\setlength{\unitlength}{3947sp}%
\begingroup\makeatletter\ifx\SetFigFont\undefined%
\gdef\SetFigFont#1#2#3#4#5{%
  \reset@font\fontsize{#1}{#2pt}%
  \fontfamily{#3}\fontseries{#4}\fontshape{#5}%
  \selectfont}%
\fi\endgroup%
\begin{picture}(174,174)(3664,-2248)
\end{picture}%
} \qquad \to \qquad
  \partial_t\widehat R
\end{equation}
\begin{equation}
  \raisebox{-4mm}{ \input dtrt.pstex_t} \qquad \to \qquad
  -\partial_t \widetilde R
\end{equation}

Let us now derive the graphical representation for the inverse of
$\Gamma^{(2)}+\mathcal R$. To this end, we first write
$\Gamma^{(2)}+\mathcal R$ as
\begin{equation}
  \begin{split}
\Gamma_k^{(2)}+\mathcal R=&({\begin{picture}(0,0)%
\includegraphics{prop.pstex}%
\end{picture}%
\setlength{\unitlength}{3947sp}%
\begingroup\makeatletter\ifx\SetFigFont\undefined%
\gdef\SetFigFont#1#2#3#4#5{%
  \reset@font\fontsize{#1}{#2pt}%
  \fontfamily{#3}\fontseries{#4}\fontshape{#5}%
  \selectfont}%
\fi\endgroup%
\begin{picture}(549,24)(2389,-1423)
\end{picture}%
 })^{-1}-\begin{picture}(0,0)%
\includegraphics{gamma_220.pstex}%
\end{picture}%
\setlength{\unitlength}{3947sp}%
\begingroup\makeatletter\ifx\SetFigFont\undefined%
\gdef\SetFigFont#1#2#3#4#5{%
  \reset@font\fontsize{#1}{#2pt}%
  \fontfamily{#3}\fontseries{#4}\fontshape{#5}%
  \selectfont}%
\fi\endgroup%
\begin{picture}(324,410)(3439,-2216)
\end{picture}%
 -\begin{picture}(0,0)%
\includegraphics{gamma_211.pstex}%
\end{picture}%
\setlength{\unitlength}{3947sp}%
\begingroup\makeatletter\ifx\SetFigFont\undefined%
\gdef\SetFigFont#1#2#3#4#5{%
  \reset@font\fontsize{#1}{#2pt}%
  \fontfamily{#3}\fontseries{#4}\fontshape{#5}%
  \selectfont}%
\fi\endgroup%
\begin{picture}(624,110)(3439,-2216)
\end{picture}%
 \\&+\frac 12\begin{picture}(0,0)%
\includegraphics{gamma_3200.pstex}%
\end{picture}%
\setlength{\unitlength}{3947sp}%
\begingroup\makeatletter\ifx\SetFigFont\undefined%
\gdef\SetFigFont#1#2#3#4#5{%
  \reset@font\fontsize{#1}{#2pt}%
  \fontfamily{#3}\fontseries{#4}\fontshape{#5}%
  \selectfont}%
\fi\endgroup%
\begin{picture}(372,335)(3565,-2366)
\end{picture}%
 +\begin{picture}(0,0)%
\includegraphics{gamma_3110.pstex}%
\end{picture}%
\setlength{\unitlength}{3947sp}%
\begingroup\makeatletter\ifx\SetFigFont\undefined%
\gdef\SetFigFont#1#2#3#4#5{%
  \reset@font\fontsize{#1}{#2pt}%
  \fontfamily{#3}\fontseries{#4}\fontshape{#5}%
  \selectfont}%
\fi\endgroup%
\begin{picture}(474,335)(3514,-2741)
\end{picture}%
    +\dots
  \end{split}
\end{equation}
In this expression, the external legs are amputated, {\it i.e.}, do
not include propagators.  We can now use the inversion formula
$(A-B)^{-1}=A^{-1}+A^{-1}BA^{-1}+A^{-1}BA^{-1}BA^{-1}+\dots$
graphically:
\begin{equation}
\begin{split}
\Big(&\Gamma^{(2)}+\mathcal R\Big)^{-1}=\underbrace{\begin{picture}(0,0)%
\includegraphics{prop.pstex}%
\end{picture}%
\setlength{\unitlength}{3947sp}%
\begingroup\makeatletter\ifx\SetFigFont\undefined%
\gdef\SetFigFont#1#2#3#4#5{%
  \reset@font\fontsize{#1}{#2pt}%
  \fontfamily{#3}\fontseries{#4}\fontshape{#5}%
  \selectfont}%
\fi\endgroup%
\begin{picture}(549,24)(2389,-1423)
\end{picture}%
 }_{\widehat P_0}  +\underbrace{\begin{picture}(0,0)%
\includegraphics{pg211p.pstex}%
\end{picture}%
\setlength{\unitlength}{3947sp}%
\begingroup\makeatletter\ifx\SetFigFont\undefined%
\gdef\SetFigFont#1#2#3#4#5{%
  \reset@font\fontsize{#1}{#2pt}%
  \fontfamily{#3}\fontseries{#4}\fontshape{#5}%
  \selectfont}%
\fi\endgroup%
\begin{picture}(849,110)(3364,-2216)
\end{picture}%

}_{\widetilde P_1} +\underbrace{\begin{picture}(0,0)%
\includegraphics{pg220p.pstex}%
\end{picture}%
\setlength{\unitlength}{3947sp}%
\begingroup\makeatletter\ifx\SetFigFont\undefined%
\gdef\SetFigFont#1#2#3#4#5{%
  \reset@font\fontsize{#1}{#2pt}%
  \fontfamily{#3}\fontseries{#4}\fontshape{#5}%
  \selectfont}%
\fi\endgroup%
\begin{picture}(474,410)(3364,-2216)
\end{picture}%
 }_{\widehat
  P_1}\\&+\underbrace{\begin{picture}(0,0)%
\includegraphics{pg220pg211p.pstex}%
\end{picture}%
\setlength{\unitlength}{3947sp}%
\begingroup\makeatletter\ifx\SetFigFont\undefined%
\gdef\SetFigFont#1#2#3#4#5{%
  \reset@font\fontsize{#1}{#2pt}%
  \fontfamily{#3}\fontseries{#4}\fontshape{#5}%
  \selectfont}%
\fi\endgroup%
\begin{picture}(774,410)(3739,-2216)
\end{picture}%
 +\begin{picture}(0,0)%
\includegraphics{pg211pg220p.pstex}%
\end{picture}%
\setlength{\unitlength}{3947sp}%
\begingroup\makeatletter\ifx\SetFigFont\undefined%
\gdef\SetFigFont#1#2#3#4#5{%
  \reset@font\fontsize{#1}{#2pt}%
  \fontfamily{#3}\fontseries{#4}\fontshape{#5}%
  \selectfont}%
\fi\endgroup%
\begin{picture}(774,410)(2989,-2216)
\end{picture}%
 +\begin{picture}(0,0)%
\includegraphics{pg211pg211p.pstex}%
\end{picture}%
\setlength{\unitlength}{3947sp}%
\begingroup\makeatletter\ifx\SetFigFont\undefined%
\gdef\SetFigFont#1#2#3#4#5{%
  \reset@font\fontsize{#1}{#2pt}%
  \fontfamily{#3}\fontseries{#4}\fontshape{#5}%
  \selectfont}%
\fi\endgroup%
\begin{picture}(1074,110)(2989,-2216)
\end{picture}%
 -\begin{picture}(0,0)%
\includegraphics{pg3110p.pstex}%
\end{picture}%
\setlength{\unitlength}{3947sp}%
\begingroup\makeatletter\ifx\SetFigFont\undefined%
\gdef\SetFigFont#1#2#3#4#5{%
  \reset@font\fontsize{#1}{#2pt}%
  \fontfamily{#3}\fontseries{#4}\fontshape{#5}%
  \selectfont}%
\fi\endgroup%
\begin{picture}(624,335)(3439,-2741)
\end{picture}%
 }_{\widetilde P_2}\\&+ \underbrace{\begin{picture}(0,0)%
\includegraphics{pg220pg220p.pstex}%
\end{picture}%
\setlength{\unitlength}{3947sp}%
\begingroup\makeatletter\ifx\SetFigFont\undefined%
\gdef\SetFigFont#1#2#3#4#5{%
  \reset@font\fontsize{#1}{#2pt}%
  \fontfamily{#3}\fontseries{#4}\fontshape{#5}%
  \selectfont}%
\fi\endgroup%
\begin{picture}(549,410)(3739,-2216)
\end{picture}%
 -\frac 12 \begin{picture}(0,0)%
\includegraphics{pg320p.pstex}%
\end{picture}%
\setlength{\unitlength}{3947sp}%
\begingroup\makeatletter\ifx\SetFigFont\undefined%
\gdef\SetFigFont#1#2#3#4#5{%
  \reset@font\fontsize{#1}{#2pt}%
  \fontfamily{#3}\fontseries{#4}\fontshape{#5}%
  \selectfont}%
\fi\endgroup%
\begin{picture}(474,335)(3514,-2366)
\end{picture}%
  }_{\widehat P_2}+\dots
\end{split}
\end{equation}
In these diagrams the external legs now come with propagators.

We now easily obtain the flow equation for $\Gamma$ by contracting the previous expression with $\partial_t \mathcal R$. We readily find
\begin{equation}
  \begin{split}
\partial_t\Gamma[\{\phi_a\}]=\frac 12\bigg(&\raisebox{-3mm}{\begin{picture}(0,0)%
\includegraphics{dtg1c.pstex}%
\end{picture}%
\setlength{\unitlength}{3947sp}%
\begingroup\makeatletter\ifx\SetFigFont\undefined%
\gdef\SetFigFont#1#2#3#4#5{%
  \reset@font\fontsize{#1}{#2pt}%
  \fontfamily{#3}\fontseries{#4}\fontshape{#5}%
  \selectfont}%
\fi\endgroup%
\begin{picture}(316,394)(3293,-2318)
\end{picture}%
}-\raisebox{-3mm}{\begin{picture}(0,0)%
\includegraphics{dtg1a1.pstex}%
\end{picture}%
\setlength{\unitlength}{3947sp}%
\begingroup\makeatletter\ifx\SetFigFont\undefined%
\gdef\SetFigFont#1#2#3#4#5{%
  \reset@font\fontsize{#1}{#2pt}%
  \fontfamily{#3}\fontseries{#4}\fontshape{#5}%
  \selectfont}%
\fi\endgroup%
\begin{picture}(292,410)(3664,-2366)
\end{picture}%
 }+ \raisebox{-3mm}{\begin{picture}(0,0)%
\includegraphics{dtg1a2.pstex}%
\end{picture}%
\setlength{\unitlength}{3947sp}%
\begingroup\makeatletter\ifx\SetFigFont\undefined%
\gdef\SetFigFont#1#2#3#4#5{%
  \reset@font\fontsize{#1}{#2pt}%
  \fontfamily{#3}\fontseries{#4}\fontshape{#5}%
  \selectfont}%
\fi\endgroup%
\begin{picture}(339,410)(3695,-2366)
\end{picture}%
 }+2\raisebox{-3mm}{\begin{picture}(0,0)%
\includegraphics{dtg2c1.pstex}%
\end{picture}%
\setlength{\unitlength}{3947sp}%
\begingroup\makeatletter\ifx\SetFigFont\undefined%
\gdef\SetFigFont#1#2#3#4#5{%
  \reset@font\fontsize{#1}{#2pt}%
  \fontfamily{#3}\fontseries{#4}\fontshape{#5}%
  \selectfont}%
\fi\endgroup%
\begin{picture}(412,442)(3695,-2366)
\end{picture}%
}-\raisebox{-3mm}{\begin{picture}(0,0)%
\includegraphics{dtg2c2.pstex}%
\end{picture}%
\setlength{\unitlength}{3947sp}%
\begingroup\makeatletter\ifx\SetFigFont\undefined%
\gdef\SetFigFont#1#2#3#4#5{%
  \reset@font\fontsize{#1}{#2pt}%
  \fontfamily{#3}\fontseries{#4}\fontshape{#5}%
  \selectfont}%
\fi\endgroup%
\begin{picture}(441,410)(3695,-2366)
\end{picture}%
}\\&+ \raisebox{-3mm}{\begin{picture}(0,0)%
\includegraphics{dtg2e1.pstex}%
\end{picture}%
\setlength{\unitlength}{3947sp}%
\begingroup\makeatletter\ifx\SetFigFont\undefined%
\gdef\SetFigFont#1#2#3#4#5{%
  \reset@font\fontsize{#1}{#2pt}%
  \fontfamily{#3}\fontseries{#4}\fontshape{#5}%
  \selectfont}%
\fi\endgroup%
\begin{picture}(412,442)(3695,-2366)
\end{picture}%
}- \raisebox{-3mm}{\begin{picture}(0,0)%
\includegraphics{dtg2e2.pstex}%
\end{picture}%
\setlength{\unitlength}{3947sp}%
\begingroup\makeatletter\ifx\SetFigFont\undefined%
\gdef\SetFigFont#1#2#3#4#5{%
  \reset@font\fontsize{#1}{#2pt}%
  \fontfamily{#3}\fontseries{#4}\fontshape{#5}%
  \selectfont}%
\fi\endgroup%
\begin{picture}(443,410)(3664,-2366)
\end{picture}%
}- \raisebox{-3mm}{\begin{picture}(0,0)%
\includegraphics{dtg2a1.pstex}%
\end{picture}%
\setlength{\unitlength}{3947sp}%
\begingroup\makeatletter\ifx\SetFigFont\undefined%
\gdef\SetFigFont#1#2#3#4#5{%
  \reset@font\fontsize{#1}{#2pt}%
  \fontfamily{#3}\fontseries{#4}\fontshape{#5}%
  \selectfont}%
\fi\endgroup%
\begin{picture}(372,517)(3565,-2366)
\end{picture}%
}+\raisebox{-5mm}{\begin{picture}(0,0)%
\includegraphics{dtg2f1.pstex}%
\end{picture}%
\setlength{\unitlength}{3947sp}%
\begingroup\makeatletter\ifx\SetFigFont\undefined%
\gdef\SetFigFont#1#2#3#4#5{%
  \reset@font\fontsize{#1}{#2pt}%
  \fontfamily{#3}\fontseries{#4}\fontshape{#5}%
  \selectfont}%
\fi\endgroup%
\begin{picture}(316,742)(3593,-2366)
\end{picture}%
}\bigg)+\dots
  \end{split}
\end{equation}
We are now in a position to write the flow equations for the cumulants
by retaining the diagram with 1 copy, 2 copies, etc, and appropriately 
symmetrizing the results. These expressions simplify if we rewrite
them in terms of the $\widetilde \partial_t$ operator. 
Recall that the operation $\tilde\partial_t$ corresponds to a
derivation of the $t$ dependence of the cutoff functions
only. When performing this operation, we must use the rules
\begin{equation}
  \tilde \partial_t\quad\begin{picture}(0,0)%
\includegraphics{prop.pstex}%
\end{picture}%
\setlength{\unitlength}{3947sp}%
\begingroup\makeatletter\ifx\SetFigFont\undefined%
\gdef\SetFigFont#1#2#3#4#5{%
  \reset@font\fontsize{#1}{#2pt}%
  \fontfamily{#3}\fontseries{#4}\fontshape{#5}%
  \selectfont}%
\fi\endgroup%
\begin{picture}(549,24)(2389,-1423)
\end{picture}%
\quad =-\quad  \raisebox{-1mm}{\begin{picture}(0,0)%
\includegraphics{dtprop.pstex}%
\end{picture}%
\setlength{\unitlength}{3947sp}%
\begingroup\makeatletter\ifx\SetFigFont\undefined%
\gdef\SetFigFont#1#2#3#4#5{%
  \reset@font\fontsize{#1}{#2pt}%
  \fontfamily{#3}\fontseries{#4}\fontshape{#5}%
  \selectfont}%
\fi\endgroup%
\begin{picture}(624,174)(3439,-2248)
\end{picture}%
}
\end{equation}
\begin{equation}
  \tilde \partial_t\quad\raisebox{0mm}{\begin{picture}(0,0)%
\includegraphics{gamma_211.pstex}%
\end{picture}%
\setlength{\unitlength}{3947sp}%
\begingroup\makeatletter\ifx\SetFigFont\undefined%
\gdef\SetFigFont#1#2#3#4#5{%
  \reset@font\fontsize{#1}{#2pt}%
  \fontfamily{#3}\fontseries{#4}\fontshape{#5}%
  \selectfont}%
\fi\endgroup%
\begin{picture}(624,110)(3439,-2216)
\end{picture}%
}\quad =\quad\raisebox{0mm}{\begin{picture}(0,0)%
\includegraphics{gamma_211c.pstex}%
\end{picture}%
\setlength{\unitlength}{3947sp}%
\begingroup\makeatletter\ifx\SetFigFont\undefined%
\gdef\SetFigFont#1#2#3#4#5{%
  \reset@font\fontsize{#1}{#2pt}%
  \fontfamily{#3}\fontseries{#4}\fontshape{#5}%
  \selectfont}%
\fi\endgroup%
\begin{picture}(624,174)(3439,-2248)
\end{picture}%
}\; ,
\end{equation}
This leads to
\begin{equation}
\label{eq_diagg1}  
\partial_t\Gamma_1[\phi]=-\frac 12\tilde \partial_{ t}\quad
    \raisebox{-3mm}{\begin{picture}(0,0)%
\includegraphics{dtg1a.pstex}%
\end{picture}%
\setlength{\unitlength}{3947sp}%
\begingroup\makeatletter\ifx\SetFigFont\undefined%
\gdef\SetFigFont#1#2#3#4#5{%
  \reset@font\fontsize{#1}{#2pt}%
  \fontfamily{#3}\fontseries{#4}\fontshape{#5}%
  \selectfont}%
\fi\endgroup%
\begin{picture}(261,410)(3695,-2366)
\end{picture}%
}+\frac 12 \quad \raisebox{-2mm}{\begin{picture}(0,0)%
\includegraphics{dtg1c.pstex}%
\end{picture}%
\setlength{\unitlength}{3947sp}%
\begingroup\makeatletter\ifx\SetFigFont\undefined%
\gdef\SetFigFont#1#2#3#4#5{%
  \reset@font\fontsize{#1}{#2pt}%
  \fontfamily{#3}\fontseries{#4}\fontshape{#5}%
  \selectfont}%
\fi\endgroup%
\begin{picture}(316,394)(3293,-2318)
\end{picture}%
} \, ,
 \end{equation}
\begin{equation}
\label{eq_diagg2}
\begin{split}
\partial_t&\Gamma_2[\phi_1,\phi_2]=\frac 12\tilde\partial_{ t}\bigg( \raisebox{-3mm}{\input dtg2c.pstex_t}+ \raisebox{-3mm}{\input dtg2d.pstex_t}+ \raisebox{-3mm}{\input dtg2e.pstex_t}\\&- \raisebox{-3mm}{\input dtg2a.pstex_t}- \raisebox{-2mm}{\input dtg2b.pstex_t}+\raisebox{-4mm}{\input dtg2f.pstex_t}
    +\raisebox{-4mm}{\input dtg2g.pstex_t}    \bigg)\, ,    
\end{split}
\end{equation}
and
\begin{equation}
\label{eq_diagg3}
\begin{split}
\partial_t\Gamma_3=\frac 12&\tilde\partial_{ t}\bigg(-2 \raisebox{-3mm}{\begin{picture}(0,0)%
\includegraphics{dtg3g.pstex}%
\end{picture}%
\setlength{\unitlength}{3947sp}%
\begingroup\makeatletter\ifx\SetFigFont\undefined%
\gdef\SetFigFont#1#2#3#4#5{%
  \reset@font\fontsize{#1}{#2pt}%
  \fontfamily{#3}\fontseries{#4}\fontshape{#5}%
  \selectfont}%
\fi\endgroup%
\begin{picture}(562,390)(3770,-2421)
\end{picture}%
}-6 \raisebox{-3mm}{\begin{picture}(0,0)%
\includegraphics{dtg3h.pstex}%
\end{picture}%
\setlength{\unitlength}{3947sp}%
\begingroup\makeatletter\ifx\SetFigFont\undefined%
\gdef\SetFigFont#1#2#3#4#5{%
  \reset@font\fontsize{#1}{#2pt}%
  \fontfamily{#3}\fontseries{#4}\fontshape{#5}%
  \selectfont}%
\fi\endgroup%
\begin{picture}(596,390)(3953,-2421)
\end{picture}%
}-6 \raisebox{-3mm}{\begin{picture}(0,0)%
\includegraphics{dtg3i.pstex}%
\end{picture}%
\setlength{\unitlength}{3947sp}%
\begingroup\makeatletter\ifx\SetFigFont\undefined%
\gdef\SetFigFont#1#2#3#4#5{%
  \reset@font\fontsize{#1}{#2pt}%
  \fontfamily{#3}\fontseries{#4}\fontshape{#5}%
  \selectfont}%
\fi\endgroup%
\begin{picture}(501,427)(3489,-2516)
\end{picture}%
}\\&+3 \raisebox{-3mm}{\begin{picture}(0,0)%
\includegraphics{dtg3c.pstex}%
\end{picture}%
\setlength{\unitlength}{3947sp}%
\begingroup\makeatletter\ifx\SetFigFont\undefined%
\gdef\SetFigFont#1#2#3#4#5{%
  \reset@font\fontsize{#1}{#2pt}%
  \fontfamily{#3}\fontseries{#4}\fontshape{#5}%
  \selectfont}%
\fi\endgroup%
\begin{picture}(412,485)(3545,-2516)
\end{picture}%
}+6\,\, \raisebox{-3mm}{\begin{picture}(0,0)%
\includegraphics{dtg3d.pstex}%
\end{picture}%
\setlength{\unitlength}{3947sp}%
\begingroup\makeatletter\ifx\SetFigFont\undefined%
\gdef\SetFigFont#1#2#3#4#5{%
  \reset@font\fontsize{#1}{#2pt}%
  \fontfamily{#3}\fontseries{#4}\fontshape{#5}%
  \selectfont}%
\fi\endgroup%
\begin{picture}(787,370)(3545,-2421)
\end{picture}%
}+6\raisebox{-3mm}{\begin{picture}(0,0)%
\includegraphics{dtg3e.pstex}%
\end{picture}%
\setlength{\unitlength}{3947sp}%
\begingroup\makeatletter\ifx\SetFigFont\undefined%
\gdef\SetFigFont#1#2#3#4#5{%
  \reset@font\fontsize{#1}{#2pt}%
  \fontfamily{#3}\fontseries{#4}\fontshape{#5}%
  \selectfont}%
\fi\endgroup%
\begin{picture}(562,370)(3770,-2421)
\end{picture}%
}
\\&-3\raisebox{-5mm}{\begin{picture}(0,0)%
\includegraphics{dtg3f.pstex}%
\end{picture}%
\setlength{\unitlength}{3947sp}%
\begingroup\makeatletter\ifx\SetFigFont\undefined%
\gdef\SetFigFont#1#2#3#4#5{%
  \reset@font\fontsize{#1}{#2pt}%
  \fontfamily{#3}\fontseries{#4}\fontshape{#5}%
  \selectfont}%
\fi\endgroup%
\begin{picture}(412,514)(3545,-2170)
\end{picture}%
}-3\, \raisebox{-3mm}{\begin{picture}(0,0)%
\includegraphics{dtg3a.pstex}%
\end{picture}%
\setlength{\unitlength}{3947sp}%
\begingroup\makeatletter\ifx\SetFigFont\undefined%
\gdef\SetFigFont#1#2#3#4#5{%
  \reset@font\fontsize{#1}{#2pt}%
  \fontfamily{#3}\fontseries{#4}\fontshape{#5}%
  \selectfont}%
\fi\endgroup%
\begin{picture}(985,314)(3272,-2618)
\end{picture}%
}+3\raisebox{-5mm}{\begin{picture}(0,0)%
\includegraphics{dtg3b.pstex}%
\end{picture}%
\setlength{\unitlength}{3947sp}%
\begingroup\makeatletter\ifx\SetFigFont\undefined%
\gdef\SetFigFont#1#2#3#4#5{%
  \reset@font\fontsize{#1}{#2pt}%
  \fontfamily{#3}\fontseries{#4}\fontshape{#5}%
  \selectfont}%
\fi\endgroup%
\begin{picture}(372,587)(3565,-2618)
\end{picture}%
}    \bigg) \, ,  
\end{split}
\end{equation}
where it is understood that the right-hand side of the above
expression should be symmetrized with respect to the copy fields.
Similar expressions can be derived for the higher cumulants.

In the case where ``ultralocality'' is satisfied (or imposed), the last
diagram of Eq.~(\ref{eq_diagg1}), the last two diagrams of
Eq.~(\ref{eq_diagg2}), and the last two of Eq.~(\ref{eq_diagg3}) give
no contribution because the propagator, which is proportional to
$\delta_{\underline\theta_1\underline\theta_2}$, is contracted with a
single copy, \textit{i.e.}, with
$\underline\theta_1=\underline\theta_2$, or because two propagators
connect the same copy, giving a contribution
$(\delta_{\underline\theta_1\underline\theta_2})^2=0$. Note however that
these diagrams give nonzero contributions if ``non-ultralocal'' terms are present in the action.

\section{``Non-ultralocal'' contributions; Grassmann structure of the 2-point vertex functions}
\label{appendix:nonultralocal}

We derive in this appendix the dependence of the 2-point vertex
functions on the Grassmann variables. We show in particular that
the ``ultralocal'' and ``non-ultralocal'' parts of the effective average action
lead to distinct Grassmannian structures.

\subsection{Consequences of the WT identities}

Let us start with $\widehat \Gamma^{(2)}_1$. We derive Eq.~(\ref{eq_ward_multicopycurved2_Gamma}) with respect to $\Phi_a(x,\underline \theta_1)$ and $\Phi_b(y,\underline \theta_2)$ and evaluate the result in a field configuration
uniform in the Grassmann variables, \textit{i.e.} $\Phi_e=\phi_e, \forall e \in \{1,...,n\}$. Focusing on the
part proportional to $\delta_{ab}$ and observing that the term with
Grassmannian derivatives of the field vanishes, we get the identity
\begin{equation}
\label{eq_ward_appendix}
\partial_{\theta_1}\left((1-\beta\bar \theta_1\theta_1)\widehat\Gamma^{(2)}_{a;\underline x_1,\underline x_2}\right)+\partial_{\theta_2}\left((1-\beta\bar \theta_2\theta_2)\widehat\Gamma^{(2)}_{a;\underline x_1,\underline x_2}\right)=0.
\end{equation}
By using the symmetry of the 2-point vertex function under permutation of the indices and
the fact that it has a vanishing number of ghosts, we can parametrize $\widehat \Gamma^{(2)}_{a}$ as
\begin{equation}
\begin{split}
\widehat \Gamma^{(2)}_{a;(\underline\theta_1 x_1),(\underline\theta_2 x_2)}&[\{\phi_e\}]=A_{a;x_1x_2}+B_{a;x_1x_2}(\bar\theta_1\theta_1+\bar\theta_2\theta_2)\\&+C_{a;x_1x_2}(\bar\theta_1\theta_2+\bar\theta_2\theta_1)+D_{a;x_1x_2}\bar\theta_1\theta_1\bar\theta_2\theta_2,
\end{split}
\end{equation}
where $A_{a;x_1x_2}$, $B_{a;x_1x_2}$, $C_{a;x_1x_2}$ and $D_{a;x_1x_2}$ are functionals of the $\phi_e$'s. Pluging this expression in Eq.~(\ref{eq_ward_appendix}) leads to a set of constraints,
\begin{align}
B_a+C_a&=\beta A_a\\
D_a&=\beta B_a,
\end{align}
where we have dropped the explicit dependence on the Euclidean coordinates. From the solution of the above equations, we derive the following form of the 2-point vertex function:
\begin{equation}
\label{eq_annex_sol1}
\widehat \Gamma^{(2)}_{a;\underline\theta_1,\underline\theta_2}[\{\phi_e\}]=A_a(1+\beta\bar\theta_1\theta_2+\beta\bar\theta_2\theta_1)+B_a(1+\beta \bar\theta_1\theta_1)\delta_{\underline\theta_1,\underline\theta_2}.
\end{equation}

The same method can be used to constrain the form of  $\widetilde
\Gamma^{(1,1)}_2[\{\phi_e\}]$. We now derive Eq.~(\ref{eq_ward_multicopycurved2_Gamma}) with respect to
$\Phi_a(x,\underline \theta_1)$ and $\Phi_b(y,\underline \theta_2)$
and evaluate the result in a field configuration $\Phi_e=\phi_e$. The part
that comes with no kronecker $\delta_{ab}$ satisfies
\begin{equation}
\label{eq_ward2_appendix}
\partial_{\theta_1}\left[(1-\beta\bar \theta_1\theta_1)\widetilde\Gamma^{(2)}_{(a\underline x_1),(b\underline x_2)}\right]=0.
\end{equation}
By using the symmetry under permutation of the 2-points vertex
function, we can parametrize $\widetilde\Gamma^{(2)}_{a,b}$ in the following manner:
\begin{equation}
\begin{split}
\widetilde\Gamma^{(11)}_{(a\underline\theta_1),(b\underline\theta_2)}[\{\phi_e\}]=&A_{ab}+B_{ab}\bar\theta_1\theta_1+B_{ba}\bar\theta_2\theta_2\\&+C_{ab}\bar\theta_1\theta_2+C_{ba}\bar\theta_2\theta_1+D_{ab}\bar\theta_1\theta_1\bar\theta_2\theta_2
\end{split}
\end{equation}
where we have again dropped the explicit dependence on the Euclidean coordinates. After inserting this parametrization in Eq.~(\ref{eq_ward2_appendix}), we end up
with a set of constraints that leads to the following form for the
2-point vertex:
\begin{equation}
\label{eq_annexe_sol2}
\widetilde\Gamma^{(11)}_{(a,\underline \theta_1),(b,\underline \theta_2)}[\{\phi_e\}]=A_{ab}(1+\beta\bar\theta_1\theta_1)(1+\beta\bar\theta_2\theta_2)
\end{equation}
with $A_{ab}=A_{ba}$. 

\subsection{Contribution of the ``ultralocal'' parts of the effective action to the 2-point vertices}

A generic ``ultralocal'' term with one copy can be written as
\begin{equation}
\sum_a\int_{\underline\theta}\int  \prod_{i=1}^p(d^dx_i\Phi_a(x_i,\underline\theta))R(\{x_i\})
\end{equation}
where the function $R$ is symmetric under permutation of any pair of
arguments.  The second derivative of the previous term with respect to
$\Phi_a(\underline x_1)$ and $\Phi_b(\underline x_2)$, evaluated for
$\Phi_e=\phi_e,  \forall e \in \{1,...,n\}$, yields a contribution to $\widehat\Gamma^{(2)}_{a;\underline\theta_1,\underline\theta_2}[\{\phi_e\}]$ of the form
\begin{equation}
\label{eq_gam2hat}p(p-1)(1+\beta\bar\theta_1\theta_1)\delta_{\underline\theta_1,\underline\theta_2}\int
\prod_{i=3}^p(d^dx_i\phi_a(x_i))R(\{x_i\})
\end{equation}
which contributes to $B_a$ in Eq.~(\ref{eq_annex_sol1}). 

Consider now a generic 2-copy ``ultralocal'' term,
\begin{equation}
\label{eq_ultra_2rep}
\begin{split}
\frac 12\sum_{ab}\int_{\underline\theta_1}\int_{\underline\theta_2}\int & \prod_{i=1}^q(d^dx_i\Phi_a(x_i,\underline\theta_1))\times\\&\prod_{j=1}^r(d^dy_j\Phi_b(y_j,\underline\theta_2))S(\{x_i\},\{y_j\})
\end{split}
\end{equation}
in the effective action, with $S$ a symmetric function under permutation of the arguments in curly brackets. After deriving with respect to $\Phi_a(x_1,\underline \theta_1)$ and $\Phi_b(y_1,\underline
\theta_2)$ and evaluating this derivative in a field configuration
$\Phi_e=\phi_e,  \forall e \in \{1,...,n\}$, we obtain a contribution to $\widetilde \Gamma^{(2)}_{(a,\underline \theta_1),(b,\underline \theta_2)}$ of the form:
\begin{equation}
\label{eq_gam2tilde}\begin{split}
q\, r(&1+\beta\bar\theta_1\theta_1)(1+\beta\bar\theta_2\theta_2)\times\\&\int  \prod_{i=2}^q(d^dx_i\phi_a(x_i))\prod_{j=2}^r(d^dy_j\phi_b(y_j))S(\{x_i\},\{y_j\})
\end{split}
\end{equation}
which contributes to $A_{ab}$ in Eq.~(\ref{eq_annexe_sol2}). 

It is easy to show that the generic ``ultralocal'' terms with two or
more copies lead to the same Grassmann structure for
$\widehat\Gamma^{(2)}$ as that found in Eq.~(\ref{eq_gam2hat}). For
instance, the generic 2-copy ``ultralocal'' term
(\ref{eq_ultra_2rep}) also contributes to $\widehat\Gamma^{(2)}$, with
the a Grassmann structure proportional to
$(1+\beta\bar\theta_1\theta_1)\delta_{\underline\theta_1,\underline\theta_2}$. Similarly,
generic ``ultralocal'' terms with three or more copies lead to the same
Grassmann structure for $\widetilde\Gamma^{(2)}$ as that found in
Eq.~(\ref{eq_gam2tilde}).

\subsection{Contribution of ``non-ultralocal'' parts of the effective action to the 2-point vertices}

We now add derivatives with respect to the Grassmann variables in the
effective action. No more than two fields can come with Grassmannian
derivatives, otherwise the derivative would vanish upon evaluating in
a uniform field configuration $\Phi_e=\phi_e,  \forall e \in \{1,...,n\}$ and the term would not
contribute to the 2-point vertex function (in such configurations). Without loss of
generality, we only need to consider the three following 1-copy terms in
the effective action:
\begin{equation}
\begin{split}
\sum_a&\int_{\underline\theta}\int
\left(\prod_{i=1}^p d^dx_i\Phi_a(x_i,\theta)\right)\int
d^dy\Delta_{\underline\theta}\Phi_a(y,\underline\theta)T(y,\{x_i\}),\\ 
\sum_a&\int_{\underline\theta}\int
\left(\prod_{i=1}^p d^dx_i\Phi_a(x_i,\theta)\right)\int
d^dy_1d^dy_2(1+\beta\bar\theta\theta)\times\\&\nonumber\partial_{\bar\theta}\Phi_a(y_1,\underline\theta)\partial_{\theta}\Phi_a(y_2,\underline\theta)U(\{y_1,y_2\},\{x_i\})
\end{split}
\end{equation}
and
\begin{equation}
\begin{split}
\sum_a&\int_{\underline\theta}\int
\left(\prod_{i=1}^p d^dx_i\Phi_a(x_i,\theta)\right)\int
d^dy_1d^dy_2\Delta_{\underline\theta}\Phi_a(y_1,\underline\theta)\times\\&\nonumber\Delta_{\underline\theta}\Phi_a(y_2,\underline\theta)V(\{y_1,y_2\},\{x_i\})),
\end{split}
\end{equation}
where $T,U,V$ are symmetric functions under permutation of the arguments in curly brackets. When deriving with respect to
$\Phi_a(y_1,\underline \theta_1)$ and $\Phi_b(y_2,\underline
\theta_2)$ and evaluating this derivative in a field configuration
$\{ \Phi_e=\phi_e\}$, these terms lead to the following contributions to $\widehat\Gamma^{(2)}_{a,\underline \theta_1 \underline \theta_2}[\{\phi_e\}]$:
\begin{align}
&4p(1+\beta\bar\theta_1\theta_2+\beta\bar\theta_2\theta_1)\int
\left(\prod_{i=2}^p d^dx_i\phi_a(x_i)\right)\times\\&\nonumber\qquad T(y_1,\{y_2,x_i\}),\\
&2(1+\beta\bar\theta_1\theta_2+\beta\bar\theta_2\theta_1)\int
\left(\prod_{i=1}^p d^dx_i\phi_a(x_i)\right)\times\\&\nonumber\qquad U(\{y_1,y_2\},\{x_i\})
\end{align}
and
\begin{align}
&-8\beta(1+\beta\bar\theta_1\theta_2+\beta\bar\theta_2\theta_1)\int
\left(\prod_{i=1}^p d^dx_i\phi_a(x_i)\right)\times\\&\nonumber\qquad V(\{y_1,y_2\},\{x_i\}),
\end{align}
all of them coming with the same Grassmann structure $(1+\beta\bar\theta_1\theta_2+\beta\bar\theta_2\theta_1)$.

Finally we study the contributions  to $\widetilde \Gamma^{(2)}_{(a,\underline\theta_1),(b,\underline\theta_2)}[\{\phi_e\}]$ of the ``non-ultralocal'' terms in the effective action. Only terms with
Grassmannian derivatives on at most two fields can contribute. Moreover,
in the case where two fields are derived, these derivatives must act on fields in
different copies. Consequently, we only need to consider two kinds of
terms:
\begin{align}
\sum_{ab}&\int_{\underline\theta_1}\int_{\underline\theta_2}\int
\left(\prod_{i=1}^q d^dx_i\Phi_a(x_i,\underline\theta_1)\right)\left(\prod_{j=1}^r d^dy_j\Phi_b(y_j,\underline\theta_2)\right)\times\\&\nonumber\int
d^dz\Delta_{\underline\theta_1}\Phi_a(z,\underline\theta_1)W(\{x_i\},\{y_j\},z)
\end{align}
and
\begin{align}
\nonumber \sum_{ab}&\int_{\underline\theta_1}\int_{\underline\theta_2}\int
\left(\prod_{i=1}^q d^dx_i\Phi_a(x_i,\underline\theta_1)\right)\left(\prod_{j=1}^r d^dy_j\Phi_b(y_j,\underline\theta_2)\right)\times\\&\nonumber
\int d^dz_1d^dz_2\Delta_{\underline\theta_1}\Phi_a(z_1,\underline\theta_1)\Delta_{\underline\theta_2}\Phi_b(z_2,\underline\theta_2)\times\\&\qquad X(\{x_i\},\{y_j\},z_1,z_2).
\end{align}
After deriving with respect to $\Phi_a(x_1,\underline\theta_1)$ and
$\Phi_b(y_1,\underline\theta_2)$ and evaluating in a uniform
configuration $\Phi_e=\phi_e,  \forall e \in \{1,...,n\}$, one can readily check that these terms give no
contribution to $\widetilde \Gamma^{(2)}$.

This concludes our proof that:
\begin{itemize}
\item ``ultralocal'' parts of the effective action
contribute to $\widehat \Gamma^{(2)}_{a,\underline\theta_1\underline\theta_2}$ by terms proportional to
$(1+\beta
\bar\theta_1\theta_1)\delta_{\underline\theta_1,\underline\theta_2}$
and to $\widetilde \Gamma^{(2)}_{(a,\underline\theta_1),(b \underline\theta_2)}$ by terms proportional to $(1+\beta
\bar\theta_1\theta_1)(1+\beta \bar\theta_2\theta_2)$.
\item ``non-ultralocal'' parts of the effective action contribute to
  $\widehat \Gamma^{(2)}_{a,\underline\theta_1 \underline\theta_2}$ by terms proportional to $(1+\beta
  \bar\theta_1\theta_2+\beta \bar\theta_2\theta_1)$ and do not
  contribute to $\widetilde \Gamma^{(2)}_{(a,\underline\theta_1),(b\underline\theta_2)}$.
\end{itemize}


\begin{thebibliography}{15}
\bibitem[{\citenamefont{Footnote}(01)}]{footnote01}
On the other hand, a ``flat'' weighting of all solutions corresponds to Eq.~(\ref{eq_generating_func1}) in which one keeps the absolute value for the determinant  and does not restrict the solutions. Then, when dropping the fermionic sources for simplicity, one finds
\begin{equation}
\label{eq_ultralocal_flat}
\mathcal W_h(\hat J,J)= \log \sum_{\alpha} e^{\hat J \sum_i \mathcal C_{\alpha,i}(h+J) \phi_{\alpha,i}(h+J)} ,
\end{equation}
which is  not ``ultralocal'' in the Grassmann coordinates. In this case, $\mathcal Z_h(\hat J=0,J)$ is equal to the total number of solutions and $\mathcal W_h(\hat J=0,J)$ is the associated ``complexity''.
\bibitem[{\citenamefont{Footnote}(1)}]{footnote1}
If $\mathcal R_{k,aa}$ is a function of the superlaplacian $\Delta_{SS}$ only, say $R_{k,aa}= f_k(\Delta_{SS})$, its Fourier transform with respect to both the Euclidean and the Grassmann coordinates is equal to
\begin{equation}
\label{eq_FT_SUSY_regulator}
\begin{split}
\mathcal R_{k,aa}(q,\eta, \bar{\eta}) &= f_k(q^2 + \Delta_B \eta \bar{\eta}) \\& = f_k(q^2) + \Delta_B \eta \bar{\eta} f_k'(q^2),
\end{split}
\end{equation}
where $q$ is the Euclidean momentum, $\eta, \bar{\eta}$ the Grassmann ``momenta'' conjugate to the coordinates $\bar \theta,\theta$, respectively, and a standard definition of the Fourier transform in Grassmann space has been used.\cite{zinnjustin89} On the other hand, following Eq.~(\ref{eq_replicatedregulator}), $\mathcal R_{k,aa}(\underline{x}_1,\underline{x}_2) = \delta_{\underline{\theta}_1 \underline{\theta}_2}\widehat{R}_k(|x_1-x_2|)+\widetilde{R}_k(|x_1-x_2|)$, so that Fourier transforming this equation gives
\begin{equation}
\label{eq_FT_regulator}
\mathcal R_{k,aa}(q,\eta, \bar{\eta}) = \widehat{R}_k(q^2) - \eta \bar{\eta} \widetilde{R}_k(q^2).
 \end{equation}
Combining Eqs.~(\ref{eq_FT_SUSY_regulator}) and (\ref{eq_FT_regulator}) then leads to the relation in Eq.~(\ref{eq_susy_regulator}).
\bibitem[{\citenamefont{Footnote}(3)}]{footnote3}
See for instance the detailed connection worked out in a series of studies by Balents and Le Doussal in the simpler case of the random-manifold model that describes elastic manifolds pinned by a disordered environment.\cite{BLbalents04,BLbalents05,BLledoussal10}.
\end{thebibliography}
\end{document}